\documentclass[]{elsarticle}

\usepackage[margin=1.0in]{geometry}
\usepackage{array}
\usepackage{subcaption}
\usepackage{booktabs}
\usepackage{multirow}
\usepackage{tabularx}

\newcolumntype{Y}{>{\centering\arraybackslash}X}

\usepackage{pdflscape}

\usepackage{multibib}
\newcites{supp}{Primary Studies}

\usepackage{amsmath,bm}
\usepackage{relsize} 

\usepackage{hyperref}
\usepackage{xcolor}
\hypersetup{
    colorlinks,
    linkcolor={red!50!black},
    citecolor={blue!50!black},
    urlcolor={blue!80!black}
}

\usepackage{tikz}
\usetikzlibrary{positioning}    
\usetikzlibrary{automata}       
\usetikzlibrary{calc}           
\usetikzlibrary{fit}            
\usetikzlibrary{matrix}         
\usetikzlibrary{arrows.meta}    
\usetikzlibrary{shapes.geometric, arrows}
\usetikzlibrary{patterns}
\usetikzlibrary{backgrounds}

\usepgflibrary{patterns}

\tikzset{>={Latex[length=4,width=4]}} 
\usetikzlibrary{calc}

\usepackage{pgfplots}
\usepackage{pgfplotstable}
    \pgfplotsset{
        compat=1.9,
        compat/bar nodes=1.8,
    }

\tikzstyle{startstop} = [rectangle, minimum width=1.75cm, minimum height=1cm, text centered, draw=black, fill=gray!10]
\tikzstyle{process} = [rectangle, minimum width=1.75cm, minimum height=1cm, text centered, draw=black, fill=white]
\tikzstyle{arrow} = [thick,->,>=stealth]

\definecolor{other}{HTML}{cdd1e2}
\definecolor{rdbms}{HTML}{ffbcb1}
\definecolor{nosql}{HTML}{aeddaf}
\definecolor{newsql}{HTML}{f0ce82}

\usepackage{color, colortbl}
\definecolor{Gray}{gray}{0.925}

\usepackage{listings} 
\usepackage{xcolor}
\definecolor{codegreen}{rgb}{0,0.6,0}
\definecolor{codegray}{rgb}{0.5,0.5,0.5}
\definecolor{codepurple}{rgb}{0.58,0,0.82}
\definecolor{backcolour}{rgb}{0.95,0.95,0.92}
\definecolor{light-gray}{gray}{0.95} 

\lstdefinestyle{mystyle}{
    basicstyle=\scriptsize\ttfamily,       
    backgroundcolor=\color{light-gray},   
    stepnumber=1,                   
    numbers=left,
    numbersep=5pt,                  
    backgroundcolor=\color{light-gray},  
    showspaces=false,               
    showstringspaces=false,         
    showtabs=false,                 
    frame=single,                   
    tabsize=2,                      
    literate={\ \ }{{\ }}1,         
    captionpos=b,                   
    breaklines=true,                
    breakatwhitespace=false,        
    xleftmargin=0.5em,              
    breakindent=0pt,
    breakatwhitespace,
    columns=fullflexible,
}

\journal{Elsevier}

\makeatletter
\def\ps@pprintTitle{%
  \let\@oddhead\@empty
  \let\@evenhead\@empty
  \let\@oddfoot\@empty
  \let\@evenfoot\@oddfoot
}
\makeatother

\usepackage{numcompress}\biboptions{authoryear}

\begin{document}

\begin{frontmatter}

\title{Database management system performance comparisons: \\A systematic literature review}

\author{Toni Taipalus}

\fntext[myfootnote]{To cite this article, please refer to the peer-reviewed, published version in The Journal of Systems and Software \url{https://www.sciencedirect.com/science/article/pii/S0164121223002674}}

\begin{abstract}
Efficiency has been a pivotal aspect of the software industry since its inception, as a system that serves the end-user fast, and the service provider cost-efficiently benefits all parties. A database management system (DBMS) is an integral part of effectively all software systems, and therefore it is logical that different studies have compared the performance of different DBMSs in hopes of finding the most efficient one. This study systematically synthesizes the results and approaches of studies that compare DBMS performance and provides recommendations for industry and research. The results show that performance is usually tested in a way that does not reflect real-world use cases, and that tests are typically reported in insufficient detail for replication or for drawing conclusions from the stated results.
\end{abstract}


\end{frontmatter}

\section{Introduction}
\label{sec-introduction}

Efficiency is important in effectively all software systems, whether efficiency is measured by response times, how many concurrent users the system can serve, or how energy-efficient the system is \cite{Toffola_2018}. Despite its importance, many software systems suffer from efficiency problems \cite{Jin_2012}, as optimization has been largely recognized as a complex task \cite{Toffola_2018,Difallah_2013}. The more a system holds and handles data, the more the system's performance depends on the database, and the database is often one of the first suspects when a performance issue is detected. The domain of database management systems (DBMS) saw rapid advancements in performance especially in the 1980s and 1990s, as benchmarking competitions between DBMS and hardware vendors led to innovations in DBMS technology that significantly improved DBMS performance \cite{DeWitt_2008}. Performance improvements are related to DBMS aspects such as different supporting data structures \cite{Valduriez_1987}, and algorithms for sorting \cite{Estivill-Castro_1992,Do_2022} and joining \cite{Schneider_1989,Patel_1996}. Given that DBMSs are annually a multi-billion dollar industry, the performance of a DBMS is one of the most crucial aspects when a company chooses a DBMS for their product or service \cite{Dietrich_1992}. As different DBMS performance comparison studies and DBMS vendor white-papers highlight the performance gains of one DBMS over another, it may seem tempting to either consider choosing the fastest DBMS for a business domain or to migrate from one DBMS to another for performance gains. However, as we show and argue in this study, performance is typically tested in very specific contexts which are not necessarily generalizable, and there are other aspects besides performance to consider.

This study was inspired by a study by Raasveldt \textit{et al.} \cite{Raasveldt_2018}, which claimed that ``\textit{[...] we will explore the common pitfalls in database performance comparisons that are present in a large number of scientific works [...]}'' while consciously refraining from citing example studies. While we agree with their claim based on our personal experiences, we wanted to systematically explore whether this phenomenon is common among performance benchmarks, and whether such studies show performance gains of one DBMS over another in a setting that can be replicated. This study is not an attempt to criticize studies comparing DBMS performance, as no scientific study (ours included) is without threats to validity. Rather, based on the survey of the literature, the primary goals of our study are to propagate information on \textit{(i)} how DBMS performance has been tested, \textit{(ii)} how performance has been recommended to be tested, \textit{(iii)} how the performance comparison results should be interpreted, \textit{(iv)} what other aspects besides performance should be considered, and \textit{(v)} what other avenues might be fruitful for DBMS performance testing. Additionally, we provide \textit{(vi)} a relatively accessible background on database system performance, followed by \textit{(vii)} a systematic review of literature on DBMS performance comparisons, \textit{(viii)} describing which DBMSs and which types of DBMSs have been compared with each other, \textit{(ix)} the outcomes of the performance comparisons, and \textit{(x)} by which benchmarks the DBMSs have been compared.



The rest of this study is structured as follows. In Sections~\ref{sec-bg-dbs} and \ref{sec-bg-perf}, we provide theoretical background for understanding the results and discussion provided by this study. These background sections are deliberately presented by refraining from using unnecessary information technology-related terms, acronyms, algorithms, or mathematics, to cater to the needs of readers from various backgrounds. For readers more technically inclined or interested, we have provided further reading at the end of Sections~\ref{sec-bg-dbs} and \ref{sec-bg-perf}. Section~\ref{sec-method} details how we searched, selected, and categorized the DBMS performance comparison studies, and Section~\ref{sec-perf_results} presents a high-level overview of the results, which is complemented by \ref{app-results} detailing the performance comparison outcomes. In Section~\ref{sec-discussion}, we discuss what these findings mean, how they are applicable in industry, and present our recommendations for industry and research based on the findings. Section~\ref{sec-concl} concludes the study.

\section{Database Systems}
\label{sec-bg-dbs}

\subsection{Database System Overview}
\label{sec-bg-overview}

A database is a collection of interrelated data, typically stored according to a data model. Typically, the database is used by one or several software applications via a DBMS. Collectively, the database, the DBMS, and the software application are referred to as a \textit{database system} \cite[][p.7]{Elmasri_2016}\cite[][p.65]{Connolly_2015}. The separation of the database and the DBMS, especially in the realm of relational databases, is typically impossible without exporting the database in another format. In these situations, the database is often unusable by the DBMS, unless the database is imported back to a format understood by the DBMS. Possibly due to this inseparability, both the DBMS and the underlying database are often colloquially referred to simply as \textit{database}. It is worth noting, though, that the former is a piece of software that \textit{does}, while the other is a collection of data that \textit{is}.

\begin{figure}
\centering
\begin{tikzpicture}[->,>=stealth',auto,node distance=1.8cm,
  thick,main node/.style={circle,draw,font=\sffamily\Large\bfseries},
  database/.style={
      cylinder,
      shape border rotate=90,
      minimum width = 1.4cm,
      minimum height = 1.0cm,
      aspect=2,
      fill=white,
      draw
    }]
\node (parser) [process, xshift=0cm, align=center]                          {Parser};
\node (optimizer) [process, right of=parser, align=center, xshift=0.5cm]    {Optimizer};
\node (execution) [process, right of=optimizer, align=center, xshift=0.5cm] {Execution\\engine};
\node (logs) [process, below of=execution, align=center]                    {Transaction\\logs};
\node (locks) [process, above of=execution, align=center]                   {Locks};
DBMS
\node[draw=black,dotted, fit=(parser) (optimizer) (execution) (logs) (locks)
    ,inner sep=5mm,label={[xshift=-3cm,yshift=-0.2cm,anchor=north]:DBMS}](FIt1)       {};
\node (mem) [process, right of=execution, align=center, xshift=1cm]         {Memory};
\node (cpu) [process, above of=mem, align=center]                           {CPU};
\node (db)  [database, below of=mem, align=center]                          {}; 
\begin{scope}[on background layer]
    \node[startstop, fit=(FIt1) (db) (cpu),inner sep=5mm](FIt2) {};
\end{scope}
\node (software) [startstop, left of=parser, align=center, xshift=-2cm]     {Software\\application};
\node (user) [startstop, below of=software, align=center]                   {User};
\draw [<->,out=0,in=0,looseness=0] (user.north) to node[midway,above,inner sep=5pt]{} (software.south);
\draw [->,out=0,in=0,looseness=0] (software.east) to node[midway,above,inner sep=5pt]{} (parser.west);
\draw [->,out=0,in=0,looseness=0] (parser.east) to node[midway,above,inner sep=5pt]{} (optimizer.west);
\draw [->,out=0,in=0,looseness=0] (optimizer.east) to node[midway,above,inner sep=5pt]{} (execution.west);
\draw [<->,out=0,in=0,looseness=0] (execution.east) to node[midway,above,inner sep=5pt]{} (mem.west);
\draw [<->,out=0,in=0,looseness=0] (execution.north) to node[midway,above,inner sep=5pt]{} (locks.south);
\draw [<->,out=0,in=0,looseness=0] (execution.south) to node[midway,above,inner sep=5pt]{} (logs.north);
\draw [<->,out=0,in=0,looseness=0] (mem.north) to node[midway,above,inner sep=5pt]{} (cpu.south);
\draw [<->,out=0,in=0,looseness=0] (mem.south) to node[midway,above,inner sep=5pt]{} (db.north);
\end{tikzpicture}
\caption{A simplified view of a database system and the end-user with the emphasis on components relevant to this study; the arrows represent the flow of information from the end-user's device to the database residing in persistent storage; the flow of information back to the software application is not illustrated here; gray rectangles represent boundaries of physical devices}
\label{fig-dbs} 
\end{figure}
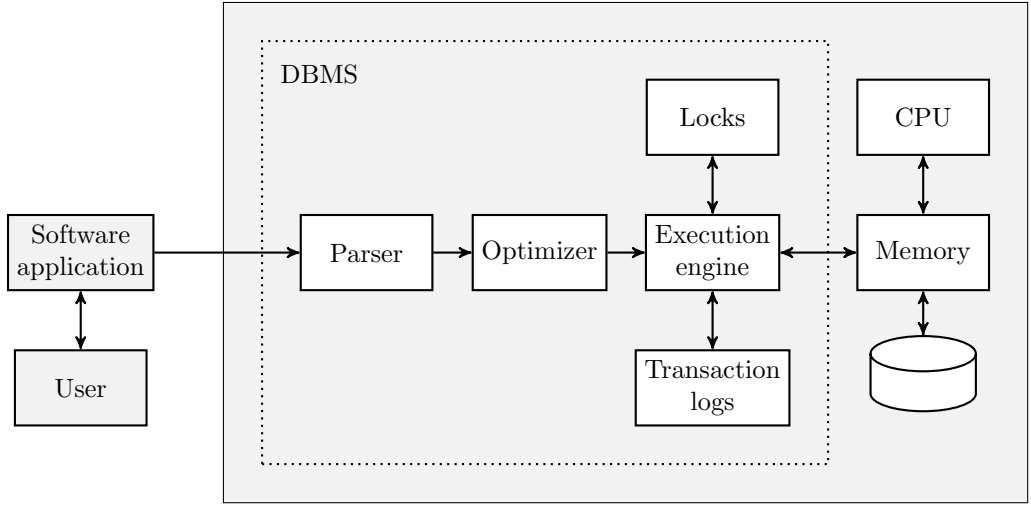

Fig.\ref{fig-dbs} shows a simplified example of a system where the components crucial for a database system and the scope of this study are emphasized. We refer to the components in the figure throughout this study. Several things are worth noting in considering the figure, as we have traded technical precision and comprehensiveness for ease of presentation by depicting only a single end-user, a single software application (some parts typically reside on the end-user's device, while others reside on a separate server), a single DBMS, single hardware components, and a single database. Furthermore, we have not illustrated other DBMS components such as access control, data structures such as metadata, or outputs such as query execution plans. The figure also adopts the view that the database resides in persistent storage --- this is not always the case. Additionally, we have depicted merely a centralized database system in which neither the DBMS nor the database has been distributed across multiple nodes. These are willful omissions given the scope of this study.

\subsection{Data Models}
\label{sec-bg-dbms}

Databases follow one or several data models, i.e., definitions of how and what data can be stored, and sometimes, what operations are available for data retrieval and manipulation. Data models may be conceptual, logical, or physical. Conceptual models such as the Entity-Relationship model \cite{Chen_1976} do not dictate how data should be stored, but are rather used to describe the interrelations and characteristics of the data. Logical data models such as the relational model \cite{Codd_1970} are related to how data is stored and presented, but often without describing how the data is physically stored, e.g., which computing node is responsible for storing the data, where the data is located on a disk, and what types of indices (i.e., redundant data structures which facilitate query performance) and physical data retrieval operators are available. One DBMS is not limited to using a single data model \cite{Forresi_2022}.

There are several popular logical data models, some of which are inseparable from their underlying physical data models. One of the most prominent logical data models is the relational data model rooted in set theory  \cite{Codd_1970}. Relational DBMSs (RDBMS) follow many of the concepts introduced in the relational model. Many of the popular RDBMSs such as PostgreSQL and Oracle Database have adopted data structures from other logical data models as well \cite{Lu_2019}. What is common for effectively all modern RDBMSs is that they utilize Structured Query Language (SQL) \cite{SQL_2016p1,SQL_2016p2} to define data structures and to retrieve and manipulate data. Typically, RDBMSs also implement a strong data consistency model which dictates or allows that database operations grouped into a transaction must all succeed or all fail, data must follow defined business logic, successful transactions persist in storage, and concurrent transactions \cite[cf.][]{Bernstein_1981} must result in the same data as if the transactions were serial. At least the last rule can often be loosened in modern implementations to various degrees. These constraints are collectively referred to as the ACID consistency model \cite{Haerder_1983}.

NoSQL is an umbrella term for several data models, typically developed or popularized in the first decade of the 2000s \cite{Grolinger_2013}. Contrary to the relational model, the data models within NoSQL typically have no formal definitions, and different NoSQL DBMSs implement different data models such as key-value (e.g., Redis), document (e.g., MongoDB), wide-column (e.g., Cassandra) and graph (e.g., Neo4J) \cite{Davoudian_2018,Reniers_2017}. Furthermore, these DBMSs often have a distinct query language developed to cater to the particular data structures available in the DBMS's implementation of a data model. While RDBMSs have favored data consistency \cite{Chaudhry_2020} by eliminating redundant data through logical database design, and through a strong consistency model, NoSQL DBMSs have generally adopted the opposite approach. In several NoSQL data models such as key-value pairs and documents, redundant data is stored at the cost of storage space \cite{Hecht_2011}. This approach enables query languages to be simple \cite{Dey_2014}, avoiding complex and potentially slow queries. Furthermore, consistency models are typically less strict than in RDBMSs \cite{Stonebraker_2010}, which facilitates higher performance demanded by, e.g., web applications with a large number of concurrent users \cite{Ramakrishnan_2012}.

Although NoSQL DBMSs popularized several database-related approaches such as non-strict database structures, data availability over data consistency, and relatively effortless database replication (i.e., data is copied over computing nodes) and sharding (i.e., data is divided between computing nodes) \cite{Grolinger_2013}, some industry leaders such as Google deemed a strong consistency model and an expressive query language important enough to design a DBMS which incorporates features from both RDBMSs and NoSQL DBMSs \cite{Corbett_2013}. These so-called NewSQL DBMSs use the relational model, often with extensions, SQL as their primary query language, and a distributed database architecture \cite{Pavlo_2016}. In addition to these three \textit{main} categories of RDBMS, NoSQL, and NewSQL data models, others such as object stores \citesupp{kulshrestha_2014} and GPU-intensive \citesupp{suh_2022} systems are used in specific contexts.

\subsection{Query Execution}
\label{sec-bg-query}

The word \textit{query} typically refers to query language statements that retrieve some data from the database. However, in this study, we use the word query to refer to any data retrieval and manipulation statement for brevity. In times it is necessary to differentiate between data retrieval and manipulation, we use appropriate terms such as \textit{read operations} for data retrieval, and \textit{write operations} for data insertion, updates, and deletes. In this subsection, we describe how queries are executed, using mainly general (i.e., not specific to a single DBMS) literature from the domain of RDBMS query execution.

When a user --- were it a human actor directly using a terminal, a transaction processing software application, or a database benchmark software --- submits a query to a DBMS, a multitude of events must take place before the user receives feedback. Illustrated in a general fashion in Fig.~\ref{fig-dbs}, the query parser checks, among other things, that the query is syntactically valid \cite{Hellerstein_2007}. If the query passes these (and other) checks, the query is translated to a lower-level presentation and passed to the query optimizer. The optimizer generates one or several query execution plans. These plans consist of physical operators for implementing, e.g., which physical data structures will be utilized in executing the query, and in RDBMSs in particular, how tables are joined together \cite{Graefe_1993}. If several plans are generated, the optimizer evaluates which of these plans is the most effective in regards to, e.g., query execution time \cite{Hellerstein_2007}. The accuracy of the optimizer relies on aspects such as database metadata \cite{Christodoulakis_1984}, statistics of previous query executions, and the indices available \cite{Chaudhuri_1998_PODS}. Generating effective query execution plans is a complex effort and takes time \cite{Graefe_1993,Chaudhuri_1998_PODS}, but once formulated, the plans can be re-used to a degree.

Next, the query execution engine implements the query execution plan, using the physical operators therein. Simplified, the data objects required by the query are typically first searched from a memory area called the buffer pool which is allocated and maintained by the DBMS. If some or all data is not found, the data is requested from disk. Before accessing the disk, many systems may additionally utilize other areas of memory to avoid disk access \cite{Yang_2018}.

Effectively all database systems function in an environment where multiple concurrent end-users use the database. This concurrency presents challenges particularly when the users execute write operations on the same database, e.g., when two or more users withdraw money from the same bank account, concurrently updating the balance \cite{Bernstein_1981}. To guarantee that the write operations do not interfere with each other in a way that would cause the data to not represent the real world, DBMSs typically implement concurrency control through locking or versioning data. Effectively, the simpler implementations of locking restrict data objects from being accessed by other operations while the data objects are being modified \cite{Hellerstein_2007}. These locking mechanisms may be implemented to ensure that no anomalies happen, or with implementations that theoretically allow some anomalies \cite{Berenson_1995}. Typically, the business domain dictates what types of anomalies are tolerated.

Finally, as strong consistency models often require that transactions persist in the database and that all of the operations in a transaction either succeed or fail, locking is typically complemented by transaction logs. These logs are written before write operations are committed to the database, and can be used in reversing earlier write operations if a later write operation in the same transaction fails. All these considerations discussed in this section play a significant role from a performance perspective, which is discussed in the next section.

\textbf{Further reading on database systems:} for readers interested in the basics of database systems, either the undergraduate level textbook by Connolly and Begg \cite{Connolly_2015}, or Elmasri and Navathe \cite{Elmasri_2016} are excellent albeit lengthy introductions covering the topic from several points of view and with the focus on RDBMSs. For readers interested in query processing, we point to studies by Chaudhury \cite{Chaudhuri_1998_PODS}, and Hellerstein, Stonebraker and Hamilton \cite{Hellerstein_2007}. If you are interested in logical relational database design, the book by Date \cite{Date_2019} is an in-depth resource covering both formal and informal approaches. For a survey of literature on NoSQL data models, the study by Davoudian, Chen and Liu \cite{Davoudian_2018} is an accessible starting point.

\section{Performance}
\label{sec-bg-perf}

\subsection{Performance Measurement}
\label{sec-bg-measurement}

In general, performance is a measurement of how efficiently a software system completes its tasks. Performance is typically measured in response time, throughput \cite{Hellerstein_2007}, or in some cases, utilization of computing resources \cite[][p.4]{Cortellessa_2011}. Response time is the time taken for a call in the system to traverse to some other part of the system and back. This is also sometimes called latency \cite[][p.10]{Gunther_2011}, and in the context of database systems, the response time may be measured as the response time to the first or the last result item \cite{Graefe_1993}. In a broad perspective described in Fig.~\ref{fig-dbs}, the response time might be the time taken after the end-user sends a request to the software application (e.g., an online store), which passes the request to a DBMS, which returns a set of data to the software application, which finally presents the data to the end-user's device. In database benchmarking, however, response time might be measured by running the benchmark on the same device the DBMS and the database reside, effectively eliminating inter-device-induced performance drawbacks such as network latency \cite{Patounas_2020,Delis_1993} and firewalls, and mitigating the effects of other software running on the devices. Although DBMSs perform other tasks besides querying, querying is typically what is measured in DBMS performance testing \cite{Dietrich_1992}. While response time is perhaps the least arduous performance metric to measure, it is not often enough for reliable measurement of transaction processing environments \cite{Dietrich_1992} (often dubbed online transaction processing, OLTP). That is, response time might be a metric better suited for long-running queries in decision support environments (often dubbed online analytical processing, OLAP), but as transaction processing environments often process a large number of concurrent transactions, response time alone might not reliably account for the effects of concurrent transactions, unless response time is measured as an average of multiple concurrent transactions.

Performance can also be measured by throughput, i.e., how many transactions the DBMS can execute in a given time frame. Throughput is often expressed as transactions per second \cite{Dietrich_1992} and requires a more sophisticated approach, e.g., benchmarking software. Again, throughput may be measured either locally (i.e., using only the hardware the DBMS and the database reside on), or over a network in case the database is distributed. Alternatively, throughput may be measured by connecting the benchmarking software to the software application, which simulates the throughput of the whole database system by accounting for, e.g., network and the software application \cite[e.g.,][]{Kumar_2022,Sundaresan_2013}. Such an approach arguably requires significantly more investment, but provides a holistic perspective on the performance of the whole system, also uncovering potential performance issues unrelated to the DBMS and the database. Finally, performance may be measured by resource utilization, either CPU time, I/O, memory allocation, or energy consumption \cite{Graefe_1993} in systems striving for energy-efficiency due to, e.g., limited battery power, or due to environmental concerns \cite{Guo_2022}.

In summary, we might consider the measurement of throughput a process that typically requires a simulation of some level, and the measurement of response time as an exact or approximated mathematical method. The former approach requires relatively high investments into the development of such simulations \cite[][p.142]{Cortellessa_2011}, while the latter often relies on a set of assumptions that do not necessarily reflect real-world scenarios due to inaccuracies in predicting what the real-world scenario ultimately is and how it can change.

\subsection{Factors Affecting Performance}
\label{sec-bg-factors}

\textit{Hardware:} An intuitive factor in performance is the power of hardware \cite[][p.1]{Osterhage_2013}, and while it is true that most of the local response time is attributed to time taken by CPU processing, memory and disk access, and software waiting for other tasks to complete \cite[][p.5]{Cortellessa_2011}, first investing in software performance rather than hardware performance is often more cost-effective. That being said, it is generally accepted that memory access is at least four orders of magnitude faster than disk access \cite[e.g.,][p.42]{Gunther_2011}. That is, if memory access takes minutes (nanoseconds), disk access takes months (milliseconds). These numbers are largely dependent on the speed of memory and the type of disk, but paint a picture of how zealously DBMS optimization strives to minimize disk access. Since memory is typically more expensive than disk storage, keeping the whole database in memory is often not feasible. Additionally, the underlying hardware is important, as, e.g., some DBMSs have been shown to utilize multi-processor or multi-core environments more effectively than others \cite{Tu_2013}. Intuitively, how well a DBMS can exploit parallelism affects the performance of query execution \cite{Tallent_2009,Tozun_2013}. Ultimately, performance measurement is about gains or losses in percentages, not in, e.g., response times. 

\textit{Data models:} Data models described in Section~\ref{sec-bg-dbms} have indirect effects on DBMS performance. Relational databases often follow design guidelines that strive to minimize redundancy to eliminate potential data anomalies caused by redundant data \cite{Codd_1972,Codd_1975}, and to minimize the need for storage space, which in turn typically causes queries to run slower due to a larger number of table joins. In contrast, different NoSQL data models --- especially key-value, document, and wide-column --- follow design guidelines according to which data structures are designed to efficiently satisfy predetermined business logic queries, with the elimination of redundant data being a secondary concern \cite{Davoudian_2018}. It follows that because many NoSQL data structures are designed to serve queries, queries are typically simple \cite{Dey_2014}, and their execution requires less computational resources than complex queries in relational databases. As discussed in Section~\ref{sec-bg-query}, locking data objects (both on disk and in memory, and both primary data structures as well as indices), logging write operations, and how memory is managed by the DBMS all play a significant role in DBMS performance \cite{Hellerstein_2007,Stonebraker_2010}. For example, preventing write operation-induced anomalies is a costly action, and the level of granularity of database locks presents significant considerations on write operation performance, which is largely dictated by the ratio of read and write operations.

\textit{Distribution:} Write operations in distributed configurations pose non-trivial challenges to both performance and data consistency \cite{Delis_1993}. In distributed database systems, effectively all transactions must choose either data consistency or data availability \cite{Brewer_2012,Gilbert_2002}. The former guarantees that the data the end-user receives are not stale, with the cost of performance, while the latter guarantees to a degree that the end-user receives data faster, but with no guarantees that the dataset received is the most recent. The preferred approach is largely dictated by business logic.

\textit{DBMS and OS parameters:} Moving from data models and database system distribution to lower levels of abstraction, operating system (OS) and DBMS parameters and their interrelationships (e.g., page size) can have direct or indirect effects on performance \cite{Dietrich_1992}. Additionally, DBMS parameters such as the amount of memory the DBMS is allowed to use for data processing is typically closely related to the amount of memory available. Furthermore, as a query is sent to the optimizer (cf. Fig.~\ref{fig-dbs}), it depends on the DBMS internals how efficiently the optimizer can select the most efficient physical operations to implement the query, and what physical operations are available to the optimizer in the first place \cite{Chaudhuri_1998_PODS}. For example, MySQL implemented only one physical operation for table joins until 2018\footnote{https://dev.mysql.com/doc/refman/5.6/en/explain-output.html}, limiting the number of options the optimizer could choose from. Regarding query optimization, the optimizers of RDBMSs in particular are relatively mature and can spot some unnecessary complications in queries, while overlooking others \cite{Brass_2006}. Despite the benefits brought by the optimizers, some queries are inherently slow and can only be optimized through query rewrites.

\textit{Physical database design:} Last, but definitely not least, physical database design plays a key role in DBMS performance. It has been argued that performance bottlenecks are difficult to find in large systems \cite{Ammons_2014}, and that efficiency is gained by focusing on the vital few areas instead of the trivial many \cite[][p.450]{Juran_2010}. One of the most vital areas in database systems is physical design. In relational databases, efficient physical design is largely achieved through indices, and in NoSQL databases, typically through database distribution over computing nodes. In contrast to a holistic system overview, performance bottlenecks may be easier to find in queries, since many DBMSs provide detailed information on query execution (Fig.~\ref{fig-qep}). PostgreSQL (Fig.~\ref{fig-qep-a}) lists the physical operations used to execute the query, which of the operations took the most time units, and which indices, if any, were used. For example, it can be seen in Fig.~\ref{fig-qep-a} that the sequential scan on line 12 accounted for approximately 94\% of the execution time of the whole query (178 \textit{time units} out of 189 \textit{ms}), probably because the query fetched a large number of records from the database. The query could be optimized by, e.g., selecting a smaller number of records, and showing the results to the end-user by paging them, i.e., showing a subset of results first, and fetching more later if necessary. In NoSQL systems, the query optimizer plays a smaller role due to typically less expressive query languages (cf. Fig.~\ref{fig-qep-b}). Some NoSQL systems such as Cassandra do not permit the execution of queries that do not utilize the physical structures effectively.

\begin{figure}
\centering
\input{fig-qep}
\caption{Query execution plans illustrating the physical operators such as \textit{hash join} and \textit{seq scan} chosen by the optimizer}
\label{fig-qep} 
\end{figure}

\subsection{Database Performance Benchmarks}
\label{sec-bg-benchmarks}

There are several database performance benchmarks available, each typically consisting of a sample database and a workload that simulates how the database could be used \cite{Difallah_2013,Qu_2022}. The benchmarks usually measure the efficiency of querying while taking into account factors such as concurrency but disregarding other DBMS tasks such as efficiency in data structure definition or bulk loading \cite{Dietrich_1992}. 

In the domain of relational databases, the Transaction Processing Council (TPC) benchmarks \cite[e.g.,][]{Gray_1992} are perhaps the most utilized \cite{Dreseler_2020,Tozun_2013}, and test the throughput of the DBMS with various parameters. For example, the TPC-A benchmark simulates a database of a bank with four tables and with one transaction, the TPC-B benchmark a database of a wholesale supplier with nine tables and with five transactions, and the TPC-E benchmark a brokerage database with 33 tables and 12 transactions. All these benchmarks have the option for simulating strong consistency, and while TPC-A and TPC-B have transactions typical for transaction processing, TPC-E includes also decision support transactions \cite{Tozun_2013}. TPC-A simulates human end-user thinking by waiting between transactions, as a human arguably would wait between clicks in an online bank. TPC-B, on the other hand, does not wait and can be used as a precursor for TPC-A in adjusting DBMS parameters \cite{Dietrich_1992}. Alternatively to transaction processing, TPC-H benchmark measures the performance of a DBMS in decision support \cite{Barata_2015,Dreseler_2020}.

In the more general DBMS domain, the Yahoo! Cloud Serving Benchmark (YCSB) is a framework for benchmarking transaction processing in systems with different data models and architectures \cite{Cooper_2010}. Due to its extensibility, YCSB can be adapted to different NoSQL data models. YCSB contains different workloads, each with a different ratio of read and write operations. YCSB and its extensions such as YCSB+T typically utilize transactions which consist of single operations and do not enforce strong consistency \cite{Qu_2022,Dey_2014}. The benchmarks described above are by no means an exhaustive list but cover the most popular benchmarks (cf. Section~\ref{sec-bg-overview}). Other benchmarks include LUBM \cite{Guo_2005}, OLTP-Bench \cite{Difallah_2013}, and JOB \cite{Leis_2015}. Regardless of the data model and DBMS, transaction processing benchmarks have typically been the de facto method of comparing different DBMSs and hardware \cite{Tozun_2013}.

\textbf{Further reading on performance:} for readers interested in physical database operations and query execution from a performance perspective, Graefe \cite{Graefe_1993} provides an in-depth, DBMS-independent survey. For more information on physical database design, especially indices and how they work, the book by Lightstone, Teorey and Nadeau \cite{Lightstone_2010} is a detailed and descriptive source. For a practical and concise guide on SQL query optimization, we point readers towards Winand's book \cite{Winand_2011}. Regarding NoSQL DBMS optimization, we suggest referring to the manual of the DBMS of your choice, and always making sure that the source of information is current, as NoSQL systems tend to evolve rapidly.

\section{Study Selection}
\label{sec-method}

\subsection{Process and Criteria}
\label{sec-met-process}

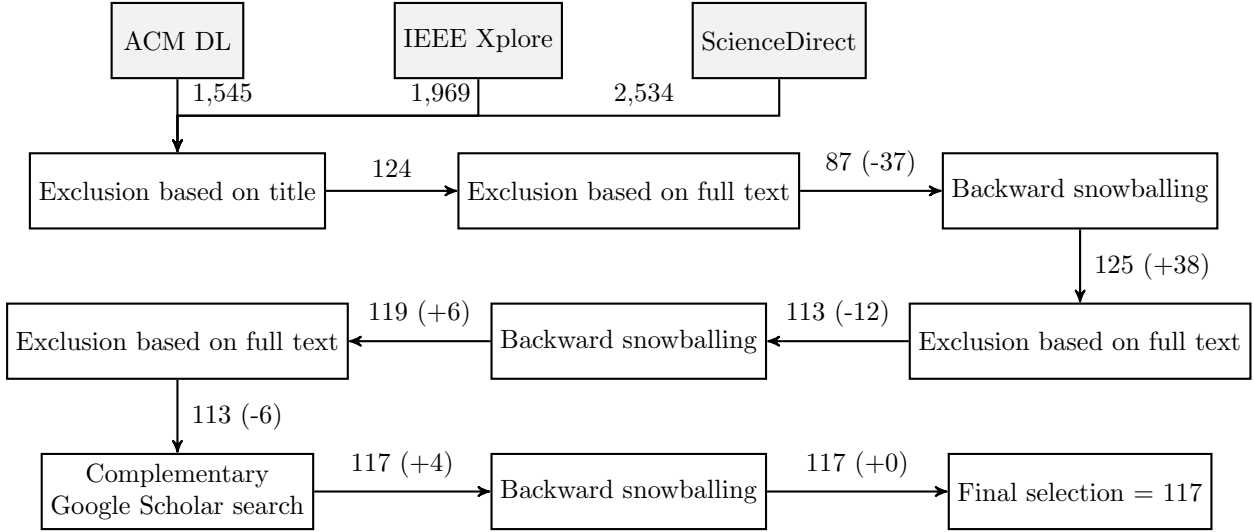
\begin{figure}
\centering
\begin{tikzpicture}[->,>=stealth',auto,node distance=3.0cm,
  thick,main node/.style={circle,draw,font=\sffamily\Large\bfseries}]
\node (acm) [startstop, xshift=0cm] {ACM DL};
\node (ieee) [startstop, right of=acm, xshift=1cm] {IEEE Xplore};
\node (sc) [startstop, right of=ieee, xshift=1cm] {ScienceDirect};
\node (ex1) [process, below of=acm, xshift=0cm,yshift=1cm] {Exclusion based on title};
\node (ex2) [process, right of=ex1, xshift=3cm,yshift=0cm] {Exclusion based on full text};
\node (sb1) [process, right of=ex2, xshift=3cm] {Backward snowballing};
\node (ex3) [process, below of=sb1, xshift=0cm,yshift=1cm] {Exclusion based on full text};
\node (sb2) [process, left of=ex3,  xshift=-3cm] {Backward snowballing};
\node (ex4) [process, left of=sb2,  xshift=-3cm] {Exclusion based on full text};
\node (sb3) [process, below of=ex4, xshift=0cm, yshift=1cm, align=center] {Complementary\\Google Scholar search};
\node (sch) [process, right of=sb3, xshift=3cm] {Backward snowballing};
\node (fin) [process, right of=sch, xshift=3cm] {Final selection = 117};
\draw[->,thick] (acm) --++ (0,-1.2) -| (ex1) 
    node[midway,above, xshift=0.6cm, yshift=0.2cm] {1,545};
\draw[->,thick] (ieee) --++ (0,-1) -| (ex1) 
    node[midway,above, xshift=3.5cm, yshift=0cm] {1,969};
\draw[->,thick] (sc) --++ (0,-1) -| (ex1) 
    node[midway,above, xshift=6.2cm, yshift=0cm] {2,534};
\draw [->,out=0,in=0,looseness=0] (ex1.east) to node[midway,above,inner sep=5pt]{124} (ex2.west);
\draw [->,out=0,in=0,looseness=0] (ex2.east) to node[midway,above,inner sep=5pt]{87 (-37)} (sb1.west);
\draw [->,out=0,in=0,looseness=0] (sb1.south) to node[midway,right,inner sep=5pt]{125 (+38)} (ex3.north);
\draw [->,out=0,in=0,looseness=0] (ex3.west) to node[midway,above,inner sep=5pt]{113 (-12)} (sb2.east);
\draw [->,out=0,in=0,looseness=0] (sb2.west) to node[midway,above,inner sep=5pt]{119 (+6)} (ex4.east);
\draw [->,out=0,in=0,looseness=0] (ex4.south) to node[midway,right,inner sep=5pt]{113 (-6)} (sb3.north);
\draw [->,out=0,in=0,looseness=0] (sb3.east) to node[midway,above,inner sep=5pt]{117 (+4)} (sch.west);
\draw [->,out=0,in=0,looseness=0] (sch.east) to node[midway,above,inner sep=5pt]{117 (+0)} (fin.west);
\end{tikzpicture}
\caption{The study selection process; the numbers refer to the number of primary studies selected in each stage of the process}
\label{fig-process} 
\end{figure}

The DBMSs in this study were selected based on the selected primary studies. That is, we did not choose, e.g., the most popular DBMSs to include, but reported the DBMSs yielded by the primary studies. The results herein may be considered the most popular DBMSs in terms of benchmarking reported in scientific literature. Fig.~\ref{fig-process} describes the primary study selection process starting from ACM Digital Library, IEEE Xplore, and ScienceDirect, complemented by subsequent Google Scholar searches. The search strings are detailed in Table~\ref{tab-terms}. To account for potentially missing relevant studies, we conducted three rounds of backward snowballing (i.e., following the lists of references in selected studies), until snowballing revealed no additional studies. A total of 117 primary studies comparing DBMS performance were selected.

\begin{table}
  \caption{Search strings}
  \label{tab-terms}
  \begin{tabularx}{\linewidth}{lX}
    \toprule
    Database & Search string \\
    \midrule
    ACM DL & 
    [Abstract: performance] AND [Abstract: comparison] AND [[Abstract: database] OR [Abstract: \\
           & dbms]] AND [Publication Date: (01/01/2000 TO 03/31/2022)] \\
    IEEE Xplore &
    ("Abstract":performance AND "Abstract":comparison AND ("Abstract":database OR "Abstract":dbms)) \\
    ScienceDirect &
    Title, abstract, keywords: performance AND comparison AND (database OR dbms) \\
    Google Scholar &
    database performance comparison \\
    \bottomrule
\end{tabularx}
\end{table}

Table~\ref{tab-criteria} describes our inclusion criteria applied in the primary study selection. The first four criteria are related to bibliographic details, while the last three criteria are concerned with article focus and content. Regarding criterion \textit{\#3}, we excluded academic theses and dissertations \cite[e.g.,][]{Coates_2009} due to the fact that they are typically not peer-reviewed. We also excluded white and gray literature for the same reason, and because those studies are often written or published by partial parties, e.g., DBMS vendors. 

We only selected studies that compared query (i.e., retrieving or modifying data) execution performance, not regarding e.g., database replication performance \cite{Elnikety_2006} or performance of different join operations \cite{Jung_2009}. We also excluded studies that compared a single DBMS performance in different configurations such as hardware, replication strategy, database structure, or query language \cite{Holzschuher_2013} and studies that compared a DBMS with different data-related platforms \cite{Purbo_2020}. Studies that reported pseudonymized DBMS names were also excluded. Finally, we only included studies that reported results based on at least seemingly objective metrics and empirical results. That is, studies simply stating the opinions of the authors such as ``based on our experiences, we believe MySQL is faster than SQL Server'' were not considered.

\begin{table}
  \caption{Primary study selection criteria}
  \centering
  \label{tab-criteria}
  \begin{tabular}{ll}
    \toprule
    \# & Inclusion criterion \\
    \midrule
    1 & Article is written in English. \\
    2 & Full article can be accessed. \\
    3 & Article is published in a scientific journal, or conference or workshop proceedings. \\
    4 & Article is published between 2000 and March 2022. \\
    \midrule
    5 & Article focus is on query language statement execution performance comparison. \\
    6 & Article focus is on comparing the performance of two or more different DBMSs. \\
    7 & Article is based on at least seemingly objective metrics. \\
    \bottomrule
\end{tabular}
\end{table}

\subsection{Selected Studies}
\label{sec-met-selected}

The selected 117 primary studies compared the performance of a total of 44 different DBMSs. We categorized these DBMSs into three top-level types defined and discussed in Section~\ref{sec-bg-dbms}: RDBMSs, NoSQL systems, and NewSQL systems. Five DBMSs not clearly pertaining to any of these three categories were categorized under \textit{other} systems (Table~\ref{tab-db_types}). It is worth noting that these DBMS types are not always clear-cut due to the lack of specificity and changing nature of the definitions, and should be interpreted as merely means to compartmentalize the results of this study into a more readable form. Five selected primary studies did not report results implying the performance of one DBMS over another \citesupp{padhy_2019,Schmid_2015_wms,Dwivedi_2012,faraj_2014,Jing_2009}.

\begin{table}
  \caption{DBMSs discussed in this study divided into four types}
  \label{tab-db_types}
  \begin{tabularx}{\linewidth}{lX}
    \toprule
    DBMS type & DBMSs  \\
    \midrule
    \midrule
    RDBMS & Access, Azure SQL, Interbase, DB/2, H2, Hive, MariaDB, MySQL Cluster, MySQL, Oracle Database, PostgreSQL, PostgresXL, SQL Server, SQLite \\
    \midrule
    NoSQL & ArangoDB, Azure Document Database, Cassandra, Couchbase, CouchDB, Elasticsearch, Firebase, HBase, Hypertable, memcached, MongoDB, Neo4J, Oracle NoSQL, OrientDB, RavenDB, Redis, RethinkDB, Riak, Scalaris, Tarantool, Voldemort \\
    \midrule
    NewSQL & CockroachDB, MemSQL (now known as SingleStoreDB), NuoDB, VoltDB \\
    \midrule
    Other  & BlazingSQL, Caché, Db4o, OmniSciDB, PG-Strom \\
  \bottomrule
\end{tabularx}
\end{table}

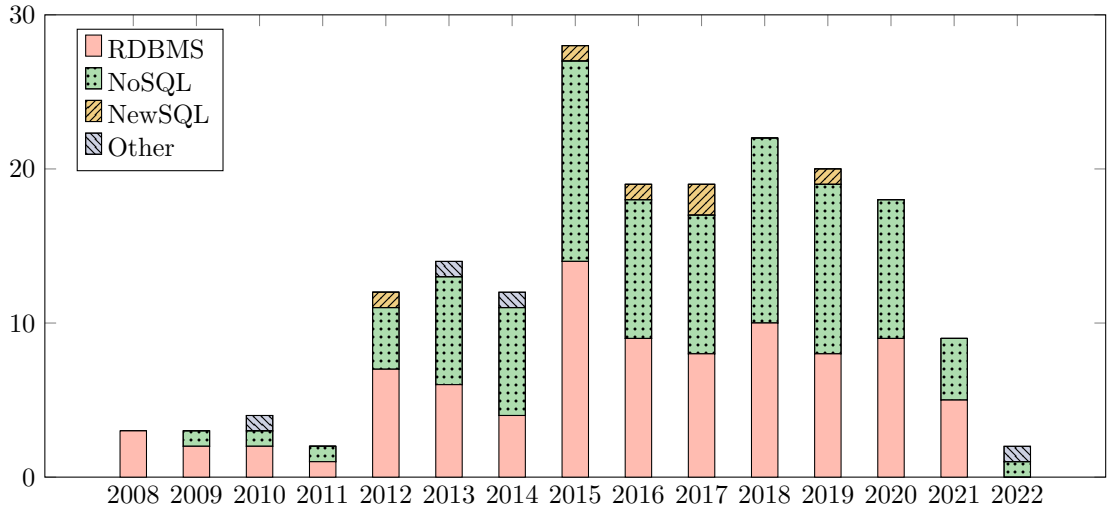
\begin{figure}[ht]
   \centering
      \begin{tikzpicture}
   \pgfplotstableread{
Label   RDBMS NoSQL NewSQL Other
2008	3	0	0	0
2009	2	1	0	0
2010	2	1	0	1
2011	1	1	0	0
2012	7	4	1	0
2013	6	7	0	1
2014	4	7	0	1
2015	14	13	1	0
2016	9	9	1	0
2017	8	9	2	0
2018	10	12	0	0
2019	8	11	1	0
2020	9	9	0	0
2021	5	4	0	0
2022	0	1	0	1
    }\testdata
        \begin{axis}[
            ybar stacked,
            ymin=0,
            ymax=30,
            xtick=data,
            legend style={
                cells={anchor=west},
                legend pos=north west,
            },
            reverse legend=false,
            xticklabels from table={\testdata}{Label},
            xticklabel style={text width=3cm,align=center},
            width=0.95\textwidth,
            height=220,
        ]
            \addplot [fill=rdbms]
                table [y=RDBMS, x expr=\coordindex]
                    {\testdata};
                        \addlegendentry{RDBMS}
            \addplot [fill=nosql, postaction={pattern=dots}]
                table [y=NoSQL, x expr=\coordindex]
                    {\testdata};
                        \addlegendentry{NoSQL}
            \addplot [fill=newsql, postaction={pattern=north east lines}]
                table [y=NewSQL, x expr=\coordindex]
                    {\testdata};
                        \addlegendentry{NewSQL}
            \addplot [fill=other, postaction={pattern=north west lines}]
                table [y=Other, x expr=\coordindex]
                    {\testdata};
                        \addlegendentry{Other}
        \end{axis}
    \end{tikzpicture}
   \caption{The number of publications by publication year and DBMS type; the year 2022 was only considered until March}
   \label{fig-results-years} 
\end{figure}

Fig.~\ref{fig-results-years} shows the distribution of publication years and the types of DBMSs discussed in the selected studies. Although our criteria allowed for studies from the year 2000, the first studies selected were published in 2008. The figure shows that generally, there is a somewhat constant number of DBMS performance comparison studies each year. It is worth noting that one study may pertain to several types of DBMSs.

\section{Performance Comparison Results}
\label{sec-perf_results}



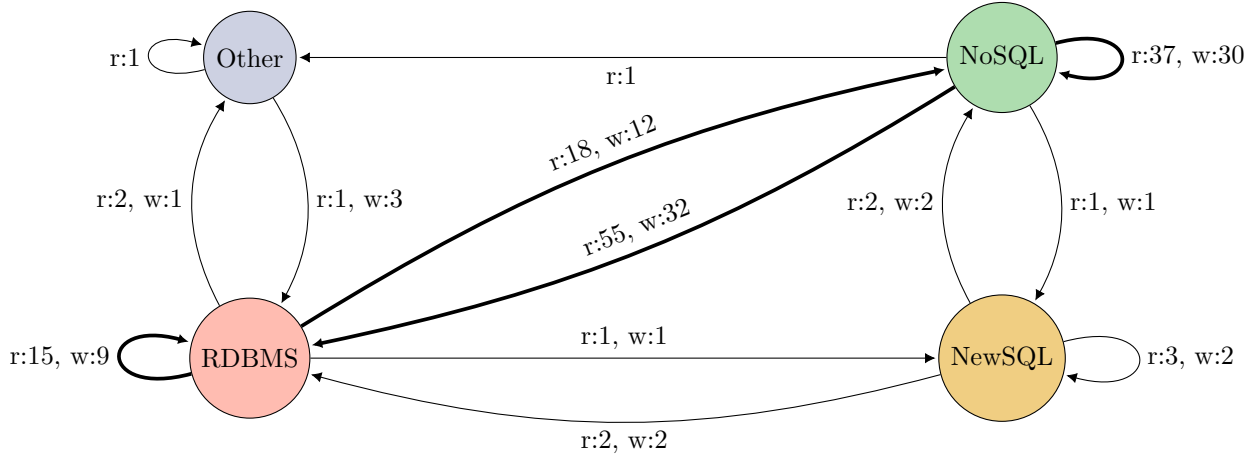
\begin{figure}
    \centering
    \begin{tikzpicture}[shorten >=1pt,node distance=4cm,on grid,auto]
  \tikzstyle{every state}=[fill={rgb:black,1;white,10}]
\node[state,fill=other]  at (0, 0)  (q_1) {Other};
\node[state,fill=rdbms]  at (0, -4) (q_2) {RDBMS};
\node[state,fill=nosql]  at (10, 0) (q_3) {NoSQL};
\node[state,fill=newsql] at (10, -4) (q_4) {NewSQL};

\path[->]
(q_1) edge [loop left]  node {r:1}                                          (   )
      edge [bend left]  node {r:1, w:3}                                     (q_2)
(q_2) edge [bend left=10, line width = 1.4pt] node[sloped] {r:18, w:12}     (q_3)
      edge [bend left]  node {r:2, w:1}                                     (q_1)
      edge []           node {r:1, w:1}                                     (q_4)
      edge [loop left, line width = 1.4pt] node {r:15, w:9}                 (   )
(q_3) edge [bend left=10, line width = 1.4pt] node[sloped] {r:55, w:32}     (q_2)
      edge []  node {r:1}                                                   (q_1)
      edge [bend left]  node {r:1, w:1}                                     (q_4)
      edge [loop right, line width = 1.4pt] node {r:37, w:30}               (   )
(q_4) edge [bend left=15]  node {r:2, w:2}                                  (q_2)
      edge [bend left]  node {r:2, w:2}                                     (q_3)
      edge [loop right]  node {r:3, w:2}                                    (   );
\end{tikzpicture}
    \caption{DBMS performance comparisons overview; a directed edge from node \textit{a} to node \textit{b} represents the number of studies according to which a system of type \textit{a} outperformed a system or systems of type \textit{b} in (r)ead and (w)rite operations, e.g., a NoSQL system outperformed a NewSQL system in read operations in one study, and in write operations in one study; thicker edges visualize the most popular comparisons}
    \label{fig-results-overview} 
\end{figure}

The most popular DBMS performance comparisons compared one or several RDBMSs to one or several NoSQL systems, one NoSQL system to another NoSQL system, or one RDBMS to another RDBMS, respectively. A total of 48 studies compared solely read performance, while 6 studies compared solely write performance. The rest of the studies compared both read and write performance, with the exception of two studies \citesupp{cheng_2019,nepaliya_2015} which were unclear whether they compared write operations. All comparisons and their results per DBMS type are summarized in Fig.~\ref{fig-results-overview}.

\begin{figure}
  \includegraphics[width=\linewidth,keepaspectratio]{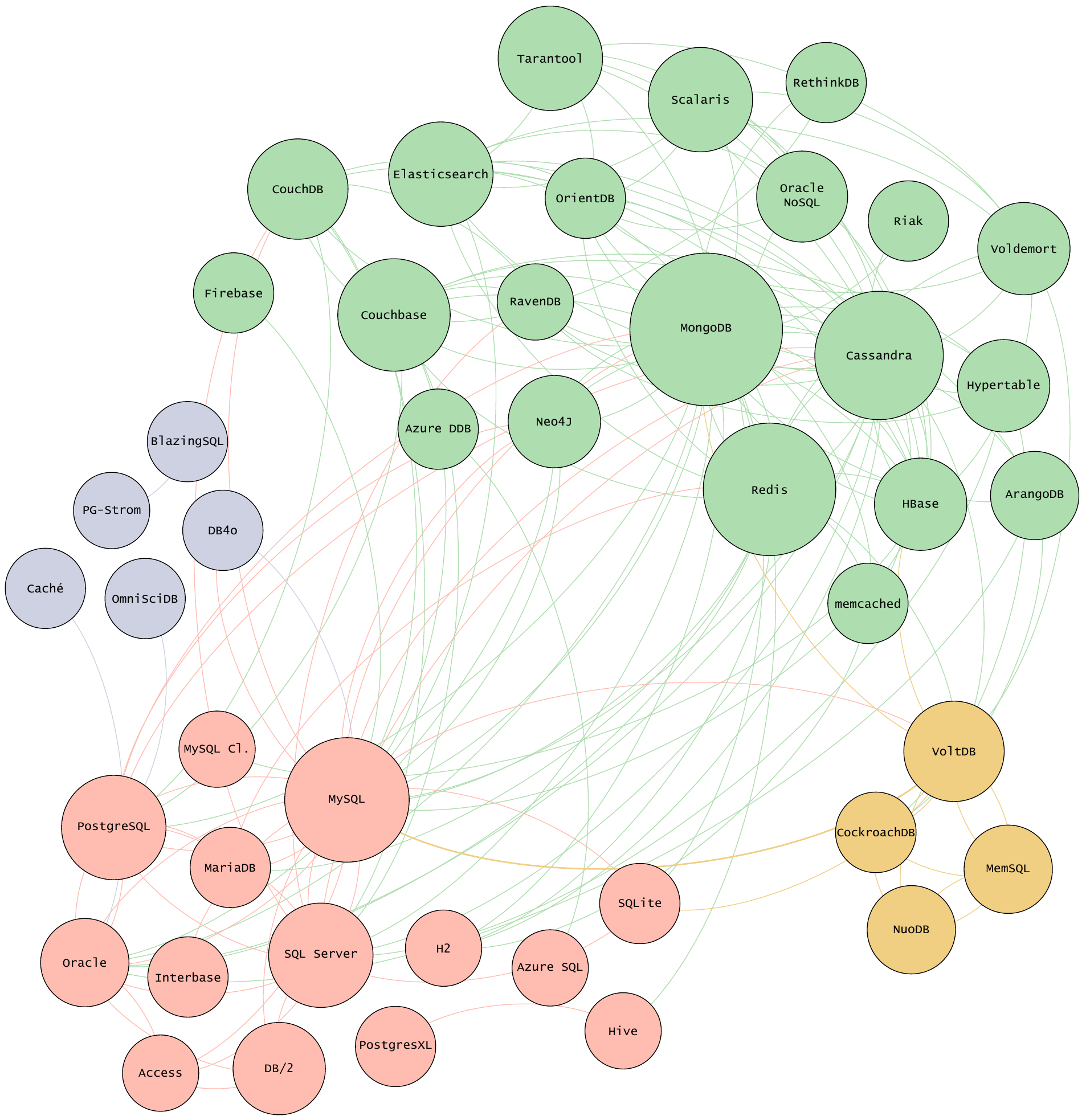}
  \caption{An overview of read operation performance comparisons between NoSQL systems (green, upper right), NewSQL systems (yellow, lower right), RDBMSs (red, lower left), and other systems (blue, upper left); a clockwise turning edge from node \textit{a} to node \textit{b} depicts node \textit{a} outperforming node \textit{b}, and the color of the edge corresponds to the type of the outperforming node, e.g., Caché outperforms PostgreSQL according to one or several studies; the size of a node represents out-degree, i.e., larger nodes have outperformed more systems than smaller nodes}
  \label{fig-results-overview-reads} 
\end{figure}

Fig.~\ref{fig-results-overview-reads} presents an overview of which DBMSs and DBMS types the primary studies compared. The figure perhaps conveys how both \textit{other} and NewSQL systems are typically compared within their respective DBMS type groups, while RDBMS and NoSQL systems are both compared within their respective groups as well as with each other. Additionally, the size of the nodes such as MongoDB, Redis, Cassandra, and MySQL show that these DBMSs typically outperform the DBMSs they are compared to. Due to their length, the detailed results from the primary study comparisons are presented in Appendix~\ref{app-results}, which includes tables detailing which DBMSs outperformed which.

Regarding the benchmarks defined in earlier scientific literature, the most popular was YCSB, which was utilized by 15 primary studies (approximately 13\%) \citesupp{abramova_2013,abramova_2014_exp,abramova_2014_which,Gandini_2014,schreiner_2019,seghier_2021,yassien_2016,abubakar_2014,kashyap_2013,swaminathan_2016,tang_2016,klein_2015,araujo_2021,hendawi_2018,rabl_2012}. The second most popular benchmark was the TPC-H benchmark and its variations, utilized by five primary studies (4\%) \citesupp{Almeida_2015,fotache_2016,oliveira_2017,suh_2022,vershinin_2021}. It is worth noting, though, that two of the studies \citesupp{oliveira_2017,vershinin_2021} seemed to have executed the \textit{queries} of TPC-H, instead of running the benchmark and accounting for, e.g., the effects of concurrent transactions. One primary study utilized the OLTP-Bench benchmark \citesupp{Tongkaw_2016}, one the LUBM benchmark \citesupp{franke_2013}, and one, in addition to TPC-H, the JOB benchmark \citesupp{suh_2022}. Regarding the benchmarks formulated by the primary study authors, 25 primary studies (21\%) reported using ad hoc queries instead of earlier defined benchmarks to compare the performance of DBMSs. These queries were defined verbatim in the primary studies. In contrast, 70 of the primary studies (60\%) compared DBMS performance using undisclosed ad hoc queries, likely formulated by the study authors. In other words, 22 primary studies (19\%) used some type of earlier defined database benchmarking suite. The performance tests of these 22 primary studies and what aspects of the environment they reported are detailed in Table~\ref{tab-studies_with_benchmarks}.

\begin{table}
  \caption{An overview of primary studies using previously defined benchmark software and which aspects of the testing environment they explicitly disclosed; performance measurements abbreviated as ET (execution time) and TP (throughput); \textsuperscript{1}the YCSB benchmark defines a single-table with \textit{n} columns (or loose equivalents in non-relational data models)}
  \scriptsize
  \label{tab-studies_with_benchmarks}
  \begin{tabularx}{\textwidth}{cccccccc}
    \toprule
    \multirow{2}{*}{Study} & \multicolumn{4}{c}{Explicitly reported} & \multirow{2}{*}{Benchmark} & \multirow{2}{*}{Measurement} & \multirow{2}{*}{Nodes} \\ \cline{2-5}
                           & \multicolumn{1}{c}{DBMS versions}  & \multicolumn{1}{c}{Hardware}  & \multicolumn{1}{c}{DB structure}  & DBMS parameters & & & \\
    \midrule
    
    \citesupp{abramova_2013} & \multicolumn{1}{c}{yes}  & \multicolumn{1}{c}{yes}  & \multicolumn{1}{c}{no\textsuperscript{1}} & no & YCSB &  ET & 1  \\

    \citesupp{abramova_2014_which} & \multicolumn{1}{c}{yes}  & \multicolumn{1}{c}{yes}  & \multicolumn{1}{c}{no\textsuperscript{1}} & no & YCSB &  ET & 1  \\

    \citesupp{abramova_2014_exp} & \multicolumn{1}{c}{yes}  & \multicolumn{1}{c}{yes}  & \multicolumn{1}{c}{no\textsuperscript{1}} & no & YCSB &  ET & 1  \\

    \citesupp{abubakar_2014} & \multicolumn{1}{c}{no}  & \multicolumn{1}{c}{no}  & \multicolumn{1}{c}{no\textsuperscript{1}} & no & YCSB &  ET & 1  \\

    \citesupp{Almeida_2015} & \multicolumn{1}{c}{no}  & \multicolumn{1}{c}{yes}  & \multicolumn{1}{c}{logical only} & no & Star Schema Benchmark &  ET & 1  \\

    \citesupp{araujo_2021} & \multicolumn{1}{c}{yes}  & \multicolumn{1}{c}{yes}  & \multicolumn{1}{c}{no\textsuperscript{1}} & no & YCSB & ET, TP & 2 \\

    \citesupp{fotache_2016} & \multicolumn{1}{c}{no}  & \multicolumn{1}{c}{yes}  & \multicolumn{1}{c}{logical only} & no & TPC-H & ET & 5 \\

    \citesupp{franke_2013} & \multicolumn{1}{c}{yes}  & \multicolumn{1}{c}{yes}  & \multicolumn{1}{c}{no} & no & LUBM-based & ET & 9 \\

    \citesupp{Gandini_2014} & \multicolumn{1}{c}{no}  & \multicolumn{1}{c}{yes}  & \multicolumn{1}{c}{no\textsuperscript{1}} & no & YCSB & ET, TP & 2-9 \\

    \citesupp{hendawi_2018} & \multicolumn{1}{c}{yes}  & \multicolumn{1}{c}{yes}  & \multicolumn{1}{c}{no\textsuperscript{1}} & no & YCSB & ET, TP & 8 \\

    \citesupp{kashyap_2013} & \multicolumn{1}{c}{yes}  & \multicolumn{1}{c}{yes}  & \multicolumn{1}{c}{no\textsuperscript{1}} & no & YCSB & ET, TP & up to 5 \\

    \citesupp{klein_2015} & \multicolumn{1}{c}{yes}  & \multicolumn{1}{c}{no}  & \multicolumn{1}{c}{no\textsuperscript{1}} & no & YCSB & ET, TP & 9 \\

    \citesupp{oliveira_2017} & \multicolumn{1}{c}{no}  & \multicolumn{1}{c}{yes}  & \multicolumn{1}{c}{logical only} & no & TPC-H & ET & 1 \\

    \citesupp{rabl_2012} & \multicolumn{1}{c}{yes}  & \multicolumn{1}{c}{yes}  & \multicolumn{1}{c}{no\textsuperscript{1}} & no & YCSB & ET, TP & 16 and 24  \\

    \citesupp{schreiner_2019} & \multicolumn{1}{c}{no}  & \multicolumn{1}{c}{yes}  & \multicolumn{1}{c}{no\textsuperscript{1}} & yes (default) & YCSB, Voter & ET, TP & 3 \\

    \citesupp{seghier_2021} & \multicolumn{1}{c}{yes}  & \multicolumn{1}{c}{yes}  & \multicolumn{1}{c}{no\textsuperscript{1}} & no & YCSB & ET & 1 \\

    \citesupp{suh_2022} & \multicolumn{1}{c}{yes}  & \multicolumn{1}{c}{yes}  & \multicolumn{1}{c}{no\textsuperscript{1}} & yes (default) & TPC-H & ET & 3 \\

    \citesupp{swaminathan_2016} & \multicolumn{1}{c}{yes}  & \multicolumn{1}{c}{yes}  & \multicolumn{1}{c}{no\textsuperscript{1}} & no & YCSB & TP & up to 14  \\

    \citesupp{tang_2016} & \multicolumn{1}{c}{yes}  & \multicolumn{1}{c}{yes}  & \multicolumn{1}{c}{no\textsuperscript{1}} & no & YCSB & ET, TP & 4 \\

    \citesupp{Tongkaw_2016} & \multicolumn{1}{c}{yes}  & \multicolumn{1}{c}{yes}  & \multicolumn{1}{c}{logical only} & no & Sysbench, OLTP-Bench &  TP & 1 \\

    \citesupp{vershinin_2021} & \multicolumn{1}{c}{no}  & \multicolumn{1}{c}{yes}  & \multicolumn{1}{c}{logical only} & no & TPC-H & ET & 1 \\

    \citesupp{yassien_2016} & \multicolumn{1}{c}{yes}  & \multicolumn{1}{c}{yes}  & \multicolumn{1}{c}{no\textsuperscript{1}} & no & YCSB & ET, TP & 1 \\
    
    \bottomrule
\end{tabularx}
\end{table}

\section{Discussion}
\label{sec-discussion}

\subsection{General Discussion}
\label{sec-disc-general}

The difficulty of rigorous performance testing is perhaps one of the root causes of why optimization is difficult, and several studies have highlighted the complexity of performance testing due to, e.g., the effects of DBMS parameters \cite{Purohith_2017}, testing environment settings \cite{Wang_2022}, and how well the data in the performance test database reflects the real application data \cite{Qu_2022_workload}. Is it also important whether an impartial actor has carried out the performance test, or whether the test results are published e.g., by a DBMS vendor \cite{DeWitt_2008}. However, this is sometimes difficult to assess and can be mitigated by simply explicitly reporting the test so that it can be replicated and verified by others.

Despite the fact that we were aware of some DBMS performance comparison studies as they have been touched on in previous works, we were surprised by the extent the few examples presented in the previous works \cite{Raasveldt_2018,Wang_2022} generalize to so many studies on the subject. For example, in read operations, MongoDB outperforms Cassandra according to ten studies, Cassandra outperforms Redis according to four studies, and Redis outperforms MongoDB according to six studies (cf. Appendix~\ref{app-results}), leading to a situation of $Mo > Ca > Re > Mo$, where MongoDB is both the best and the worst performing DBMS. Furthermore, as discussed in Section~\ref{sec-perf_results}, few of the selected studies reported the test setting in enough detail for replication. Unfortunately, without sufficient details for replicating an experiment, such experimental results can claim any outcome \cite{Raasveldt_2018}. One aspect that was typically reported was some details about the hardware the test was run on, i.e., processor make and model, clock rate, memory size, and disk size. Without other details about the DBMS parameters, parallel execution, etc., these details are inconsequential. Despite the importance of the topic of DBMS performance comparisons, with the exception of one study \citesupp{rabl_2012}, no primary studies were published in major data management fora such as ACM SIGMOD or VLDB. 

\subsection{Considerations for Industry}
\label{sec-disc-industry}

\subsubsection{Consider the Environments in Performance Testing Studies}

If the environment in which the performance testing was carried out does not provide sufficient details, whatever the study states, you may interpret the results as if they do not generalize to other environments. That is, if you are in the process of deciding on a DBMS for your application, or perhaps considering changing one DBMS to another, consider whether the performance comparison study you are reading presents a similar use case. Compare your business domain to that presented in performance comparison studies, remembering that a single, sometimes even a seemingly inconsequential parameter (cf. e.g., data types in SQLite \cite{Purohith_2017}) may change the results. DeWitt and Levine \cite{DeWitt_2008} aptly describe performance comparisons as the \textit{maximum} potential performance gain of one DBMS over another. The performance gain in your particular environment might be less, or it might be that the DBMS that performed better in the comparison performs worse in your environment.

One important aspect of the environment is the physical setup. Different hardware has been shown to affect DBMS performance, as some DBMSs exploit parallelism more efficiently than others \cite{Marek_1992,Jiang_2010}, effectively meaning that if a test was performed on one single-core CPU, the results might not generalize to distributed environments. Additionally, different hardware aspects such as the relative sizes of different CPU memory caches may significantly affect DBMS performance, making performance comparisons between different hardware a complex task \cite{Ailamaki_1999,Wang_2022}. In distributed environments, which were rarely tested in the primary studies, it is worth considering whether data availability is prioritized over data consistency, as the latter setup is typically significantly slower. Benchmarks that simulate concurrent users should also be considered separately from performance tests that merely execute queries sequentially. Concurrency introduces several challenges, many of which severely affect performance \cite{Wang_2022}. For example, SQLite uses database locking on a level of granularity which makes concurrent writes slow, but this has no negative effects on single-user writes \cite{Obradovic_2019}. Unfortunately, some studies have shown that developers do not widely understand concurrency-related security aspects \cite{Warszawski_2017}, and that concurrency-related performance problems are understudied \cite{Yu_2018}. Some have even stated that the research has not been focusing on relevant issues \cite{Pavlo_2017}.

Intuitively, different business domains have different databases and they are used in different ways. For example, in some domains, the end-users typically read data, while in others, write operations are more common. The ratio of read and write operations in a performance test plays a crucial role, as some DBMSs are specifically designed for specific workloads \cite{Cooper_2010}. The credibility of testing results is also related to how well the test database and data therein represent the target environment \cite{Qu_2022_workload}. Furthermore, in business domains such as online stores, there are typically popular products, and thus the data related to them are targets of a relatively large number of database operations. For generalizable benchmarking results, the benchmark must account for such skewness in database use, rather than, e.g., randomly querying data objects. It is also worth considering how the performance tests have tested performance. For example, is your application about inserting 10,000 rows in bulk, but one row at a time randomly generated by the application? If it is not, you should not consider this type of benchmark results as an indication of how well one DBMS performs compared to another in your particular business context. It is also worth considering that decision support benchmarks such as TPC-H test performance in environments that can be fundamentally different from transaction processing environments. Finally, even similar business domains can have a myriad of different technical implementations.

We have discussed some of the particulars involved in database system design in this subsection, and in Sections~\ref{sec-bg-dbs} and \ref{sec-bg-perf}, from which one can infer what has often been repeated in database system research: the environments and their optimization is a task so complex \cite{Graefe_1993,Cooper_2010} that DBMS optimization is a whole profession \cite{Raasveldt_2018}. It follows that there are several threats to rigorous DBMS benchmarking. Even though RDBMS optimization is widely and deeply studied in both academic and industry contexts, RDBMS optimization remains a complex task. In the domain of NoSQL DBMSs, there exist far fewer scientific studies simply due to the age of the NoSQL DBMSs, and the heterogeneity of NoSQL data models. Additionally, there are several querying anti-patterns to avoid, such as performing joins in the software application instead of the DBMS, or paging query results by utilizing ordering, limiting and offsetting. All these points considered, a reader of a performance comparison study must trust that the performance comparison study writers have been able to optimize the database systems to a similar degree for the performance comparison results to be credible. This requires particularly specific, in-depth expertise when DBMSs with more than one data model are compared. Furthermore, decades of benchmarking software development by entire councils (e.g., TPC) cannot simply be skipped by writing a set of (often arbitrary) queries, running them on two or more DBMSs in a single-user environment, recording response times, and consequently stating that one DBMS is faster than another. Although this was the case in over 80\% of the selected primary studies, we do not consider this sufficient.

In summary, if it is possible that changing even one of the environmental aspects discussed above may affect the performance test results significantly, it seems reasonable to argue that, no matter how many DBMS performance comparison studies state that one DBMS outperforms another, these DBMSs were not tested in an environment that is the same as your environment, and thus have little concern in the decision of which DBMS is performance-wise the best fit for your environment.

\subsubsection{Consider Other Aspects Besides Performance}

There are other aspects besides response time or throughput to consider when choosing a DBMS. Performance gains, such as those provided by many NoSQL systems, rely heavily on redundant data to minimize the complexity of queries, thus providing faster response times. Naturally, storing redundant data increases the cost of storage, and may lead to data inconsistencies. Another comparison perspective is related to the features provided by the DBMSs compared. Intuitively, a DBMS that is tailored for a specific purpose outperforms a general-purpose DBMS \cite{Raasveldt_2018,Stonebraker_2007}. For example, one primary study \citesupp{Bartoszewski_2019} noted that while MongoDB outperformed PostgreSQL/PostGIS in most of the tests, MongoDB provides only a subset of the geospatial operations provided by PostGIS. If the rest of the operations needed by the business domain need to be implemented in the software application, it is not realistic to assume that such task is either trivial to implement, nor trivial to implement in a way that outperforms the solutions offered by existing DBMS features.

Another consideration is the availability of suitable workforce, which is closely related to the DBMS technology and its maturity. It is not surprising that as query languages such as SQL have been a topic of effectively all information technology-related curricula in higher education for several years \cite{CS_2013,SE_2015}, there is a relatively large number of professionals fluent in SQL, as opposed to new query languages. Some studies have also shown that strong consistency models \cite{Corbett_2013} and the SQL language \cite{Stephen_2022} are desired as skills as well as features in a DBMS. That is, it is worth considering how feasible it is to implement a database system with each specific technology, and DBMS performance is only one of the important aspects to consider.

Finally, as the primary studies typically considered performance in terms of response time or throughput, we have approached the topic from a similar viewpoint. However, as discussed in Section~\ref{sec-bg-measurement}, performance may be measured by the usage of computing resources, which can be a goal conflicting with response time \cite{Chaudhuri_1998_PODS}. It is typical that increasing parallelism through multiple CPUs lowers response time, but increases the total amount of work due to the parallelism overhead \cite[][p.13]{Osterhage_2013}. Finally, it has been shown that migrating data from one DBMS to another is all but trivial, and prone to fail due to a lack of clear methodologies \cite{Thalheim_2013} --- especially when the DBMSs differ in data models and query languages \cite{Kim_2018}. Therefore, migrations such as RDBMS $\Leftrightarrow$ RDBMS or RDBMS $\Leftrightarrow$ NewSQL are arguably less complex than migrations such as NoSQL $\Leftrightarrow$ NoSQL, RDBMS $\Leftrightarrow$ NoSQL or NewSQL $\Leftrightarrow$ NoSQL.

\subsection{Considerations for Research}
\label{sec-disc-research}

\subsubsection{Consider Using Existing Guidelines for Testing and Reporting}

Database benchmarking guidelines are not a novel invention in database system research and have been described in detail \cite{Gray_1992} and in short \cite{Dietrich_1992} in the early 1990s, and as a reader-friendly checklist later \cite{Raasveldt_2018}. Additionally, benchmarking pitfalls have been discussed in numerous studies in respected database systems fora \cite{Wang_2022,Dreseler_2020}. Based on the primary studies, however, neither of these lines of research has been widely applied in practice. Database benchmarking has been argued to be difficult \cite{Raasveldt_2018}, as environmental parameters such as the nature of data \cite{Qu_2022_workload}, DBMS parameters \cite{Wang_2022}, and data types \cite{Purohith_2017} can all have significant impacts on performance testing results. Furthermore, benchmarking tools have received critique \cite{Reniers_2017,Grolinger_2013} despite the fact that some of the tools have been under development for decades. Therefore, we urge researchers, at the very least, to consider whether using a performance test suite of one's ad hoc queries is credible when well-known performance benchmarks are freely available.

As for reporting, Raasveldt \textit{et al.} \cite{Raasveldt_2018} provide a 24-point checklist for fair benchmarking. Some of the points are concerned about how performance is tested, and others about how the testing is reported. A performance comparison that cannot be replicated may present whatever results \cite{Raasveldt_2018}. Furthermore, an empirical study without reproducible evidence should be considered an opinion of the authors, rather than an empirical study. Indeed, at the start of the NoSQL movement, we have witnessed several studies with high praise for the strengths of different NoSQL products, yet with little or no critical notions addressing the acknowledged shortcoming of such DBMSs. Therefore, we would caution the reader from inferring from these results that one DBMS performs better than another. Rather, each such argument should be carefully scrutinized and interpreted in a specific context, like in the primary study assessing the performance of GPU DBMSs \citesupp{suh_2022}, in which performance between DBMSs was compared, but the comparison was merely one aspect of the study.

\subsubsection{Consider a Different Approach to DBMS-DBMS Testing}

Especially for a junior researcher, comparing the performance of one DBMS to another may seem like a relatively simple research setting to both carry out, and also justify based on the prevalence of the DBMS industry. We hope that the arguments presented in previous studies as well as here have highlighted that neither of these points are as clear-cut. Following the guidelines \cite[e.g.,][]{Raasveldt_2018} can make performance testing a time-consuming task, and in many cases, perhaps overly time-consuming, and given the considerations on the generalizability of the results, the results may not be of interest in other environments. Alternatively, not following guidelines introduces significant threats to validity. While generalizability is hardly an intrinsic value, concluding that, e.g., MySQL outperforms PostgreSQL in ``my webstore'' but not in others unless they have similar data, hardware, number of end-users, etc., does not carry the implication of being as a scientifically impactful result as saying that, e.g., MySQL will always outperform PostgreSQL. Therefore, we must either perform the performance comparisons with rigor and accept that the results do not probably generalize, or perform the comparisons without scientific rigor and state sophisms. Since the latter is hardly ethically sound, DBMS performance comparisons should be limited to domains where the goal of a study is not the generalizability of the results, but the betterment of the very particular domain the study concerns \citesupp[e.g.,][]{ameri_2014}.

Given the arguments above, we propose that future studies, if inter-DBMS performance must be compared, consider taking a different approach to performance testing. First, using a wide range of database system optimization experts to ensure that all aspects of the system are fairly optimized, and avoiding situations where one system is optimized beyond diminishing returns, while the other is barely optimized. We challenge research teams to explicitly disclose which authors optimized which systems, for authors to further one's intellectual investments in the performance comparison. These solutions should be benchmarked by a party independent of both optimization teams, and fair benchmarking guidelines should be utilized. Second, after the benchmarking has been carried out, we urge researchers to consider what causes the differences in performance, and critically compare those aspects as well, as gains in performance arguably have root causes such as loosened consistency or increased storage space. Nonetheless, performance comparisons of two or more DBMS with different data models should be considered particularly complex. Unfortunately, such comparisons seem to be the most popular (cf. Fig.~\ref{fig-results-overview}).

\subsubsection{Consider Other Use Cases Besides DBMS-DBMS Testing Altogether}

It is worth noting that benchmarking software has other use cases besides inter-DBMS performance comparisons. Instead of comparing one DBMS to another, researchers might consider testing the performance effects of different hardware \cite{Do_2011}, DBMS parameters \cite{Wang_2022}, operating system parameters, query languages \cite{Holzschuher_2013}, physical configurations such as database distribution, physical structures such as different indices, or different levels of data consistency.

\subsection{Limitations and Threats to Validity}

It might be that some relevant studies are missing from this literature review. However, it was not our intention to select primary studies to quantitatively demonstrate that one DBMS outperforms another by the number of studies corroborating such an argument. Rather, the results verify previous observations \cite{Raasveldt_2018} according to which many of such comparisons are problematic and should be interpreted with care, if at all. Nevertheless, we have strived to include at least most of the primary studies that fit our criteria (Table~\ref{tab-criteria}) by several rounds of snowballing (Fig.~\ref{fig-process}) as well as a complementary literature search. Furthermore, as the DBMS classification (Table~\ref{tab-db_types}) and the interpretation of the primary study results (\ref{app-results}) involve human judgment, it is possible that another group of researchers may attain at least slightly different results.

\section{Conclusion}
\label{sec-concl}

Several database management system performance comparisons have been conducted and published as both vendor white-papers as well as in scientific fora. The approaches and reporting in such studies have been criticized in previous literature. In this study, we systematically surveyed 117 DBMS performance comparison studies. What seemed to be common among the selected primary studies is that they lack sufficient detail for reproducibility. Scientific, peer-reviewed research of high external validity concerning database management performance comparison is effectively scarce. Based on the review of literature, we presented several considerations for the industry as well as database system researchers. Namely, we argued for considering \textit{(i)} the \textit{environments} (i.e., business domain, amount of data, amount of concurrent users, hardware, database distribution, read/write operation ratio, etc.) when interpreting the results of DBMS performance comparison tests, and for considering \textit{(ii)} other aspects besides DBMS performance when choosing a DBMS or changing one DBMS to another, and for researchers to consider \textit{(iii)} using existing guidelines in performance testing and reporting the testing environments transparently, to consider \textit{(iv)} different approaches to performance testing when one DBMS is compared to another, and to consider \textit{(v)} other use cases for performance testing besides comparing the performance of one DBMS to another. The results highlight how rarely benchmarking software is used in performance testing, how often different DBMSs with different data models are compared with each other, how often performance testing results in different studies conflict with each other, and why. This study is not an attempt to argue the performance gains of one DBMS over another using primary studies. That is, please do not cite this study by consulting the Appendix and stating that \textit{DBMS\textsubscript{1} outperforms DBMS\textsubscript{2}}.

\bibliography{sample-base}

\begin{thebibliography}{187}
\expandafter\ifx\csname natexlab\endcsname\relax\def\natexlab#1{#1}\fi
\providecommand{\url}[1]{\texttt{#1}}
\providecommand{\href}[2]{#2}
\providecommand{\path}[1]{#1}
\providecommand{\DOIprefix}{doi:}
\providecommand{\ArXivprefix}{arXiv:}
\providecommand{\URLprefix}{URL: }
\providecommand{\Pubmedprefix}{pmid:}
\providecommand{\doi}[1]{\href{http://dx.doi.org/#1}{\path{#1}}}
\providecommand{\Pubmed}[1]{\href{pmid:#1}{\path{#1}}}
\providecommand{\bibinfo}[2]{#2}
\ifx\xfnm\undefined \def\xfnm[#1]{\unskip,\space#1}\fi
\bibitem[{Ailamaki et~al.(1999)Ailamaki, DeWitt, Hill and Wood}]{Ailamaki_1999}
\bibinfo{author}{Ailamaki\xfnm[ A.]}, \bibinfo{author}{DeWitt\xfnm[ D.J.]},
  \bibinfo{author}{Hill\xfnm[ M.D.]}, \bibinfo{author}{Wood\xfnm[ D.A.]}.
\newblock \bibinfo{title}{{DBMSs} on a modern processor: Where does time go?}
\newblock In: \bibinfo{booktitle}{VLDB'99, Proceedings of 25th International
  Conference on Very Large Data Bases, September 7-10, 1999, Edinburgh,
  Scotland, UK}. Number \bibinfo{number}{CONF}; \bibinfo{year}{1999}. p.
  \bibinfo{pages}{266--277}.
\bibitem[{Ammons et~al.(2004)Ammons, Choi, Gupta and Swamy}]{Ammons_2014}
\bibinfo{author}{Ammons\xfnm[ G.]}, \bibinfo{author}{Choi\xfnm[ J.D.]},
  \bibinfo{author}{Gupta\xfnm[ M.]}, \bibinfo{author}{Swamy\xfnm[ N.]}.
\newblock \bibinfo{title}{Finding and removing performance bottlenecks in large
  systems}.
\newblock In: \bibinfo{editor}{Odersky\xfnm[ M.]}, editor.
  \bibinfo{booktitle}{ECOOP 2004 -- Object-Oriented Programming}.
  \bibinfo{address}{Berlin, Heidelberg}: \bibinfo{publisher}{Springer Berlin
  Heidelberg}; \bibinfo{year}{2004}. p. \bibinfo{pages}{172--196}.
\bibitem[{Barata et~al.(2015)Barata, Bernardino and Furtado}]{Barata_2015}
\bibinfo{author}{Barata\xfnm[ M.]}, \bibinfo{author}{Bernardino\xfnm[ J.]},
  \bibinfo{author}{Furtado\xfnm[ P.]}.
\newblock \bibinfo{title}{An overview of decision support benchmarks: {TPC-DS},
  {TPC-H} and {SSB}}.
\newblock In: \bibinfo{editor}{Rocha\xfnm[ A.]}, \bibinfo{editor}{Correia\xfnm[
  A.M.]}, \bibinfo{editor}{Costanzo\xfnm[ S.]}, \bibinfo{editor}{Reis\xfnm[
  L.P.]}, editors. \bibinfo{booktitle}{New Contributions in Information Systems
  and Technologies}. \bibinfo{address}{Cham}: \bibinfo{publisher}{Springer
  International Publishing}; \bibinfo{year}{2015}. p.
  \bibinfo{pages}{619--628}.
\bibitem[{Berenson et~al.(1995)Berenson, Bernstein, Gray, Melton, O'Neil and
  O'Neil}]{Berenson_1995}
\bibinfo{author}{Berenson\xfnm[ H.]}, \bibinfo{author}{Bernstein\xfnm[ P.]},
  \bibinfo{author}{Gray\xfnm[ J.]}, \bibinfo{author}{Melton\xfnm[ J.]},
  \bibinfo{author}{O'Neil\xfnm[ E.]}, \bibinfo{author}{O'Neil\xfnm[ P.]}.
\newblock \bibinfo{title}{A critique of {ANSI SQL} isolation levels}.
\newblock In: \bibinfo{booktitle}{Proceedings of the 1995 ACM SIGMOD
  International Conference on Management of Data}. \bibinfo{address}{New York,
  NY, USA}: \bibinfo{publisher}{Association for Computing Machinery}; SIGMOD
  '95; \bibinfo{year}{1995}. p. \bibinfo{pages}{1–10}.
\newblock \URLprefix \url{https://doi.org/10.1145/223784.223785}.
  \DOIprefix\doi{10.1145/223784.223785}.
\bibitem[{Bernstein and Goodman(1981)}]{Bernstein_1981}
\bibinfo{author}{Bernstein\xfnm[ P.A.]}, \bibinfo{author}{Goodman\xfnm[ N.]}.
\newblock \bibinfo{title}{Concurrency control in distributed database systems}.
\newblock \bibinfo{journal}{ACM Computing Surveys}
  \bibinfo{year}{1981};\bibinfo{volume}{13}(\bibinfo{number}{2}):\bibinfo{pages}{185–221}.
\newblock \URLprefix \url{https://doi.org/10.1145/356842.356846}.
  \DOIprefix\doi{10.1145/356842.356846}.
\bibitem[{Brass and Goldberg(2006)}]{Brass_2006}
\bibinfo{author}{Brass\xfnm[ S.]}, \bibinfo{author}{Goldberg\xfnm[ C.]}.
\newblock \bibinfo{title}{Semantic errors in {SQL} queries: A quite complete
  list}.
\newblock \bibinfo{journal}{Journal of Systems and Software}
  \bibinfo{year}{2006};\bibinfo{volume}{79}(\bibinfo{number}{5}):\bibinfo{pages}{630--644}.
\newblock \URLprefix
  \url{https://www.sciencedirect.com/science/article/pii/S016412120500124X}.
  \DOIprefix\doi{https://doi.org/10.1016/j.jss.2005.06.028};
  \bibinfo{note}{quality Software}.
\bibitem[{Brewer(2012)}]{Brewer_2012}
\bibinfo{author}{Brewer\xfnm[ E.]}.
\newblock \bibinfo{title}{{CAP} twelve years later: How the "rules" have
  changed}.
\newblock \bibinfo{journal}{Computer}
  \bibinfo{year}{2012};\bibinfo{volume}{45}(\bibinfo{number}{2}):\bibinfo{pages}{23--29}.
\newblock \DOIprefix\doi{10.1109/MC.2012.37}.
\bibitem[{Cass(2022)}]{Stephen_2022}
\bibinfo{author}{Cass\xfnm[ S.]}.
\newblock \bibinfo{title}{{SQL} should be your second language}.
\newblock \bibinfo{journal}{IEEE Spectrum}
  \bibinfo{year}{2022};\bibinfo{volume}{59}(\bibinfo{number}{10}):\bibinfo{pages}{20--21}.
\newblock \DOIprefix\doi{10.1109/MSPEC.2022.9915547}.
\bibitem[{Chaudhry and Yousaf(2020)}]{Chaudhry_2020}
\bibinfo{author}{Chaudhry\xfnm[ N.]}, \bibinfo{author}{Yousaf\xfnm[ M.M.]}.
\newblock \bibinfo{title}{Architectural assessment of {NoSQL} and {NewSQL}
  systems}.
\newblock \bibinfo{journal}{Distributed and Parallel Databases}
  \bibinfo{year}{2020};\bibinfo{volume}{38}(\bibinfo{number}{4}):\bibinfo{pages}{881--926}.
\newblock \URLprefix \url{https://doi.org/10.1007\%2Fs10619-020-07310-1}.
  \DOIprefix\doi{10.1007/s10619-020-07310-1}.
\bibitem[{Chaudhuri(1998)}]{Chaudhuri_1998_PODS}
\bibinfo{author}{Chaudhuri\xfnm[ S.]}.
\newblock \bibinfo{title}{An overview of query optimization in relational
  systems}.
\newblock In: \bibinfo{booktitle}{Proceedings of the Seventeenth ACM
  SIGACT-SIGMOD-SIGART Symposium on Principles of Database Systems}.
  \bibinfo{address}{New York, NY, USA}: \bibinfo{publisher}{Association for
  Computing Machinery}; PODS '98; \bibinfo{year}{1998}. p.
  \bibinfo{pages}{34–43}.
\newblock \URLprefix \url{https://doi.org/10.1145/275487.275492}.
  \DOIprefix\doi{10.1145/275487.275492}.
\bibitem[{Chen(1976)}]{Chen_1976}
\bibinfo{author}{Chen\xfnm[ P.P.S.]}.
\newblock \bibinfo{title}{The {Entity-relationship} model - toward a unified
  view of data}.
\newblock \bibinfo{journal}{ACM Transactions on Database Systems}
  \bibinfo{year}{1976};\bibinfo{volume}{1}(\bibinfo{number}{1}):\bibinfo{pages}{9--36}.
\newblock \DOIprefix\doi{10.1145/320434.320440}.
\bibitem[{Christodoulakis(1984)}]{Christodoulakis_1984}
\bibinfo{author}{Christodoulakis\xfnm[ S.]}.
\newblock \bibinfo{title}{Implications of certain assumptions in database
  performance evauation}.
\newblock \bibinfo{journal}{ACM Transactions on Database Systems}
  \bibinfo{year}{1984};\bibinfo{volume}{9}(\bibinfo{number}{2}):\bibinfo{pages}{163–186}.
\newblock \URLprefix \url{https://doi.org/10.1145/329.318578}.
  \DOIprefix\doi{10.1145/329.318578}.
\bibitem[{Coates(2009)}]{Coates_2009}
\bibinfo{author}{Coates\xfnm[ S.S.]}.
\newblock \bibinfo{title}{Comparing the Performance of Open Source and
  Proprietary Relational Database Management Systems}.
\newblock Ph.D. thesis; \bibinfo{address}{Northcentral University};
  \bibinfo{year}{2009}.
\bibitem[{Codd(1970)}]{Codd_1970}
\bibinfo{author}{Codd\xfnm[ E.F.]}.
\newblock \bibinfo{title}{A relational model of data for large shared data
  banks}.
\newblock \bibinfo{journal}{Communications of the ACM}
  \bibinfo{year}{1970};\bibinfo{volume}{13}(\bibinfo{number}{6}):\bibinfo{pages}{377--387}.
\newblock \DOIprefix\doi{10.1145/362384.362685}.
\bibitem[{Codd(1972)}]{Codd_1972}
\bibinfo{author}{Codd\xfnm[ E.F.]}.
\newblock \bibinfo{title}{Further normalization of the data base relational
  model}.
\newblock \bibinfo{journal}{Data base systems}
  \bibinfo{year}{1972};\bibinfo{volume}{6}:\bibinfo{pages}{33--64}.
\bibitem[{Codd(1975)}]{Codd_1975}
\bibinfo{author}{Codd\xfnm[ E.F.]}.
\newblock \bibinfo{title}{Recent investigations in relational data base
  systems} \bibinfo{year}{1975};.
\bibitem[{Connolly and Begg(2015)}]{Connolly_2015}
\bibinfo{author}{Connolly\xfnm[ T.]}, \bibinfo{author}{Begg\xfnm[ C.]}.
\newblock \bibinfo{title}{Database Systems (6th. ed.)}.
\newblock \bibinfo{publisher}{Pearson}, \bibinfo{year}{2015}.
\bibitem[{Cooper et~al.(2010)Cooper, Silberstein, Tam, Ramakrishnan and
  Sears}]{Cooper_2010}
\bibinfo{author}{Cooper\xfnm[ B.F.]}, \bibinfo{author}{Silberstein\xfnm[ A.]},
  \bibinfo{author}{Tam\xfnm[ E.]}, \bibinfo{author}{Ramakrishnan\xfnm[ R.]},
  \bibinfo{author}{Sears\xfnm[ R.]}.
\newblock \bibinfo{title}{Benchmarking cloud serving systems with {YCSB}}.
\newblock In: \bibinfo{booktitle}{Proceedings of the 1st ACM Symposium on Cloud
  Computing}. \bibinfo{address}{New York, NY, USA}:
  \bibinfo{publisher}{Association for Computing Machinery}; SoCC '10;
  \bibinfo{year}{2010}. p. \bibinfo{pages}{143–154}.
\newblock \URLprefix \url{https://doi.org/10.1145/1807128.1807152}.
  \DOIprefix\doi{10.1145/1807128.1807152}.
\bibitem[{Corbett et~al.(2013)Corbett, Dean, Epstein, Fikes, Frost, Furman,
  Ghemawat, Gubarev, Heiser, Hochschild, Hsieh, Kanthak, Kogan, Li, Lloyd,
  Melnik, Mwaura, Nagle, Quinlan, Rao, Rolig, Saito, Szymaniak, Taylor, Wang
  and Woodford}]{Corbett_2013}
\bibinfo{author}{Corbett\xfnm[ J.C.]}, \bibinfo{author}{Dean\xfnm[ J.]},
  \bibinfo{author}{Epstein\xfnm[ M.]}, \bibinfo{author}{Fikes\xfnm[ A.]},
  \bibinfo{author}{Frost\xfnm[ C.]}, \bibinfo{author}{Furman\xfnm[ J.J.]},
  \bibinfo{author}{Ghemawat\xfnm[ S.]}, \bibinfo{author}{Gubarev\xfnm[ A.]},
  \bibinfo{author}{Heiser\xfnm[ C.]}, \bibinfo{author}{Hochschild\xfnm[ P.]},
  \bibinfo{author}{Hsieh\xfnm[ W.]}, \bibinfo{author}{Kanthak\xfnm[ S.]},
  \bibinfo{author}{Kogan\xfnm[ E.]}, \bibinfo{author}{Li\xfnm[ H.]},
  \bibinfo{author}{Lloyd\xfnm[ A.]}, \bibinfo{author}{Melnik\xfnm[ S.]},
  \bibinfo{author}{Mwaura\xfnm[ D.]}, \bibinfo{author}{Nagle\xfnm[ D.]},
  \bibinfo{author}{Quinlan\xfnm[ S.]}, \bibinfo{author}{Rao\xfnm[ R.]},
  \bibinfo{author}{Rolig\xfnm[ L.]}, \bibinfo{author}{Saito\xfnm[ Y.]},
  \bibinfo{author}{Szymaniak\xfnm[ M.]}, \bibinfo{author}{Taylor\xfnm[ C.]},
  \bibinfo{author}{Wang\xfnm[ R.]}, \bibinfo{author}{Woodford\xfnm[ D.]}.
\newblock \bibinfo{title}{Spanner: {Google's} globally distributed database}.
\newblock \bibinfo{journal}{{ACM} Transactions on Computer Systems}
  \bibinfo{year}{2013};\bibinfo{volume}{31}(\bibinfo{number}{3}):\bibinfo{pages}{1--22}.
\newblock \URLprefix \url{https://doi.org/10.1145\%2F2491245}.
  \DOIprefix\doi{10.1145/2491245}.
\bibitem[{Cortellessa et~al.(2011)Cortellessa, Di~Marco and
  Inverardi}]{Cortellessa_2011}
\bibinfo{author}{Cortellessa\xfnm[ V.]}, \bibinfo{author}{Di~Marco\xfnm[ A.]},
  \bibinfo{author}{Inverardi\xfnm[ P.]}.
\newblock \bibinfo{title}{Model-based software performance analysis}.
\newblock volume \bibinfo{volume}{980}.
\newblock \bibinfo{publisher}{Springer}, \bibinfo{year}{2011}.
\bibitem[{Date(2019)}]{Date_2019}
\bibinfo{author}{Date\xfnm[ C.J.]}.
\newblock \bibinfo{title}{Database design and relational theory: normal forms
  and all that jazz}.
\newblock \bibinfo{publisher}{Apress}, \bibinfo{year}{2019}.
\bibitem[{Davoudian et~al.(2018)Davoudian, Chen and Liu}]{Davoudian_2018}
\bibinfo{author}{Davoudian\xfnm[ A.]}, \bibinfo{author}{Chen\xfnm[ L.]},
  \bibinfo{author}{Liu\xfnm[ M.]}.
\newblock \bibinfo{title}{A survey on {NoSQL} stores}.
\newblock \bibinfo{journal}{ACM Computing Surveys}
  \bibinfo{year}{2018};\bibinfo{volume}{51}(\bibinfo{number}{2}).
\newblock \URLprefix \url{https://doi.org/10.1145/3158661}.
  \DOIprefix\doi{10.1145/3158661}.
\bibitem[{Delis and Roussopoulos(1993)}]{Delis_1993}
\bibinfo{author}{Delis\xfnm[ A.]}, \bibinfo{author}{Roussopoulos\xfnm[ N.]}.
\newblock \bibinfo{title}{Performance comparison of three modern {DBMS}
  architectures}.
\newblock \bibinfo{journal}{IEEE Transactions on Software Engineering}
  \bibinfo{year}{1993};\bibinfo{volume}{19}(\bibinfo{number}{2}):\bibinfo{pages}{120--138}.
\newblock \DOIprefix\doi{10.1109/32.214830}.
\bibitem[{DeWitt and Levine(2008)}]{DeWitt_2008}
\bibinfo{author}{DeWitt\xfnm[ D.J.]}, \bibinfo{author}{Levine\xfnm[ C.]}.
\newblock \bibinfo{title}{Not just correct, but correct and fast: A look at one
  of jim gray's contributions to database system performance}.
\newblock \bibinfo{journal}{SIGMOD Rec}
  \bibinfo{year}{2008};\bibinfo{volume}{37}(\bibinfo{number}{2}):\bibinfo{pages}{45–49}.
\newblock \URLprefix \url{https://doi.org/10.1145/1379387.1379403}.
  \DOIprefix\doi{10.1145/1379387.1379403}.
\bibitem[{Dey et~al.(2014)Dey, Fekete, Nambiar and Röhm}]{Dey_2014}
\bibinfo{author}{Dey\xfnm[ A.]}, \bibinfo{author}{Fekete\xfnm[ A.]},
  \bibinfo{author}{Nambiar\xfnm[ R.]}, \bibinfo{author}{Röhm\xfnm[ U.]}.
\newblock \bibinfo{title}{{YCSB+T}: Benchmarking web-scale transactional
  databases}.
\newblock In: \bibinfo{booktitle}{2014 IEEE 30th International Conference on
  Data Engineering Workshops}. \bibinfo{year}{2014}. p.
  \bibinfo{pages}{223--230}.
\newblock \DOIprefix\doi{10.1109/ICDEW.2014.6818330}.
\bibitem[{Dietrich et~al.(1992)Dietrich, Brown, Cortes-Rello and
  Wunderlin}]{Dietrich_1992}
\bibinfo{author}{Dietrich\xfnm[ S.W.]}, \bibinfo{author}{Brown\xfnm[ M.]},
  \bibinfo{author}{Cortes-Rello\xfnm[ E.]}, \bibinfo{author}{Wunderlin\xfnm[
  S.]}.
\newblock \bibinfo{title}{A practitioner's introduction to database performance
  benchmarks and measurements}.
\newblock \bibinfo{journal}{The Computer Journal}
  \bibinfo{year}{1992};\bibinfo{volume}{35}(\bibinfo{number}{4}):\bibinfo{pages}{322--331}.
\bibitem[{Difallah et~al.(2013)Difallah, Pavlo, Curino and
  Cudre-Mauroux}]{Difallah_2013}
\bibinfo{author}{Difallah\xfnm[ D.E.]}, \bibinfo{author}{Pavlo\xfnm[ A.]},
  \bibinfo{author}{Curino\xfnm[ C.]}, \bibinfo{author}{Cudre-Mauroux\xfnm[
  P.]}.
\newblock \bibinfo{title}{{OLTP-Bench}: An extensible testbed for benchmarking
  relational databases}.
\newblock \bibinfo{journal}{Proc VLDB Endow}
  \bibinfo{year}{2013};\bibinfo{volume}{7}(\bibinfo{number}{4}):\bibinfo{pages}{277–288}.
\newblock \URLprefix \url{https://doi.org/10.14778/2732240.2732246}.
  \DOIprefix\doi{10.14778/2732240.2732246}.
\bibitem[{Do et~al.(2011)Do, Zhang, Patel, DeWitt, Naughton and
  Halverson}]{Do_2011}
\bibinfo{author}{Do\xfnm[ J.]}, \bibinfo{author}{Zhang\xfnm[ D.]},
  \bibinfo{author}{Patel\xfnm[ J.M.]}, \bibinfo{author}{DeWitt\xfnm[ D.J.]},
  \bibinfo{author}{Naughton\xfnm[ J.F.]}, \bibinfo{author}{Halverson\xfnm[
  A.]}.
\newblock \bibinfo{title}{Turbocharging {DBMS} buffer pool using {SSDs}}.
\newblock In: \bibinfo{booktitle}{Proceedings of the 2011 ACM SIGMOD
  International Conference on Management of Data}. \bibinfo{address}{New York,
  NY, USA}: \bibinfo{publisher}{Association for Computing Machinery}; SIGMOD
  '11; \bibinfo{year}{2011}. p. \bibinfo{pages}{1113–1124}.
\newblock \URLprefix \url{https://doi.org/10.1145/1989323.1989442}.
  \DOIprefix\doi{10.1145/1989323.1989442}.
\bibitem[{Do et~al.(2022)Do, Graefe and Naughton}]{Do_2022}
\bibinfo{author}{Do\xfnm[ T.]}, \bibinfo{author}{Graefe\xfnm[ G.]},
  \bibinfo{author}{Naughton\xfnm[ J.]}.
\newblock \bibinfo{title}{Efficient sorting, duplicate removal, grouping, and
  aggregation}.
\newblock \bibinfo{journal}{ACM Transactions on Database Systems}
  \bibinfo{year}{2022};\URLprefix \url{https://doi.org/10.1145/3568027}.
  \DOIprefix\doi{10.1145/3568027}.
\bibitem[{Dreseler et~al.(2020)Dreseler, Boissier, Rabl and
  Uflacker}]{Dreseler_2020}
\bibinfo{author}{Dreseler\xfnm[ M.]}, \bibinfo{author}{Boissier\xfnm[ M.]},
  \bibinfo{author}{Rabl\xfnm[ T.]}, \bibinfo{author}{Uflacker\xfnm[ M.]}.
\newblock \bibinfo{title}{Quantifying {TPC-H} choke points and their
  optimizations}.
\newblock \bibinfo{journal}{Proc VLDB Endow}
  \bibinfo{year}{2020};\bibinfo{volume}{13}(\bibinfo{number}{8}):\bibinfo{pages}{1206–1220}.
\newblock \URLprefix \url{https://doi.org/10.14778/3389133.3389138}.
  \DOIprefix\doi{10.14778/3389133.3389138}.
\bibitem[{Elmasri and Navathe(2016)}]{Elmasri_2016}
\bibinfo{author}{Elmasri\xfnm[ R.]}, \bibinfo{author}{Navathe\xfnm[ S.B.]}.
\newblock \bibinfo{title}{Fundamentals of Database Systems (7th. ed.)}.
\newblock \bibinfo{publisher}{Pearson}, \bibinfo{year}{2016}.
\bibitem[{Elnikety et~al.(2006)Elnikety, Dropsho and Pedone}]{Elnikety_2006}
\bibinfo{author}{Elnikety\xfnm[ S.]}, \bibinfo{author}{Dropsho\xfnm[ S.]},
  \bibinfo{author}{Pedone\xfnm[ F.]}.
\newblock \bibinfo{title}{Tashkent: Uniting durability with transaction
  ordering for high-performance scalable database replication}.
\newblock \bibinfo{journal}{SIGOPS Oper Syst Rev}
  \bibinfo{year}{2006};\bibinfo{volume}{40}(\bibinfo{number}{4}):\bibinfo{pages}{117–130}.
\newblock \DOIprefix\doi{10.1145/1218063.1217947}.
\bibitem[{Estivill-Castro and Wood(1992)}]{Estivill-Castro_1992}
\bibinfo{author}{Estivill-Castro\xfnm[ V.]}, \bibinfo{author}{Wood\xfnm[ D.]}.
\newblock \bibinfo{title}{A survey of adaptive sorting algorithms}.
\newblock \bibinfo{journal}{ACM Computing Surveys}
  \bibinfo{year}{1992};\bibinfo{volume}{24}(\bibinfo{number}{4}):\bibinfo{pages}{441–476}.
\newblock \URLprefix \url{https://doi.org/10.1145/146370.146381}.
  \DOIprefix\doi{10.1145/146370.146381}.
\bibitem[{Forresi et~al.(2022)Forresi, Francia, Gallinucci and
  Golfarelli}]{Forresi_2022}
\bibinfo{author}{Forresi\xfnm[ C.]}, \bibinfo{author}{Francia\xfnm[ M.]},
  \bibinfo{author}{Gallinucci\xfnm[ E.]}, \bibinfo{author}{Golfarelli\xfnm[
  M.]}.
\newblock \bibinfo{title}{Cost-based optimization of multistore query plans}.
\newblock \bibinfo{journal}{Information Systems Frontiers}
  \bibinfo{year}{2022};:\bibinfo{pages}{1--27}.
\bibitem[{Gilbert and Lynch(2002)}]{Gilbert_2002}
\bibinfo{author}{Gilbert\xfnm[ S.]}, \bibinfo{author}{Lynch\xfnm[ N.]}.
\newblock \bibinfo{title}{Brewer's conjecture and the feasibility of
  consistent, available, partition-tolerant web services}.
\newblock \bibinfo{journal}{SIGACT News}
  \bibinfo{year}{2002};\bibinfo{volume}{33}(\bibinfo{number}{2}):\bibinfo{pages}{51–59}.
\newblock \URLprefix \url{https://doi.org/10.1145/564585.564601}.
  \DOIprefix\doi{10.1145/564585.564601}.
\bibitem[{Graefe(1993)}]{Graefe_1993}
\bibinfo{author}{Graefe\xfnm[ G.]}.
\newblock \bibinfo{title}{Query evaluation techniques for large databases}.
\newblock \bibinfo{journal}{ACM Computing Surveys}
  \bibinfo{year}{1993};\bibinfo{volume}{25}(\bibinfo{number}{2}):\bibinfo{pages}{73–169}.
\newblock \URLprefix \url{https://doi.org/10.1145/152610.152611}.
  \DOIprefix\doi{10.1145/152610.152611}.
\bibitem[{Gray(1992)}]{Gray_1992}
\bibinfo{author}{Gray\xfnm[ J.]}.
\newblock \bibinfo{title}{Benchmark Handbook: For Database and Transaction
  Processing Systems}.
\newblock \bibinfo{address}{San Francisco, CA, USA}: \bibinfo{publisher}{Morgan
  Kaufmann Publishers Inc.}, \bibinfo{year}{1992}.
\bibitem[{Grolinger et~al.(2013)Grolinger, Higashino, Tiwari and
  Capretz}]{Grolinger_2013}
\bibinfo{author}{Grolinger\xfnm[ K.]}, \bibinfo{author}{Higashino\xfnm[ W.A.]},
  \bibinfo{author}{Tiwari\xfnm[ A.]}, \bibinfo{author}{Capretz\xfnm[ M.A.]}.
\newblock \bibinfo{title}{Data management in cloud environments: {NoSQL} and
  {NewSQL} data stores}.
\newblock \bibinfo{journal}{Journal of Cloud Computing: Advances, Systems and
  Applications}
  \bibinfo{year}{2013};\bibinfo{volume}{2}(\bibinfo{number}{1}):\bibinfo{pages}{22}.
\newblock \URLprefix \url{https://doi.org/10.1186\%2F2192-113x-2-22}.
  \DOIprefix\doi{10.1186/2192-113x-2-22}.
\bibitem[{Gunther(2011)}]{Gunther_2011}
\bibinfo{author}{Gunther\xfnm[ N.J.]}.
\newblock \bibinfo{title}{Analyzing Computer System Performance with
  {Perl::PDQ}}.
\newblock \bibinfo{publisher}{Springer}, \bibinfo{year}{2011}.
\bibitem[{Guo et~al.(2022)Guo, Yu, Yang, Leng and Liao}]{Guo_2022}
\bibinfo{author}{Guo\xfnm[ B.]}, \bibinfo{author}{Yu\xfnm[ J.]},
  \bibinfo{author}{Yang\xfnm[ D.]}, \bibinfo{author}{Leng\xfnm[ H.]},
  \bibinfo{author}{Liao\xfnm[ B.]}.
\newblock \bibinfo{title}{Energy-efficient database systems: A systematic
  survey}.
\newblock \bibinfo{journal}{ACM Computing Surveys}
  \bibinfo{year}{2022};\URLprefix \url{https://doi.org/10.1145/3538225}.
  \DOIprefix\doi{10.1145/3538225}.
\bibitem[{Guo et~al.(2005)Guo, Pan and Heflin}]{Guo_2005}
\bibinfo{author}{Guo\xfnm[ Y.]}, \bibinfo{author}{Pan\xfnm[ Z.]},
  \bibinfo{author}{Heflin\xfnm[ J.]}.
\newblock \bibinfo{title}{{LUBM}: A benchmark for {OWL} knowledge base
  systems}.
\newblock \bibinfo{journal}{Journal of Web Semantics}
  \bibinfo{year}{2005};\bibinfo{volume}{3}(\bibinfo{number}{2}):\bibinfo{pages}{158--182}.
\newblock \URLprefix
  \url{https://www.sciencedirect.com/science/article/pii/S1570826805000132}.
  \DOIprefix\doi{https://doi.org/10.1016/j.websem.2005.06.005}.
\bibitem[{Haerder and Reuter(1983)}]{Haerder_1983}
\bibinfo{author}{Haerder\xfnm[ T.]}, \bibinfo{author}{Reuter\xfnm[ A.]}.
\newblock \bibinfo{title}{Principles of transaction-oriented database
  recovery}.
\newblock \bibinfo{journal}{ACM Computing Surveys}
  \bibinfo{year}{1983};\bibinfo{volume}{15}(\bibinfo{number}{4}):\bibinfo{pages}{287–317}.
\newblock \URLprefix \url{https://doi.org/10.1145/289.291}.
  \DOIprefix\doi{10.1145/289.291}.
\bibitem[{Hecht and Jablonski(2011)}]{Hecht_2011}
\bibinfo{author}{Hecht\xfnm[ R.]}, \bibinfo{author}{Jablonski\xfnm[ S.]}.
\newblock \bibinfo{title}{{NoSQL} evaluation: A use case oriented survey}.
\newblock In: \bibinfo{booktitle}{2011 International Conference on Cloud and
  Service Computing}. \bibinfo{year}{2011}. p. \bibinfo{pages}{336--341}.
\newblock \DOIprefix\doi{10.1109/CSC.2011.6138544}.
\bibitem[{Hellerstein et~al.(2007)Hellerstein, Stonebraker and
  Hamilton}]{Hellerstein_2007}
\bibinfo{author}{Hellerstein\xfnm[ J.M.]}, \bibinfo{author}{Stonebraker\xfnm[
  M.]}, \bibinfo{author}{Hamilton\xfnm[ J.]}.
\newblock \bibinfo{title}{Architecture of a database system}.
\newblock \bibinfo{journal}{Foundations and Trends in Databases}
  \bibinfo{year}{2007};\bibinfo{volume}{1}(\bibinfo{number}{2}):\bibinfo{pages}{141--259}.
\newblock \URLprefix \url{http://dx.doi.org/10.1561/1900000002}.
  \DOIprefix\doi{10.1561/1900000002}.
\bibitem[{Holzschuher and Peinl(2013)}]{Holzschuher_2013}
\bibinfo{author}{Holzschuher\xfnm[ F.]}, \bibinfo{author}{Peinl\xfnm[ R.]}.
\newblock \bibinfo{title}{Performance of graph query languages: Comparison of
  {Cypher}, {Gremlin} and native access in {Neo4j}}.
\newblock \bibinfo{address}{New York, NY, USA}: \bibinfo{publisher}{Association
  for Computing Machinery}; EDBT '13; \bibinfo{year}{2013}. p.
  \bibinfo{pages}{195–204}.
\newblock \URLprefix \url{https://doi.org/10.1145/2457317.2457351}.
  \DOIprefix\doi{10.1145/2457317.2457351}.
\bibitem[{ISO/IEC(2016{\natexlab{a}})}]{SQL_2016p1}
\bibinfo{author}{ISO/IEC\xfnm[]}.
\newblock \bibinfo{title}{{ISO/IEC 9075-1:2016 - SQL - Part 1: Framework}}.
\newblock \bibinfo{type}{Technical Report}; \bibinfo{year}{2016}{\natexlab{a}}.
\newblock \URLprefix \url{https://www.iso.org/standard/63555.html}.
\bibitem[{ISO/IEC(2016{\natexlab{b}})}]{SQL_2016p2}
\bibinfo{author}{ISO/IEC\xfnm[]}.
\newblock \bibinfo{title}{{ISO/IEC 9075-2:2016 - SQL - Part 2: Foundation}}.
\newblock \bibinfo{type}{Technical Report}; \bibinfo{year}{2016}{\natexlab{b}}.
\newblock \URLprefix \url{https://www.iso.org/standard/63556.html}.
\bibitem[{Jiang et~al.(2010)Jiang, Ooi, Shi and Wu}]{Jiang_2010}
\bibinfo{author}{Jiang\xfnm[ D.]}, \bibinfo{author}{Ooi\xfnm[ B.C.]},
  \bibinfo{author}{Shi\xfnm[ L.]}, \bibinfo{author}{Wu\xfnm[ S.]}.
\newblock \bibinfo{title}{The performance of {MapReduce}: An in-depth study}.
\newblock \bibinfo{journal}{Proc VLDB Endow}
  \bibinfo{year}{2010};\bibinfo{volume}{3}(\bibinfo{number}{1–2}):\bibinfo{pages}{472–483}.
\newblock \URLprefix \url{https://doi.org/10.14778/1920841.1920903}.
  \DOIprefix\doi{10.14778/1920841.1920903}.
\bibitem[{Jin et~al.(2012)Jin, Song, Shi, Scherpelz and Lu}]{Jin_2012}
\bibinfo{author}{Jin\xfnm[ G.]}, \bibinfo{author}{Song\xfnm[ L.]},
  \bibinfo{author}{Shi\xfnm[ X.]}, \bibinfo{author}{Scherpelz\xfnm[ J.]},
  \bibinfo{author}{Lu\xfnm[ S.]}.
\newblock \bibinfo{title}{Understanding and detecting real-world performance
  bugs}.
\newblock In: \bibinfo{booktitle}{Proceedings of the 33rd ACM SIGPLAN
  Conference on Programming Language Design and Implementation}.
  \bibinfo{address}{New York, NY, USA}: \bibinfo{publisher}{Association for
  Computing Machinery}; PLDI '12; \bibinfo{year}{2012}. p.
  \bibinfo{pages}{77–88}.
\newblock \URLprefix \url{https://doi.org/10.1145/2254064.2254075}.
  \DOIprefix\doi{10.1145/2254064.2254075}.
\bibitem[{{Joint Task Force on Computing Curricula, Association for Computing
  Machinery (ACM) and IEEE Computer Society}(2013)}]{CS_2013}
\bibinfo{author}{{Joint Task Force on Computing Curricula, Association for
  Computing Machinery (ACM) and IEEE Computer Society}\xfnm[]}.
\newblock \bibinfo{title}{Computer Science Curricula 2013: Curriculum
  Guidelines for Undergraduate Degree Programs in Computer Science}.
\newblock \bibinfo{type}{Technical Report}; \bibinfo{address}{New York, NY,
  USA}; \bibinfo{year}{2013}.
\newblock \URLprefix \url{doi.org/10.1145/2534860}.
  \DOIprefix\doi{10.1145/2534860}; \bibinfo{note}{999133}.
\bibitem[{Juran and De~Feo(2010)}]{Juran_2010}
\bibinfo{author}{Juran\xfnm[ J.M.]}, \bibinfo{author}{De~Feo\xfnm[ J.A.]}.
\newblock \bibinfo{title}{Juran's quality handbook: the complete guide to
  performance excellence (6th. ed.)}.
\newblock \bibinfo{publisher}{McGraw-Hill Education}, \bibinfo{year}{2010}.
\bibitem[{Kim et~al.(2018)Kim, Ko, Jeon and Lee}]{Kim_2018}
\bibinfo{author}{Kim\xfnm[ H.J.]}, \bibinfo{author}{Ko\xfnm[ E.J.]},
  \bibinfo{author}{Jeon\xfnm[ Y.H.]}, \bibinfo{author}{Lee\xfnm[ K.H.]}.
\newblock \bibinfo{title}{Migration from {RDBMS} to column-oriented {NoSQL}:
  Lessons learned and open problems}.
\newblock In: \bibinfo{editor}{Lee\xfnm[ W.]}, \bibinfo{editor}{Choi\xfnm[
  W.]}, \bibinfo{editor}{Jung\xfnm[ S.]}, \bibinfo{editor}{Song\xfnm[ M.]},
  editors. \bibinfo{booktitle}{Proceedings of the 7th International Conference
  on Emerging Databases}. \bibinfo{address}{Singapore}:
  \bibinfo{publisher}{Springer Singapore}; \bibinfo{year}{2018}. p.
  \bibinfo{pages}{25--33}.
\bibitem[{Kim and Patel(2010)}]{Jung_2009}
\bibinfo{author}{Kim\xfnm[ Y.J.]}, \bibinfo{author}{Patel\xfnm[ J.]}.
\newblock \bibinfo{title}{Performance comparison of the {R*-Tree} and the
  quadtree for {kNN} and distance join queries}.
\newblock \bibinfo{journal}{IEEE Transactions on Knowledge and Data
  Engineering}
  \bibinfo{year}{2010};\bibinfo{volume}{22}(\bibinfo{number}{7}):\bibinfo{pages}{1014--1027}.
\newblock \DOIprefix\doi{10.1109/TKDE.2009.141}.
\bibitem[{Kumar and Grot(2022)}]{Kumar_2022}
\bibinfo{author}{Kumar\xfnm[ R.]}, \bibinfo{author}{Grot\xfnm[ B.]}.
\newblock \bibinfo{title}{Shooting down the server front-end bottleneck}.
\newblock \bibinfo{journal}{ACM Transactions on Computer Systems}
  \bibinfo{year}{2022};\bibinfo{volume}{38}(\bibinfo{number}{3–4}).
\newblock \URLprefix \url{https://doi.org/10.1145/3484492}.
  \DOIprefix\doi{10.1145/3484492}.
\bibitem[{Leis et~al.(2015)Leis, Gubichev, Mirchev, Boncz, Kemper and
  Neumann}]{Leis_2015}
\bibinfo{author}{Leis\xfnm[ V.]}, \bibinfo{author}{Gubichev\xfnm[ A.]},
  \bibinfo{author}{Mirchev\xfnm[ A.]}, \bibinfo{author}{Boncz\xfnm[ P.]},
  \bibinfo{author}{Kemper\xfnm[ A.]}, \bibinfo{author}{Neumann\xfnm[ T.]}.
\newblock \bibinfo{title}{How good are query optimizers, really?}
\newblock \bibinfo{journal}{Proc VLDB Endow}
  \bibinfo{year}{2015};\bibinfo{volume}{9}(\bibinfo{number}{3}):\bibinfo{pages}{204–215}.
\newblock \URLprefix \url{https://doi.org/10.14778/2850583.2850594}.
  \DOIprefix\doi{10.14778/2850583.2850594}.
\bibitem[{Lightstone et~al.(2010)Lightstone, Teorey and
  Nadeau}]{Lightstone_2010}
\bibinfo{author}{Lightstone\xfnm[ S.S.]}, \bibinfo{author}{Teorey\xfnm[ T.J.]},
  \bibinfo{author}{Nadeau\xfnm[ T.]}.
\newblock \bibinfo{title}{Physical Database Design: the database professional's
  guide to exploiting indexes, views, storage, and more}.
\newblock \bibinfo{publisher}{Morgan Kaufmann}, \bibinfo{year}{2010}.
\bibitem[{Lu and Holubov\'{a}(2019)}]{Lu_2019}
\bibinfo{author}{Lu\xfnm[ J.]}, \bibinfo{author}{Holubov\'{a}\xfnm[ I.]}.
\newblock \bibinfo{title}{Multi-model databases: A new journey to handle the
  variety of data}.
\newblock \bibinfo{journal}{ACM Computing Surveys}
  \bibinfo{year}{2019};\bibinfo{volume}{52}(\bibinfo{number}{3}).
\newblock \URLprefix \url{https://doi.org/10.1145/3323214}.
  \DOIprefix\doi{10.1145/3323214}.
\bibitem[{Marek and Rahm(1992)}]{Marek_1992}
\bibinfo{author}{Marek\xfnm[ R.]}, \bibinfo{author}{Rahm\xfnm[ E.]}.
\newblock \bibinfo{title}{Performance evaluation of parallel transaction
  processing in shared nothing database systems}.
\newblock In: \bibinfo{editor}{Etiemble\xfnm[ D.]}, \bibinfo{editor}{Syre\xfnm[
  J.C.]}, editors. \bibinfo{booktitle}{PARLE '92 Parallel Architectures and
  Languages Europe}. \bibinfo{address}{Berlin, Heidelberg}:
  \bibinfo{publisher}{Springer Berlin Heidelberg}; \bibinfo{year}{1992}. p.
  \bibinfo{pages}{295--310}.
\bibitem[{Obradovic et~al.(2019)Obradovic, Kelec and Dujlovic}]{Obradovic_2019}
\bibinfo{author}{Obradovic\xfnm[ N.]}, \bibinfo{author}{Kelec\xfnm[ A.]},
  \bibinfo{author}{Dujlovic\xfnm[ I.]}.
\newblock \bibinfo{title}{Performance analysis on {Android SQLite} database}.
\newblock In: \bibinfo{booktitle}{2019 18th International Symposium
  INFOTEH-JAHORINA (INFOTEH)}. \bibinfo{year}{2019}. p. \bibinfo{pages}{1--4}.
\newblock \DOIprefix\doi{10.1109/INFOTEH.2019.8717652}.
\bibitem[{Osterhage(2013)}]{Osterhage_2013}
\bibinfo{author}{Osterhage\xfnm[ W.]}.
\newblock \bibinfo{title}{Computer Performance Optimization}.
\newblock \bibinfo{publisher}{Springer}, \bibinfo{year}{2013}.
\bibitem[{Patel and DeWitt(1996)}]{Patel_1996}
\bibinfo{author}{Patel\xfnm[ J.M.]}, \bibinfo{author}{DeWitt\xfnm[ D.J.]}.
\newblock \bibinfo{title}{Partition based spatial-merge join}.
\newblock In: \bibinfo{booktitle}{Proceedings of the 1996 ACM SIGMOD
  International Conference on Management of Data}. \bibinfo{address}{New York,
  NY, USA}: \bibinfo{publisher}{Association for Computing Machinery}; SIGMOD
  '96; \bibinfo{year}{1996}. p. \bibinfo{pages}{259–270}.
\newblock \URLprefix \url{https://doi.org/10.1145/233269.233338}.
  \DOIprefix\doi{10.1145/233269.233338}.
\bibitem[{Patounas et~al.(2020)Patounas, Foukas, Elmokashfi and
  Marina}]{Patounas_2020}
\bibinfo{author}{Patounas\xfnm[ G.]}, \bibinfo{author}{Foukas\xfnm[ X.]},
  \bibinfo{author}{Elmokashfi\xfnm[ A.]}, \bibinfo{author}{Marina\xfnm[ M.K.]}.
\newblock \bibinfo{title}{Characterization and identification of cloudified
  mobile network performance bottlenecks}.
\newblock \bibinfo{journal}{IEEE Transactions on Network and Service
  Management}
  \bibinfo{year}{2020};\bibinfo{volume}{17}(\bibinfo{number}{4}):\bibinfo{pages}{2567--2583}.
\newblock \DOIprefix\doi{10.1109/TNSM.2020.3018538}.
\bibitem[{Pavlo(2017)}]{Pavlo_2017}
\bibinfo{author}{Pavlo\xfnm[ A.]}.
\newblock \bibinfo{title}{What are we doing with our lives? nobody cares about
  our concurrency control research}.
\newblock In: \bibinfo{booktitle}{Proceedings of the 2017 ACM International
  Conference on Management of Data}. \bibinfo{address}{New York, NY, USA}:
  \bibinfo{publisher}{Association for Computing Machinery}; SIGMOD '17;
  \bibinfo{year}{2017}. p.~\bibinfo{pages}{3}.
\newblock \URLprefix \url{https://doi.org/10.1145/3035918.3056096}.
  \DOIprefix\doi{10.1145/3035918.3056096}.
\bibitem[{Pavlo and Aslett(2016)}]{Pavlo_2016}
\bibinfo{author}{Pavlo\xfnm[ A.]}, \bibinfo{author}{Aslett\xfnm[ M.]}.
\newblock \bibinfo{title}{What's really new with {NewSQL}?}
\newblock \bibinfo{journal}{{SIGMOD} Rec}
  \bibinfo{year}{2016};\bibinfo{volume}{45}(\bibinfo{number}{2}):\bibinfo{pages}{45--55}.
\newblock \URLprefix \url{https://doi.org/10.1145/3003665.3003674}.
  \DOIprefix\doi{10.1145/3003665.3003674}.
\bibitem[{Purbo et~al.(2020)Purbo, Sriyanto, Suhendro, Aziz and
  Herwanto}]{Purbo_2020}
\bibinfo{author}{Purbo\xfnm[ O.W.]}, \bibinfo{author}{Sriyanto\xfnm[ S.]},
  \bibinfo{author}{Suhendro\xfnm[ S.]}, \bibinfo{author}{Aziz\xfnm[ R.A.]},
  \bibinfo{author}{Herwanto\xfnm[ R.]}.
\newblock \bibinfo{title}{Benchmark and comparison between hyperledger and
  {MySQL}}.
\newblock \bibinfo{journal}{TELKOMNIKA (Telecommunication Computing Electronics
  and Control)}
  \bibinfo{year}{2020};\bibinfo{volume}{18}(\bibinfo{number}{2}):\bibinfo{pages}{705--715}.
\bibitem[{Purohith et~al.(2017)Purohith, Mohan and Chidambaram}]{Purohith_2017}
\bibinfo{author}{Purohith\xfnm[ D.]}, \bibinfo{author}{Mohan\xfnm[ J.]},
  \bibinfo{author}{Chidambaram\xfnm[ V.]}.
\newblock \bibinfo{title}{The dangers and complexities of {SQLite}
  benchmarking}.
\newblock In: \bibinfo{booktitle}{Proceedings of the 8th Asia-Pacific Workshop
  on Systems}. \bibinfo{address}{New York, NY, USA}:
  \bibinfo{publisher}{Association for Computing Machinery}; APSys '17;
  \bibinfo{year}{2017}. \URLprefix
  \url{https://doi.org/10.1145/3124680.3124719}.
  \DOIprefix\doi{10.1145/3124680.3124719}.
\bibitem[{Qu et~al.(2022{\natexlab{a}})Qu, Li, Zhang, Chen, Shu, Qian and
  Zhou}]{Qu_2022_workload}
\bibinfo{author}{Qu\xfnm[ L.]}, \bibinfo{author}{Li\xfnm[ Y.]},
  \bibinfo{author}{Zhang\xfnm[ R.]}, \bibinfo{author}{Chen\xfnm[ T.]},
  \bibinfo{author}{Shu\xfnm[ K.]}, \bibinfo{author}{Qian\xfnm[ W.]},
  \bibinfo{author}{Zhou\xfnm[ A.]}.
\newblock \bibinfo{title}{Application-oriented workload generation for
  transactional database performance evaluation}.
\newblock In: \bibinfo{booktitle}{2022 IEEE 38th International Conference on
  Data Engineering (ICDE)}. \bibinfo{year}{2022}{\natexlab{a}}. p.
  \bibinfo{pages}{420--432}.
\newblock \DOIprefix\doi{10.1109/ICDE53745.2022.00036}.
\bibitem[{Qu et~al.(2022{\natexlab{b}})Qu, Wang, Chen, Li, Zhang, Zhou, Xu,
  Yang, Yang, Qian and Zhou}]{Qu_2022}
\bibinfo{author}{Qu\xfnm[ L.]}, \bibinfo{author}{Wang\xfnm[ Q.]},
  \bibinfo{author}{Chen\xfnm[ T.]}, \bibinfo{author}{Li\xfnm[ K.]},
  \bibinfo{author}{Zhang\xfnm[ R.]}, \bibinfo{author}{Zhou\xfnm[ X.]},
  \bibinfo{author}{Xu\xfnm[ Q.]}, \bibinfo{author}{Yang\xfnm[ Z.]},
  \bibinfo{author}{Yang\xfnm[ C.]}, \bibinfo{author}{Qian\xfnm[ W.]},
  \bibinfo{author}{Zhou\xfnm[ A.]}.
\newblock \bibinfo{title}{Are current benchmarks adequate to evaluate
  distributed transactional databases?}
\newblock \bibinfo{journal}{BenchCouncil Transactions on Benchmarks, Standards
  and Evaluations}
  \bibinfo{year}{2022}{\natexlab{b}};\bibinfo{volume}{2}(\bibinfo{number}{1}):\bibinfo{pages}{100031}.
\newblock \URLprefix
  \url{https://www.sciencedirect.com/science/article/pii/S2772485922000187}.
  \DOIprefix\doi{https://doi.org/10.1016/j.tbench.2022.100031}.
\bibitem[{Raasveldt et~al.(2018)Raasveldt, Holanda, Gubner and
  M\"{u}hleisen}]{Raasveldt_2018}
\bibinfo{author}{Raasveldt\xfnm[ M.]}, \bibinfo{author}{Holanda\xfnm[ P.]},
  \bibinfo{author}{Gubner\xfnm[ T.]}, \bibinfo{author}{M\"{u}hleisen\xfnm[
  H.]}.
\newblock \bibinfo{title}{Fair benchmarking considered difficult: Common
  pitfalls in database performance testing}.
\newblock In: \bibinfo{booktitle}{Proceedings of the Workshop on Testing
  Database Systems}. \bibinfo{address}{New York, NY, USA}:
  \bibinfo{publisher}{Association for Computing Machinery}; DBTest'18;
  \bibinfo{year}{2018}. \URLprefix
  \url{https://doi.org/10.1145/3209950.3209955}.
  \DOIprefix\doi{10.1145/3209950.3209955}.
\bibitem[{Ramakrishnan(2012)}]{Ramakrishnan_2012}
\bibinfo{author}{Ramakrishnan\xfnm[ R.]}.
\newblock \bibinfo{title}{{CAP} and cloud data management}.
\newblock \bibinfo{journal}{Computer}
  \bibinfo{year}{2012};\bibinfo{volume}{45}(\bibinfo{number}{2}):\bibinfo{pages}{43--49}.
\newblock \DOIprefix\doi{10.1109/MC.2011.388}.
\bibitem[{Reniers et~al.(2017)Reniers, Van~Landuyt, Rafique and
  Joosen}]{Reniers_2017}
\bibinfo{author}{Reniers\xfnm[ V.]}, \bibinfo{author}{Van~Landuyt\xfnm[ D.]},
  \bibinfo{author}{Rafique\xfnm[ A.]}, \bibinfo{author}{Joosen\xfnm[ W.]}.
\newblock \bibinfo{title}{On the state of {NoSQL} benchmarks}.
\newblock In: \bibinfo{booktitle}{Proceedings of the 8th ACM/SPEC on
  International Conference on Performance Engineering Companion}.
  \bibinfo{address}{New York, NY, USA}: \bibinfo{publisher}{Association for
  Computing Machinery}; ICPE '17 Companion; \bibinfo{year}{2017}. p.
  \bibinfo{pages}{107–112}.
\newblock \URLprefix \url{https://doi.org/10.1145/3053600.3053622}.
  \DOIprefix\doi{10.1145/3053600.3053622}.
\bibitem[{Schneider and DeWitt(1989)}]{Schneider_1989}
\bibinfo{author}{Schneider\xfnm[ D.A.]}, \bibinfo{author}{DeWitt\xfnm[ D.J.]}.
\newblock \bibinfo{title}{A performance evaluation of four parallel join
  algorithms in a shared-nothing multiprocessor environment}.
\newblock In: \bibinfo{booktitle}{Proceedings of the 1989 ACM SIGMOD
  International Conference on Management of Data}. \bibinfo{address}{New York,
  NY, USA}: \bibinfo{publisher}{Association for Computing Machinery}; SIGMOD
  '89; \bibinfo{year}{1989}. p. \bibinfo{pages}{110–121}.
\newblock \URLprefix \url{https://doi.org/10.1145/67544.66937}.
  \DOIprefix\doi{10.1145/67544.66937}.
\bibitem[{Stonebraker(2010)}]{Stonebraker_2010}
\bibinfo{author}{Stonebraker\xfnm[ M.]}.
\newblock \bibinfo{title}{{SQL} databases v. {NoSQL} databases}.
\newblock \bibinfo{journal}{Communications of the ACM}
  \bibinfo{year}{2010};\bibinfo{volume}{53}(\bibinfo{number}{4}):\bibinfo{pages}{10–11}.
\newblock \URLprefix \url{https://doi.org/10.1145/1721654.1721659}.
  \DOIprefix\doi{10.1145/1721654.1721659}.
\bibitem[{Stonebraker et~al.(2007)Stonebraker, Bear, {\c{C}}etintemel,
  Cherniack, Ge, Hachem, Harizopoulos, Lifter, Rogers and
  Zdonik}]{Stonebraker_2007}
\bibinfo{author}{Stonebraker\xfnm[ M.]}, \bibinfo{author}{Bear\xfnm[ C.]},
  \bibinfo{author}{{\c{C}}etintemel\xfnm[ U.]},
  \bibinfo{author}{Cherniack\xfnm[ M.]}, \bibinfo{author}{Ge\xfnm[ T.]},
  \bibinfo{author}{Hachem\xfnm[ N.]}, \bibinfo{author}{Harizopoulos\xfnm[ S.]},
  \bibinfo{author}{Lifter\xfnm[ J.]}, \bibinfo{author}{Rogers\xfnm[ J.]},
  \bibinfo{author}{Zdonik\xfnm[ S.]}.
\newblock \bibinfo{title}{One size fits all? part 2: Benchmarking results}.
\newblock In: \bibinfo{booktitle}{Proc. CIDR}. \bibinfo{year}{2007}. .
\bibitem[{Sundaresan et~al.(2013)Sundaresan, Magharei, Feamster, Teixeira and
  Crawford}]{Sundaresan_2013}
\bibinfo{author}{Sundaresan\xfnm[ S.]}, \bibinfo{author}{Magharei\xfnm[ N.]},
  \bibinfo{author}{Feamster\xfnm[ N.]}, \bibinfo{author}{Teixeira\xfnm[ R.]},
  \bibinfo{author}{Crawford\xfnm[ S.]}.
\newblock \bibinfo{title}{Web performance bottlenecks in broadband access
  networks}.
\newblock \bibinfo{journal}{SIGMETRICS Perform Eval Rev}
  \bibinfo{year}{2013};\bibinfo{volume}{41}(\bibinfo{number}{1}):\bibinfo{pages}{383–384}.
\newblock \URLprefix \url{https://doi.org/10.1145/2494232.2465745}.
  \DOIprefix\doi{10.1145/2494232.2465745}.
\bibitem[{Tallent and Mellor-Crummey(2009)}]{Tallent_2009}
\bibinfo{author}{Tallent\xfnm[ N.R.]}, \bibinfo{author}{Mellor-Crummey\xfnm[
  J.M.]}.
\newblock \bibinfo{title}{Identifying performance bottlenecks in work-stealing
  computations}.
\newblock \bibinfo{journal}{Computer}
  \bibinfo{year}{2009};\bibinfo{volume}{42}(\bibinfo{number}{12}):\bibinfo{pages}{44--50}.
\newblock \DOIprefix\doi{10.1109/MC.2009.396}.
\bibitem[{Thalheim and Wang(2013)}]{Thalheim_2013}
\bibinfo{author}{Thalheim\xfnm[ B.]}, \bibinfo{author}{Wang\xfnm[ Q.]}.
\newblock \bibinfo{title}{Data migration: A theoretical perspective}.
\newblock \bibinfo{journal}{Data \& Knowledge Engineering}
  \bibinfo{year}{2013};\bibinfo{volume}{87}:\bibinfo{pages}{260--278}.
\newblock \URLprefix
  \url{https://www.sciencedirect.com/science/article/pii/S0169023X12001048}.
  \DOIprefix\doi{https://doi.org/10.1016/j.datak.2012.12.003}.
\bibitem[{{The Joint Task Force on Computing Curricula}(2015)}]{SE_2015}
\bibinfo{author}{{The Joint Task Force on Computing Curricula}\xfnm[]}.
\newblock \bibinfo{title}{Curriculum Guidelines for Undergraduate Degree
  Programs in Software Engineering}.
\newblock \bibinfo{type}{Technical Report}; \bibinfo{address}{New York, NY,
  USA}; \bibinfo{year}{2015}.
\newblock \URLprefix \url{https://dl.acm.org/citation.cfm?id=2965631}.
\bibitem[{Toffola et~al.(2018)Toffola, Pradel and Gross}]{Toffola_2018}
\bibinfo{author}{Toffola\xfnm[ L.D.]}, \bibinfo{author}{Pradel\xfnm[ M.]},
  \bibinfo{author}{Gross\xfnm[ T.R.]}.
\newblock \bibinfo{title}{Synthesizing programs that expose performance
  bottlenecks}.
\newblock In: \bibinfo{booktitle}{Proceedings of the 2018 International
  Symposium on Code Generation and Optimization}. \bibinfo{address}{New York,
  NY, USA}: \bibinfo{publisher}{Association for Computing Machinery}; CGO 2018;
  \bibinfo{year}{2018}. p. \bibinfo{pages}{314–326}.
\newblock \URLprefix \url{https://doi.org/10.1145/3168830}.
  \DOIprefix\doi{10.1145/3168830}.
\bibitem[{T\"{o}z\"{u}n et~al.(2013)T\"{o}z\"{u}n, Pandis, Kaynak, Jevdjic and
  Ailamaki}]{Tozun_2013}
\bibinfo{author}{T\"{o}z\"{u}n\xfnm[ P.]}, \bibinfo{author}{Pandis\xfnm[ I.]},
  \bibinfo{author}{Kaynak\xfnm[ C.]}, \bibinfo{author}{Jevdjic\xfnm[ D.]},
  \bibinfo{author}{Ailamaki\xfnm[ A.]}.
\newblock \bibinfo{title}{From {A to E}: Analyzing {TPC's OLTP} benchmarks: The
  obsolete, the ubiquitous, the unexplored}.
\newblock \bibinfo{address}{New York, NY, USA}: \bibinfo{publisher}{Association
  for Computing Machinery}; EDBT '13; \bibinfo{year}{2013}. p.
  \bibinfo{pages}{17–28}.
\newblock \URLprefix \url{https://doi.org/10.1145/2452376.2452380}.
  \DOIprefix\doi{10.1145/2452376.2452380}.
\bibitem[{Tu et~al.(2013)Tu, Zheng, Kohler, Liskov and Madden}]{Tu_2013}
\bibinfo{author}{Tu\xfnm[ S.]}, \bibinfo{author}{Zheng\xfnm[ W.]},
  \bibinfo{author}{Kohler\xfnm[ E.]}, \bibinfo{author}{Liskov\xfnm[ B.]},
  \bibinfo{author}{Madden\xfnm[ S.]}.
\newblock \bibinfo{title}{Speedy transactions in multicore in-memory
  databases}.
\newblock In: \bibinfo{booktitle}{Proceedings of the Twenty-Fourth ACM
  Symposium on Operating Systems Principles}. \bibinfo{address}{New York, NY,
  USA}: \bibinfo{publisher}{Association for Computing Machinery}; SOSP '13;
  \bibinfo{year}{2013}. p. \bibinfo{pages}{18–32}.
\newblock \URLprefix \url{https://doi.org/10.1145/2517349.2522713}.
  \DOIprefix\doi{10.1145/2517349.2522713}.
\bibitem[{Valduriez(1987)}]{Valduriez_1987}
\bibinfo{author}{Valduriez\xfnm[ P.]}.
\newblock \bibinfo{title}{Join indices}.
\newblock \bibinfo{journal}{ACM Transactions on Database Systems}
  \bibinfo{year}{1987};\bibinfo{volume}{12}(\bibinfo{number}{2}):\bibinfo{pages}{218–246}.
\newblock \URLprefix \url{https://doi.org/10.1145/22952.22955}.
  \DOIprefix\doi{10.1145/22952.22955}.
\bibitem[{Wang et~al.(2022)Wang, Yu, Hui, Zhou, Huang, Zhu, Ren, Li and
  Lu}]{Wang_2022}
\bibinfo{author}{Wang\xfnm[ Y.]}, \bibinfo{author}{Yu\xfnm[ M.]},
  \bibinfo{author}{Hui\xfnm[ Y.]}, \bibinfo{author}{Zhou\xfnm[ F.]},
  \bibinfo{author}{Huang\xfnm[ Y.]}, \bibinfo{author}{Zhu\xfnm[ R.]},
  \bibinfo{author}{Ren\xfnm[ X.]}, \bibinfo{author}{Li\xfnm[ T.]},
  \bibinfo{author}{Lu\xfnm[ X.]}.
\newblock \bibinfo{title}{A study of database performance sensitivity to
  experiment settings.}
\newblock \bibinfo{journal}{Proceedings of the VLDB Endowment}
  \bibinfo{year}{2022};\bibinfo{volume}{15}(\bibinfo{number}{7}).
\bibitem[{Warszawski and Bailis(2017)}]{Warszawski_2017}
\bibinfo{author}{Warszawski\xfnm[ T.]}, \bibinfo{author}{Bailis\xfnm[ P.]}.
\newblock \bibinfo{title}{{ACIDRain}: Concurrency-related attacks on
  database-backed web applications}.
\newblock In: \bibinfo{booktitle}{Proceedings of the 2017 ACM International
  Conference on Management of Data}. \bibinfo{address}{New York, NY, USA}:
  \bibinfo{publisher}{Association for Computing Machinery}; SIGMOD '17;
  \bibinfo{year}{2017}. p. \bibinfo{pages}{5–20}.
\newblock \URLprefix \url{https://doi.org/10.1145/3035918.3064037}.
  \DOIprefix\doi{10.1145/3035918.3064037}.
\bibitem[{Winand(2012)}]{Winand_2011}
\bibinfo{author}{Winand\xfnm[ M.]}.
\newblock \bibinfo{title}{{SQL} performance explained}.
\newblock \bibinfo{publisher}{Markus Winand}, \bibinfo{year}{2012}.
\bibitem[{Yang and Lilja(2018)}]{Yang_2018}
\bibinfo{author}{Yang\xfnm[ J.]}, \bibinfo{author}{Lilja\xfnm[ D.J.]}.
\newblock \bibinfo{title}{Reducing relational database performance bottlenecks
  using {3D XPoint} storage technology}.
\newblock In: \bibinfo{booktitle}{2018 17th IEEE International Conference On
  Trust, Security And Privacy In Computing And Communications/ 12th IEEE
  International Conference On Big Data Science And Engineering
  (TrustCom/BigDataSE)}. \bibinfo{year}{2018}. p. \bibinfo{pages}{1804--1808}.
\newblock \DOIprefix\doi{10.1109/TrustCom/BigDataSE.2018.00272}.
\bibitem[{Yu and Pradel(2018)}]{Yu_2018}
\bibinfo{author}{Yu\xfnm[ T.]}, \bibinfo{author}{Pradel\xfnm[ M.]}.
\newblock \bibinfo{title}{Pinpointing and repairing performance bottlenecks in
  concurrent programs}.
\newblock \bibinfo{journal}{Empirical Software Engineering}
  \bibinfo{year}{2018};\bibinfo{volume}{23}(\bibinfo{number}{5}):\bibinfo{pages}{3034--3071}.
\bibitem{Abdullah_2015}
Ahmad Abdullah and Qingfeng Zhuge.
\newblock From relational databases to {NoSQL} databases: Performance
  evaluation.
\newblock {\em Research Journal of Applied Sciences, Engineering and
  Technology}, 11(4):434--439, 2015.

\bibitem{aboutorabi_2015}
Seyyed~Hamid Aboutorabi, Mehdi Rezapour, Milad Moradi, and Nasser Ghadiri.
\newblock Performance evaluation of {SQL} and {MongoDB} databases for big
  e-commerce data.
\newblock In {\em 2015 {International} {Symposium} on {Computer} {Science} and
  {Software} {Engineering} ({CSSE})}, pages 1--7, Tabriz, Iran, 2015. IEEE.

\bibitem{abramova_2013}
Veronika Abramova and Jorge Bernardino.
\newblock {NoSQL} databases: {MongoDB} vs {Cassandra}.
\newblock In {\em Proceedings of the {International} {C}* {Conference} on
  {Computer} {Science} and {Software} {Engineering} - {C3S2E} '13}, pages
  14--22, Porto, Portugal, 2013. ACM Press.

\bibitem{abramova_2014_exp}
Veronika Abramova, Jorge Bernardino, and Pedro Furtado.
\newblock Experimental evaluation of {NoSQL} databases.
\newblock {\em International Journal of Database Management Systems},
  6(3):01--16, 2014.

\bibitem{abramova_2014_which}
Veronika Abramova, Jorge Bernardino, and Pedro Furtado.
\newblock Which {NoSQL} database? {A} performance overview.
\newblock {\em Open Journal of Databases (OJDB)}, 1(2):17--24, 2014.

\bibitem{abubakar_2014}
Yusuf Abubakar, Thankgod~Sani Adeyi, and Ibrahim~Gambo Auta.
\newblock Performance evaluation of {NoSQL} systems using {YCSB} in a resource
  austere environment.
\newblock {\em Performance Evaluation}, 7(8):23--27, 2014.

\bibitem{afolabi_2008}
A.O. Afolabi and A.O. Ajayi.
\newblock Performance evaluation of a database management system (a case study
  of {INTERBASE} and {MySQL}).
\newblock {\em Journal of Engineering and Applied Sciences}, 3(2):155--160,
  2008.

\bibitem{agarwal_2016}
Sarthak Agarwal and K.~S. Rajan.
\newblock Performance analysis of {MongoDB} versus {PostGIS}/{PostGreSQL}
  databases for line intersection and point containment spatial queries.
\newblock {\em Spatial Information Research}, 24(6):671--677, 2016.

\bibitem{aghi_2015}
Rajat Aghi, Sumeet Mehta, Rahul Chauhan, Siddhant Chaudhary, and Navdeep Bohra.
\newblock A comprehensive comparison of {SQL} and {MongoDB} databases.
\newblock {\em International Journal of Scientific and Research Publications},
  5(2):1--3, 2015.

\bibitem{ahmed_2017}
Nadeem Ahmed, Shakil Ahamed, Jahir~Ibna Rafiq, and Sifatur Rahim.
\newblock Data processing in {Hive} vs. {SQL Server}: {A} comparative analysis
  in the query performance.
\newblock In {\em 2017 {IEEE} 3rd {International} {Conference} on {Engineering}
  {Technologies} and {Social} {Sciences} ({ICETSS})}, pages 1--5, Bangkok,
  2017. IEEE.

\bibitem{almabdy_2018}
Soad Almabdy.
\newblock Comparative analysis of relational and graph databases for social
  networks.
\newblock In {\em 2018 1st International Conference on Computer Applications \&
  Information Security (ICCAIS)}, pages 1--4. IEEE, 2018.

\bibitem{Almeida_2015}
Rafael Almeida, Pedro Furtado, and Jorge Bernardino.
\newblock Performance evaluation {MySQL} {InnoDB} and {Microsoft SQL Server}
  2012 for decision support environments.
\newblock In {\em Proceedings of the Eighth International Conference on
  Computer Science {\&} Software Engineering - C3S2E'15}. {ACM} Press, 2015.

\bibitem{ameri_2014}
Parinaz Ameri, Udo Grabowski, Jorg Meyer, and Achim Streit.
\newblock On the application and performance of {MongoDB} for climate satellite
  data.
\newblock In {\em 2014 {IEEE} 13th {International} {Conference} on {Trust},
  {Security} and {Privacy} in {Computing} and {Communications}}, pages
  652--659, Beijing, China, 2014. IEEE.

\bibitem{andjelic_2008}
Svetlana Andjelic, Slobodan Obradovic, and Branislav Gacesa.
\newblock A performance analysis of the {DBMS} - {MySQL} {Vs} {PostgreSQL}.
\newblock {\em Communications - Scientific letters of the University of
  Zilina}, 10(4):53--57, 2008.

\bibitem{araujo_2021}
Jose Maria~A. Araujo, Alysson Cristiano~E. de~Moura, Silvia Laryssa~B.
  da~Silva, Maristela Holanda, Edward de~Oliveira Ribeiro, and Gladston~Luiz
  da~Silva.
\newblock Comparative {Performance} {Analysis} of {NoSQL} {Cassandra} and
  {MongoDB} {Databases}.
\newblock In {\em 2021 16th {Iberian} {Conference} on {Information} {Systems}
  and {Technologies} ({CISTI})}, pages 1--6, Chaves, Portugal, 2021. IEEE.

\bibitem{baralis_2017}
Elena Baralis, Andrea Dalla~Valle, Paolo Garza, Claudio Rossi, and Francesco
  Scullino.
\newblock {SQL} versus {NoSQL} databases for geospatial applications.
\newblock In {\em 2017 {IEEE} {International} {Conference} on {Big} {Data}
  ({Big} {Data})}, pages 3388--3397, Boston, MA, 2017. IEEE.

\bibitem{Bartoszewski_2019}
Dominik Bartoszewski, Adam Piorkowski, and Michal Lupa.
\newblock The comparison of processing efficiency of spatial data for {PostGIS}
  and {MongoDB} databases.
\newblock In {\em Beyond Databases, Architectures and Structures. Paving the
  Road to Smart Data Processing and Analysis}, pages 291--302. Springer
  International Publishing, 2019.

\bibitem{baruffa_2020}
Giuseppe Baruffa, Mauro Femminella, Matteo Pergolesi, and Gianluca Reali.
\newblock Comparison of {MongoDB} and {Cassandra} {Databases} for {Spectrum}
  {Monitoring} {As}-a-{Service}.
\newblock {\em IEEE Transactions on Network and Service Management},
  17(1):346--360, 2020.

\bibitem{bassil_2012}
Youssef Bassil.
\newblock A comparative study on the performance of the top {DBMS} systems.
\newblock {\em Journal of Computer Science \& Research}, 2012.

\bibitem{batra_2012}
Shalini Batra and Charu Tyagi.
\newblock Comparative analysis of relational and graph databases.
\newblock {\em International Journal of Soft Computing and Engineering
  (IJSCE)}, 2(2):509--512, 2012.

\bibitem{Boicea_2012}
Alexandru Boicea, Florin Radulescu, and Laura~Ioana Agapin.
\newblock {MongoDB vs Oracle} -- database comparison.
\newblock In {\em 2012 Third International Conference on Emerging Intelligent
  Data and Web Technologies}. {IEEE}, 2012.

\bibitem{chakraborty_2021}
Soarov Chakraborty, Shourav Paul, and K.~M. Azharul~Hasan.
\newblock Performance {Comparison} for {Data} {Retrieval} from {NoSQL} and
  {SQL} {Databases}: {A} {Case} {Study} for {COVID}-19 {Genome} {Sequence}
  {Dataset}.
\newblock In {\em 2021 2nd {International} {Conference} on {Robotics},
  {Electrical} and {Signal} {Processing} {Techniques} ({ICREST})}, pages
  324--328, DHAKA, Bangladesh, 2021. IEEE.

\bibitem{chaudhary_2018}
Anurag~Singh Chaudhary, Kanika Singh, Sanchi Kalra, and Parmeet Kaur.
\newblock An {Empirical} {Comparison} of {MongoDB} and {Hive}.
\newblock In {\em 2018 4th {International} {Conference} on {Computing}
  {Communication} and {Automation} ({ICCCA})}, pages 1--4, Greater Noida,
  India, 2018. IEEE.

\bibitem{cheng_2019}
Yinyi Cheng, Kefa Zhou, and Jinlin Wang.
\newblock Performance {Analysis} of {PostgreSQL} and {MongoDB} {Databases} for
  {Unstructured} {Data}.
\newblock In {\em Proceedings of the 2019 {International} {Conference} on
  {Mathematics}, {Big} {Data} {Analysis} and {Simulation} and {Modelling}
  ({MBDASM} 2019)}, Changsha, China, 2019. Atlantis Press.

\bibitem{chickerur_2015}
Satyadhyan Chickerur, Anoop Goudar, and Ankita Kinnerkar.
\newblock Comparison of {Relational} {Database} with {Document}-{Oriented}
  {Database} ({MongoDB}) for {Big} {Data} {Applications}.
\newblock In {\em 2015 8th {International} {Conference} on {Advanced}
  {Software} {Engineering} \& {Its} {Applications} ({ASEA})}, pages 41--47,
  Jeju Island, South Korea, 2015. IEEE.

\bibitem{chopade_2017}
Mrs. Rupali~M. Chopade and Nikhil~S. Dhavase.
\newblock {MongoDB, Couchbase}: {Performance} comparison for image dataset.
\newblock In {\em 2017 2nd {International} {Conference} for {Convergence} in
  {Technology} ({I2CT})}, pages 255--258, Mumbai, 2017. IEEE.

\bibitem{damodaran_2016}
Dipina Damodaran~B, Shirin Salim, and Surekha~Marium Vargese.
\newblock Performance {Evaluation} of {MySQL} and {MongoDB} {Databases}.
\newblock {\em International Journal on Cybernetics \& Informatics},
  5(2):387--394, 2016.

\bibitem{deari_2018}
Raif Deari, Xhemal Zenuni, Jaumin Ajdari, Florije Ismaili, and Bujar Raufi.
\newblock Analysis {And} {Comparision} of {Document}-{Based} {Databases} with
  {Relational} {Databases}: {MongoDB} vs {MySQL}.
\newblock In {\em 2018 {International} {Conference} on {Information}
  {Technologies} ({InfoTech})}, pages 1--4, Varna, 2018. IEEE.

\bibitem{Ding_2012}
Haijie Ding, Yuehui Jin, Yidong Cui, and Tan Yang.
\newblock Distributed storage of network measurement data on {HBase}.
\newblock In {\em 2012 {IEEE} 2nd International Conference on Cloud Computing
  and Intelligence Systems}. {IEEE}, 2012.

\bibitem{Eyada_2020}
Mahmoud~Moustafa Eyada, Walaa Saber, Mohammed M.~El Genidy, and Fathy Amer.
\newblock Performance evaluation of {IoT} data management using {MongoDB}
  versus {MySQL} databases in different cloud environments.
\newblock {\em {IEEE} Access}, 8:110656--110668, 2020.

\bibitem{faraj_2014}
Azhi Faraj, Bilal Rashid, and Twana Shareef.
\newblock Comparative study of relational and non-relations database
  performances using {Oracle} and {MongoDB} systems.
\newblock {\em International Journal of Computer Engineering and Technology
  (IJCET)}, 5(11):11--22, 2014.

\bibitem{Fatima_2016}
Haleemunnisa Fatima and Kumud Wasnik.
\newblock Comparison of {SQL}, {NoSQL} and {NewSQL} databases for internet of
  things.
\newblock In {\em 2016 {IEEE} Bombay Section Symposium ({IBSS})}. {IEEE}, 2016.

\bibitem{Filip_2020}
Petr Filip and Lukas Cegan.
\newblock Comparison of {MySQL} and {MongoDB} with focus on performance.
\newblock In {\em 2020 International Conference on Informatics, Multimedia,
  Cyber and Information System ({ICIMCIS})}. {IEEE}, 2020.

\bibitem{fioravanti_2016}
Sara Fioravanti, Simone Mattolini, Fulvio Patara, and Enrico Vicario.
\newblock Experimental {Performance} {Evaluation} of different {Data} {Models}
  for a {Reflection} {Software} {Architecture} over {NoSQL} {Persistence}
  {Layers}.
\newblock In {\em Proceedings of the 7th {ACM}/{SPEC} on {International}
  {Conference} on {Performance} {Engineering}}, pages 297--308, Delft The
  Netherlands, 2016. ACM.

\bibitem{fotache_2016}
Marin Fotache and Ionuț Hrubaru.
\newblock Performance {Analysis} of {Two} {Big} {Data} {Technologies} on a
  {Cloud} {Distributed} {Architecture}. {Results} for {Non}-{Aggregate}
  {Queries} on {Medium}-{Sized} {Data}.
\newblock {\em Scientific Annals of Economics and Business}, 63(s1):21--50,
  2016.

\bibitem{Fraczek_2017}
Konrad Fraczek and Malgorzata Plechawska-Wojcik.
\newblock Comparative analysis of relational and non-relational databases in
  the context of performance in web applications.
\newblock In {\em Beyond Databases, Architectures and Structures. Towards
  Efficient Solutions for Data Analysis and Knowledge Representation}, pages
  153--164. Springer International Publishing, 2017.

\bibitem{franke_2013}
Craig Franke, Samuel Morin, Artem Chebotko, John Abraham, and Pearl Brazier.
\newblock Efficient {Processing} of {Semantic} {Web} {Queries} in {HBase} and
  {MySQL} {Cluster}.
\newblock {\em IT Professional}, 15(3):36--43, 2013.

\bibitem{Gandini_2014}
Andrea Gandini, Marco Gribaudo, William~J. Knottenbelt, Rasha Osman, and Pietro
  Piazzolla.
\newblock Performance evaluation of {NoSQL} databases.
\newblock In {\em Computer Performance Engineering}, pages 16--29. Springer
  International Publishing, 2014.

\bibitem{gomes_2021}
Augusto Gomes, Vitor Lopes, Edward Ribeiro, Jorge Lima, Wagner Costa, Luis
  Garcia, and Maristela Holanda.
\newblock An {Empirical} {Performance} {Comparison} between {MySQL} and
  {MongoDB} on {Analytical} {Queries} in the {COMEX} {Database}.
\newblock In {\em 2021 16th {Iberian} {Conference} on {Information} {Systems}
  and {Technologies} ({CISTI})}, pages 1--5, Chaves, Portugal, 2021. IEEE.

\bibitem{gunawan_2019}
Rohmat Gunawan, Alam Rahmatulloh, and Irfan Darmawan.
\newblock Performance {Evaluation} of {Query} {Response} {Time} in {The}
  {Document} {Stored} {NoSQL} {Database}.
\newblock In {\em 2019 16th {International} {Conference} on {Quality} in
  {Research} ({QIR}): {International} {Symposium} on {Electrical} and
  {Computer} {Engineering}}, pages 1--6, Padang, Indonesia, 2019. IEEE.

\bibitem{gyorodi_2015}
Cornelia Gyorodi, Robert Gyorodi, George Pecherle, and Andrada Olah.
\newblock A comparative study: {MongoDB} vs. {MySQL}.
\newblock In {\em 2015 13th {International} {Conference} on {Engineering} of
  {Modern} {Electric} {Systems} ({EMES})}, pages 1--6, Oradea, Romania, 2015.
  IEEE.

\bibitem{gyorodi_2020}
Cornelia~A. Gyorodi, Diana~V. Dumse-Burescu, Doina~R. Zmaranda, Robert~s.
  Gyorodi, Gianina~A. Gabor, and George~D. Pecherle.
\newblock Performance {Analysis} of {NoSQL} and {Relational} {Databases} with
  {CouchDB} and {MySQL} for {Application}'’'s {Data} {Storage}.
\newblock {\em Applied Sciences}, 10(23):8524, 2020.

\bibitem{hairah_2021}
U~Hairah and E~Budiman.
\newblock Inner join query performance: {MariaDB} vs {PostgreSQL}.
\newblock {\em Journal of Physics: Conference Series}, 1844(1):012021, 2021.

\bibitem{Haiyan_2010}
Yu~Haiyan, Li~Jingsong, Chen Huan, Zhang Xiaoguang, Tian Yu, and Yang Yibing.
\newblock Performance evaluation of post-relational database in hospital
  information systems.
\newblock In {\em 2010 Second International Workshop on Education Technology
  and Computer Science}. {IEEE}, 2010.

\bibitem{Hajjaji_2018}
Yosra Hajjaji and Imed~Riadh Farah.
\newblock Performance investigation of selected {NoSQL} databases for massive
  remote sensing image data storage.
\newblock In {\em 2018 4th International Conference on Advanced Technologies
  for Signal and Image Processing ({ATSIP})}. {IEEE}, 2018.

\bibitem{hassan_2018}
Mahmudul Hassan and Srividya~K. Bansal.
\newblock Semantic data querying over {NoSQL} databases with {Apache} {Spark}.
\newblock In {\em 2018 {IEEE} {International} {Conference} on {Information}
  {Reuse} and {Integration} ({IRI})}, pages 364--371, Salt Lake City, UT, 2018.
  IEEE.

\bibitem{hendawi_2018}
Abdeltawab Hendawi, Jayant Gupta, Liu Jiayi, Ankur Teredesai, Ramakrishnan
  Naveen, Shah Mohak, and Mohamed Ali.
\newblock Distributed {NoSQL} {Data} {Stores}: {Performance} {Analysis} and a
  {Case} {Study}.
\newblock In {\em 2018 {IEEE} {International} {Conference} on {Big} {Data}
  ({Big} {Data})}, pages 1937--1944, Seattle, WA, USA, 2018. IEEE.

\bibitem{ilic_202}
Milo{\v{s}} Ili{\'c}, Lazar Kopanja, Dragan Zlatkovi{\'c}, Milica
  Trajkovi{\'c}, and Dejana {\'C}urguz.
\newblock {Microsoft SQL Server} and {Oracle}: Comparative performance
  analysis.
\newblock Book of proceedings of the 7th International conference Knowledge
  management, 2021.

\bibitem{Jaiswal_2013}
Garima Jaiswal.
\newblock Comparative analysis of relational and graph databases.
\newblock {\em {IOSR} Journal of Engineering}, 03(08):25--27, 2013.

\bibitem{jandaeng_2015}
Chanankorn Jandaeng.
\newblock Comparison of {RDBMS} and document oriented database in audit log
  analysis.
\newblock In {\em 2015 7th {International} {Conference} on {Information}
  {Technology} and {Electrical} {Engineering} ({ICITEE})}, pages 332--336,
  Chiang Mai, Thailand, 2015. IEEE.

\bibitem{Jing_2009}
Yinan Jing, Chunwang Zhang, and Xueping Wang.
\newblock An empirical study on performance comparison of {Lucene} and
  relational database.
\newblock In {\em 2009 International Conference on Communication Software and
  Networks}. {IEEE}, 2009.

\bibitem{jose_2020}
Benymol Jose and Sajimon Abraham.
\newblock Performance analysis of {NoSQL} and relational databases with
  {MongoDB} and {MySQL}.
\newblock {\em Materials Today: Proceedings}, 24:2036--2043, 2020.

\bibitem{jung_2015}
Min-Gyue Jung, Seon-A Youn, Jayon Bae, and Yong-Lak Choi.
\newblock A {Study} on {Data} {Input} and {Output} {Performance} {Comparison}
  of {MongoDB} and {PostgreSQL} in the {Big} {Data} {Environment}.
\newblock In {\em 2015 8th {International} {Conference} on {Database} {Theory}
  and {Application} ({DTA})}, pages 14--17, Jeju Island, South Korea, 2015.
  IEEE.

\bibitem{kabakus_2017}
Abdullah~Talha Kabakus and Resul Kara.
\newblock A performance evaluation of in-memory databases.
\newblock {\em Journal of King Saud University - Computer and Information
  Sciences}, 29(4):520--525, 2017.

\bibitem{kashyap_2013}
Suman Kashyap, Shruti Zamwar, Tanvi Bhavsar, and Snigdha Singh.
\newblock Benchmarking and analysis of {NoSQL} technologies.
\newblock {\em International Journal of Emerging Technology and Advanced
  Engineering}, 3(9):422--426, 2013.

\bibitem{Kaur_2017}
Karambir Kaur and Monika Sachdeva.
\newblock Performance evaluation of {NewSQL} databases.
\newblock In {\em 2017 International Conference on Inventive Systems and
  Control ({ICISC})}. {IEEE}, 2017.

\bibitem{khan_2019}
Wisal Khan, Waqas Ahmad, Bin Luo, and Ejaz Ahmed.
\newblock {SQL} {Database} with physical database tuning technique and {NoSQL}
  graph database comparisons.
\newblock In {\em 2019 {IEEE} 3rd {Information} {Technology}, {Networking},
  {Electronic} and {Automation} {Control} {Conference} ({ITNEC})}, pages
  110--116, Chengdu, China, 2019. IEEE.

\bibitem{khan_2017}
Wisal Khan, Ejaz ahmed, and Waseem Shahzad.
\newblock Predictive {Performance} {Comparison} {Analysis} of {Relational} \&
  {NoSQL} {Graph} {Databases}.
\newblock {\em International Journal of Advanced Computer Science and
  Applications}, 8(5), 2017.

\bibitem{khanna_2018}
Deepti Khanna, VB~Aggarwal, JIMS Director, and India~Meenu Dave.
\newblock Performance analysis for select, project and join operations of
  {Oracle}, {My-SQL} and {Microsoft Access} {DBMSS}.
\newblock {\em International Journal of Computer Engineering \& Technology
  (IJCET)}, 2018.

\bibitem{klein_2015}
John Klein, Ian Gorton, Neil Ernst, Patrick Donohoe, Kim Pham, and Chrisjan
  Matser.
\newblock Performance {Evaluation} of {NoSQL} {Databases}: {A} {Case} {Study}.
\newblock In {\em Proceedings of the 1st {Workshop} on {Performance} {Analysis}
  of {Big} {Data} {Systems}}, pages 5--10, Austin Texas USA, 2015. ACM.

\bibitem{kulshrestha_2014}
Sudhanshu Kulshrestha and Shelly Sachdeva.
\newblock Performance comparison for data storage - {Db4o} and {MySQL}
  databases.
\newblock In {\em 2014 {Seventh} {International} {Conference} on {Contemporary}
  {Computing} ({IC3})}, pages 166--170, Noida, India, 2014. IEEE.

\bibitem{kumar_2017}
K.~B.~Sundhara Kumar, {Srividya}, and S.~Mohanavalli.
\newblock A performance comparison of document oriented {NoSQL} databases.
\newblock In {\em 2017 {International} {Conference} on {Computer},
  {Communication} and {Signal} {Processing} ({ICCCSP})}, pages 1--6, Chennai,
  India, 2017. IEEE.

\bibitem{kumar_2015}
Lokesh Kumar, Shalini Rajawat, and Krati Joshi.
\newblock Comparative analysis of {NoSQL (MongoDB)} with {MySQL} database.
\newblock {\em International Journal of Modern Trends in Engineering and
  Research}, 2(5):120--127, 2015.

\bibitem{Dwivedi_2012}
Amit KumarDwivedi, C.~S. Lamba, and Shweta Shukla.
\newblock Performance analysis of column oriented database vs row oriented
  database.
\newblock {\em International Journal of Computer Applications}, 50(14):31--34,
  2012.

\bibitem{laksono_2018}
Dany Laksono.
\newblock Testing {Spatial} {Data} {Deliverance} in {SQL} and {NoSQL}
  {Database} {Using} {NodeJS} {Fullstack} {Web} {App}.
\newblock In {\em 2018 4th {International} {Conference} on {Science} and
  {Technology} ({ICST})}, pages 1--5, Yogyakarta, 2018. IEEE.

\bibitem{lazarska_2019}
Malgorzata Lazarska and Olga Siedlecka-Lamch.
\newblock Comparative study of relational and graph databases.
\newblock In {\em 2019 IEEE 15th International Scientific Conference on
  Informatics}, pages 000363--000370. IEEE, 2019.

\bibitem{lee_2018}
Chao-Hsien Lee and Zhe-Wei Shih.
\newblock A {Comparison} of {NoSQL} and {SQL} {Databases} {Over} the {Hadoop}
  and {Spark} {Cloud} {Platforms} {Using} {Machine} {Learning} {Algorithms}.
\newblock In {\em 2018 {IEEE} {International} {Conference} on {Consumer}
  {Electronics}-{Taiwan} ({ICCE}-{TW})}, pages 1--2, Taichung, 2018. IEEE.

\bibitem{Li_2013}
Yishan Li and Sathiamoorthy Manoharan.
\newblock A performance comparison of {SQL} and {NoSQL} databases.
\newblock In {\em 2013 {IEEE} Pacific Rim Conference on Communications,
  Computers and Signal Processing ({PACRIM})}. {IEEE}, 2013.

\bibitem{lorincz_2020}
Josip Lorincz, Vlatka Huljic, and Dinko Begusic.
\newblock Transforming {Product} {Catalogue} {Relational} into {Graph}
  {Database}: a {Performance} {Comparison}.
\newblock In {\em 2020 43rd {International} {Convention} on {Information},
  {Communication} and {Electronic} {Technology} ({MIPRO})}, pages 523--528,
  Opatija, Croatia, 2020. IEEE.

\bibitem{magdum_2018}
Junaid Magdum and Rahul Barhate.
\newblock Performance {Analysis} of {DML} {Operations} on {NoSQL} {Databases}
  for {Streaming} {Data}.
\newblock In {\em 2018 {Fourth} {International} {Conference} on {Computing}
  {Communication} {Control} and {Automation} ({ICCUBEA})}, pages 1--6, Pune,
  India, 2018. IEEE.

\bibitem{mahmood_2019}
Khalid Mahmood, Kjell Orsborn, and Tore Risch.
\newblock Comparison of {NoSQL} {Datastores} for {Large} {Scale} {Data}
  {Stream} {Log} {Analytics}.
\newblock In {\em 2019 {IEEE} {International} {Conference} on {Smart}
  {Computing} ({SMARTCOMP})}, pages 478--480, Washington, DC, USA, 2019. IEEE.

\bibitem{makris_2019}
Antonios Makris, Konstantinos Tserpes, Giannis Spiliopoulos, and Dimosthenis
  Anagnostopoulos.
\newblock Performance evaluation of {MongoDB} and {PostgreSQL} for
  spatio-temporal data.
\newblock In {\em EDBT/ICDT Workshops}, 2019.

\bibitem{makris_2021}
Antonios Makris, Konstantinos Tserpes, Giannis Spiliopoulos, Dimitrios Zissis,
  and Dimosthenis Anagnostopoulos.
\newblock {MongoDB} {Vs} {PostgreSQL}: {A} comparative study on performance
  aspects.
\newblock {\em GeoInformatica}, 25(2):243--268, 2021.

\bibitem{marrero_2020}
Luciano Marrero, Verena Olsowy, Fernando Tesone, Pablo Thomas, Lisandro Delia,
  and Patricia Pesado.
\newblock Performance analysis in nosql databases, relational databases and
  {NoSQL} databases as a service in the cloud.
\newblock In {\em Argentine Congress of Computer Science}, pages 157--170.
  Springer, 2020.

\bibitem{mavrogiorgos_2021}
Konstanitnos Mavrogiorgos, Athanasios Kiourtis, Argyro Mavrogiorgou, and
  Dimosthenis Kyriazis.
\newblock A {Comparative} {Study} of {MongoDB}, {ArangoDB} and {CouchDB} for
  {Big} {Data} {Storage}.
\newblock In {\em 2021 5th {International} {Conference} on {Cloud} and {Big}
  {Data} {Computing} ({ICCBDC})}, pages 8--14, Liverpool United Kingdom, 2021.
  ACM.

\bibitem{murazza_2016}
Muh.~Rafif Murazza and Arif Nurwidyantoro.
\newblock Cassandra and {SQL} database comparison for near real-time {Twitter}
  data warehouse.
\newblock In {\em 2016 {International} {Seminar} on {Intelligent} {Technology}
  and {Its} {Applications} ({ISITIA})}, pages 195--200, Lombok, Indonesia,
  2016. IEEE.

\bibitem{nepaliya_2015}
Prateek Nepaliya and Prateek Gupta.
\newblock Performance {Analysis} of {NoSQL} {Databases}.
\newblock {\em International Journal of Computer Applications}, 127(12):36--39,
  2015.

\bibitem{nyati_2013}
Suyog~S. Nyati, Shivanand Pawar, and Rajesh Ingle.
\newblock Performance evaluation of unstructured {NoSQL} data over distributed
  framework.
\newblock In {\em 2013 {International} {Conference} on {Advances} in
  {Computing}, {Communications} and {Informatics} ({ICACCI})}, pages
  1623--1627, Mysore, 2013. IEEE.

\bibitem{ohyver_2019}
Margaretha Ohyver, Jurike~V. Moniaga, Iwa Sungkawa, Bonifasius~Edwin Subagyo,
  and Ian~Argus Chandra.
\newblock The {Comparison} {Firebase} {Realtime} {Database} and {MySQL}
  {Database} {Performance} using {Wilcoxon} {Signed}-{Rank} {Test}.
\newblock {\em Procedia Computer Science}, 157:396--405, 2019.

\bibitem{oliveira_2017}
João Oliveira and Jorge Bernardino.
\newblock {NewSQL} {Databases} - {MemSQL} and {VoltDB} {Experimental}
  {Evaluation}:.
\newblock In {\em Proceedings of the 9th {International} {Joint} {Conference}
  on {Knowledge} {Discovery}, {Knowledge} {Engineering} and {Knowledge}
  {Management}}, pages 276--281, Funchal, Madeira, Portugal, 2017. SCITEPRESS -
  Science and Technology Publications.

\bibitem{padhy_2019}
Sarita Padhy and G~Mayil~Muthu Kumaran.
\newblock A quantitative performance analysis between {Mongodb} and {Oracle
  NoSQL}.
\newblock In {\em 2019 6th International Conference on Computing for
  Sustainable Global Development (INDIACom)}, pages 387--391. IEEE, 2019.

\bibitem{parker_2013}
Zachary Parker, Scott Poe, and Susan~V. Vrbsky.
\newblock Comparing {NoSQL} {MongoDB} to an {SQL} {DB}.
\newblock In {\em Proceedings of the 51st {ACM} {Southeast} {Conference} on -
  {ACMSE} '13}, page~1, Savannah, Georgia, 2013. ACM Press.

\bibitem{Patil_2017}
Mayur~M Patil, Akkamahadevi Hanni, C~H Tejeshwar, and Priyadarshini Patil.
\newblock A qualitative analysis of the performance of {MongoDB} vs {MySQL}
  database based on insertion and retriewal operations using a web/android
  application to explore load balancing {\textemdash} sharding in {MongoDB} and
  its advantages.
\newblock In {\em 2017 International Conference on I-{SMAC} ({IoT} in Social,
  Mobile, Analytics and Cloud) (I-{SMAC})}. {IEEE}, 2017.

\bibitem{pereira_2018}
Diogo~Augusto Pereira, Wagner Ourique~de Morais, and Edison Pignaton~de
  Freitas.
\newblock {NoSQL} real-time database performance comparison.
\newblock {\em International Journal of Parallel, Emergent and Distributed
  Systems}, 33(2):144--156, 2018.

\bibitem{poljak_2017}
R.~Poljak, P.~Poscic, and D.~Jaksic.
\newblock Comparative analysis of the selected relational database management
  systems.
\newblock In {\em 2017 40th {International} {Convention} on {Information} and
  {Communication} {Technology}, {Electronics} and {Microelectronics}
  ({MIPRO})}, pages 1496--1500, Opatija, Croatia, 2017. IEEE.

\bibitem{puangsaijai_2017}
Wittawat Puangsaijai and Sutheera Puntheeranurak.
\newblock A comparative study of relational database and key-value database for
  big data applications.
\newblock In {\em 2017 {International} {Electrical} {Engineering} {Congress}
  ({iEECON})}, pages 1--4, Pattaya, Thailand, 2017. IEEE.

\bibitem{rabl_2012}
Tilmann Rabl, Sergio Gómez-Villamor, Mohammad Sadoghi, Victor Muntés-Mulero,
  Hans-Arno Jacobsen, and Serge Mankovskii.
\newblock Solving big data challenges for enterprise application performance
  management.
\newblock {\em Proceedings of the VLDB Endowment}, 5(12):1724--1735, 2012.

\bibitem{rafamantanantsoa_2018}
Fontaine Rafamantanantsoa and Maherindefo Laha.
\newblock Analysis and {Neural} {Networks} {Modeling} of {Web} {Server}
  {Performances} {Using} {MySQL} and {PostgreSQL}.
\newblock {\em Communications and Network}, 10(04):142--151, 2018.

\bibitem{rautmare_2016}
Sharvari Rautmare and D.~M. Bhalerao.
\newblock {MySQL} and {NoSQL} database comparison for {IoT} application.
\newblock In {\em 2016 {IEEE} {International} {Conference} on {Advances} in
  {Computer} {Applications} ({ICACA})}, pages 235--238, Coimbatore, 2016. IEEE.

\bibitem{ribeiro_2017}
Jardel Ribeiro, Jonas Henrique, Rodrigo Ribeiro, and Rosalvo Neto.
\newblock {NoSQL} vs relational database: {A} comparative study about the
  generation of the most frequent {N}-grams.
\newblock In {\em 2017 4th {International} {Conference} on {Systems} and
  {Informatics} ({ICSAI})}, pages 1568--1572, Hangzhou, 2017. IEEE.

\bibitem{Roopak_2013}
K.E. Roopak, K.S.~Swati Rao, S.~Ritesh, and Satyadhyan Chickerur.
\newblock Performance comparison of relational database with object database
  ({DB}4o).
\newblock In {\em 2013 5th International Conference on Computational
  Intelligence and Communication Networks}. {IEEE}, 2013.

\bibitem{saikia_2015}
Amlanjyoti Saikia, Sherin Joy, Dhondup Dolma, and Roseline Mary.~R.
\newblock Comparative {Performance} {Analysis} of {MySQL} and {SQL} {Server}
  {Relational} {Database} {Management} {Systems} in {Windows} {Environment}.
\newblock {\em IJARCCE}, pages 160--164, 2015.

\bibitem{Samanta_2018}
Ashis~Kumar Samanta, Bidut~Biman Sarkar, and Nabendu Chaki.
\newblock Query performance analysis of {NoSQL} and big data.
\newblock In {\em 2018 Fourth International Conference on Research in
  Computational Intelligence and Communication Networks ({ICRCICN})}. {IEEE},
  2018.

\bibitem{schmid_2015_perf}
Stephan Schmid, Eszter Galicz, and Wolfgang Reinhardt.
\newblock Performance investigation of selected {SQL} and {NoSQL} databases.
\newblock In {\em Proceedings of the AGILE}, pages 1--5, 2015.

\bibitem{Schmid_2015_wms}
Stephan Schmid, Eszter Galicz, and Wolfgang Reinhardt.
\newblock {WMS} performance of selected {SQL} and {NoSQL} databases.
\newblock In {\em International Conference on Military Technologies ({ICMT})
  2015}. {IEEE}, 2015.

\bibitem{schreiner_2019}
Geomar~A. Schreiner, Ronan Knob, Denio Duarte, Patricia Vilain, and Ronaldo
  dos~Santos Mello.
\newblock {NewSQL} {Through} the {Looking} {Glass}.
\newblock In {\em Proceedings of the 21st {International} {Conference} on
  {Information} {Integration} and {Web}-based {Applications} \& {Services}},
  pages 361--369, Munich Germany, 2019. ACM.

\bibitem{Seda_2018}
Pavel Seda, Jiri Hosek, Pavel Masek, and Jiri Pokorny.
\newblock Performance testing of {NoSQL} and {RDBMS} for storing big data in
  e-applications.
\newblock In {\em 2018 3rd International Conference on Intelligent Green
  Building and Smart Grid ({IGBSG})}. {IEEE}, 2018.

\bibitem{seghier_2021}
Nadia~Ben Seghier and Okba Kazar.
\newblock Performance benchmarking and comparison of {NoSQL} databases: {Redis}
  vs {MongoDB} vs {Cassandra} using {YCSB} tool.
\newblock In {\em 2021 {International} {Conference} on {Recent} {Advances} in
  {Mathematics} and {Informatics} ({ICRAMI})}, pages 1--6, Tebessa, Algeria,
  2021. IEEE.

\bibitem{sharma_2018}
Monika Sharma, Vishal~Deep Sharma, and Mahesh~M. Bundele.
\newblock Performance {Analysis} of {RDBMS} and {No} {SQL} {Databases}:
  {PostgreSQL}, {MongoDB} and {Neo4j}.
\newblock In {\em 2018 3rd {International} {Conference} and {Workshops} on
  {Recent} {Advances} and {Innovations} in {Engineering} ({ICRAIE})}, pages
  1--5, Jaipur, India, 2018. IEEE.

\bibitem{sholichah_2020}
Rahmatian~Jayanty Sholichah, Mahmud Imrona, and Andry Alamsyah.
\newblock Performance {Analysis} of {Neo4j} and {MySQL} {Databases} using
  {Public} {Policies} {Decision} {Making} {Data}.
\newblock In {\em 2020 7th {International} {Conference} on {Information}
  {Technology}, {Computer}, and {Electrical} {Engineering} ({ICITACEE})}, pages
  152--157, Semarang, Indonesia, 2020. IEEE.

\bibitem{shetty_2019}
B~Sirish~Shetty and Kc~Akshay.
\newblock Performance {Analysis} of {Queries} in {RDBMS} vs {NoSQL}.
\newblock In {\em 2019 2nd {International} {Conference} on {Intelligent}
  {Computing}, {Instrumentation} and {Control} {Technologies} ({ICICICT})},
  pages 1283--1286, Kannur,Kerala, India, 2019. IEEE.

\bibitem{stancu-mara_2008}
Sorin Stancu-Mara and Peter Baumann.
\newblock A comparative benchmark of large objects in relational databases.
\newblock In {\em Proceedings of the 2008 international symposium on {Database}
  engineering \& applications - {IDEAS} '08}, page 277, Coimbra, Portugal,
  2008. ACM Press.

\bibitem{suh_2022}
Young-Kyoon Suh, Junyoung An, Byungchul Tak, and Gap-Joo Na.
\newblock A {Comprehensive} {Empirical} {Study} of {Query} {Performance}
  {Across} {GPU} {DBMSes}.
\newblock {\em Proceedings of the ACM on Measurement and Analysis of Computing
  Systems}, 6(1):1--29, 2022.

\bibitem{swaminathan_2016}
Surya~Narayanan Swaminathan and Ramez Elmasri.
\newblock Quantitative {Analysis} of {Scalable} {NoSQL} {Databases}.
\newblock In {\em 2016 {IEEE} {International} {Congress} on {Big} {Data}
  ({BigData} {Congress})}, pages 323--326, San Francisco, CA, USA, 2016. IEEE.

\bibitem{tang_2016}
Enqing Tang and Yushun Fan.
\newblock Performance {Comparison} between {Five} {NoSQL} {Databases}.
\newblock In {\em 2016 7th {International} {Conference} on {Cloud} {Computing}
  and {Big} {Data} ({CCBD})}, pages 105--109, Macau, China, 2016. IEEE.

\bibitem{Tongkaw_2016}
Sasalak Tongkaw and Aumnat Tongkaw.
\newblock A comparison of database performance of {MariaDB} and {MySQL} with
  {OLTP} workload.
\newblock In {\em 2016 {IEEE} Conference on Open Systems ({ICOS})}. {IEEE},
  2016.

\bibitem{truica_2015}
Ciprian-Octavian Truica, Florin Radulescu, Alexandru Boicea, and Ion Bucur.
\newblock Performance {Evaluation} for {CRUD} {Operations} in {Asynchronously}
  {Replicated} {Document} {Oriented} {Database}.
\newblock In {\em 2015 20th {International} {Conference} on {Control} {Systems}
  and {Computer} {Science}}, pages 191--196, Bucharest, Romania, 2015. IEEE.

\bibitem{van_der_Veen_2012}
Jan~Sipke van~der Veen, Bram van~der Waaij, and Robert~J. Meijer.
\newblock Sensor data storage performance: {SQL} or {NoSQL}, physical or
  virtual.
\newblock In {\em 2012 {IEEE} Fifth International Conference on Cloud
  Computing}. {IEEE}, 2012.

\bibitem{vershinin_2021}
I.~S. Vershinin and A.~R. Mustafina.
\newblock Performance {Analysis} of {PostgreSQL}, {MySQL}, {Microsoft} {SQL}
  {Server} {Systems} {Based} on {TPC}-{H} {Tests}.
\newblock In {\em 2021 {International} {Russian} {Automation} {Conference}
  ({RusAutoCon})}, pages 683--687, Sochi, Russian Federation, 2021. IEEE.

\bibitem{vicknair_2010}
Chad Vicknair, Michael Macias, Zhendong Zhao, Xiaofei Nan, Yixin Chen, and Dawn
  Wilkins.
\newblock A comparison of a graph database and a relational database: a data
  provenance perspective.
\newblock In {\em Proceedings of the 48th {Annual} {Southeast} {Regional}
  {Conference} on - {ACM} {SE} '10}, page~1, Oxford, Mississippi, 2010. ACM
  Press.

\bibitem{Wei_2011}
Zhu Wei-ping, Li~Ming-xin, and Chen Huan.
\newblock Using {MongoDB} to implement textbook management system instead of
  {MySQL}.
\newblock In {\em 2011 {IEEE} 3rd International Conference on Communication
  Software and Networks}. {IEEE}, 2011.

\bibitem{Wiseso_2020}
Linggis~Galih Wiseso, Mahmud Imrona, and Andry Alamsyah.
\newblock Performance analysis of {Neo4j}, {MongoDB}, and {PostgreSQL} on 2019
  national election big data management database.
\newblock In {\em 2020 6th International Conference on Science in Information
  Technology ({ICSITech})}. {IEEE}, 2020.

\bibitem{Xu_2014}
Wei Xu, Zhonghua Zhou, Hong Zhou, Wu~Zhang, and Jiang Xie.
\newblock {MongoDB} improves big data analysis performance on electric health
  record system.
\newblock In {\em Communications in Computer and Information Science}, pages
  350--357. Springer Berlin Heidelberg, 2014.

\bibitem{yassien_2016}
Amal~W. Yassien and Amr~F. Desouky.
\newblock {RDBMS}, {NoSQL}, {Hadoop}: {A} {Performance}-{Based} {Empirical}
  {Analysis}.
\newblock In {\em Proceedings of the 2nd {Africa} and {Middle} {East}
  {Conference} on {Software} {Engineering} - {AMECSE} '16}, pages 52--59,
  Cairo, Egypt, 2016. ACM Press.

\bibitem{wang_2015}
{Yinfeng Wang}, {Guiquan Zhong}, {Lin Kun}, {Longxiang Wang}, {Huang Kai},
  {Fuliang Guo}, {Chengzhe Liu}, and {Xiaoshe Dong}.
\newblock The {Performance} {Survey} of in {Memory} {Database}.
\newblock In {\em 2015 {IEEE} 21st {International} {Conference} on {Parallel}
  and {Distributed} {Systems} ({ICPADS})}, pages 815--820, Melbourne, VIC,
  2015. IEEE.

\bibitem{Zhou_2009}
Zhonghai Zhou, Bin Zhou, Wenwen Li, Brian Griglak, Carmen Caiseda, and Qunying
  Huang.
\newblock Evaluating query performance on object-relational spatial databases.
\newblock In {\em 2009 2nd {IEEE} International Conference on Computer Science
  and Information Technology}. {IEEE}, 2009.

\bibitem{ceresnak_2019}
Roman Čerešňák and Michal Kvet.
\newblock Comparison of query performance in relational a non-relation
  databases.
\newblock {\em Transportation Research Procedia}, 40:170--177, 2019.
\end{thebibliography}
\bibliographystylesupp{plain}
\bibliographysupp{sample-base}

\appendix

\pagebreak
\section{Detailed Comparison Results}
\label{app-results}

\begin{table}[h!]
  \centering
  \scriptsize
  \caption{DBMS performance comparisons in \textit{read} operations; a table cell shows that according to the study or studies cited, the DBMS in the row outperformed the DBMS in the column, e.g., Azure DocumentDB (DDB) outperformed Azure SQL DB [103], and DB/2 outperformed Access [106]}
  \label{table-results-read-1}
    \setlength\tabcolsep{3pt} 
  \begin{tabularx}{\linewidth}{r|c!{\color{Gray}\vrule}c!{\color{Gray}\vrule}c!{\color{Gray}\vrule}c!{\color{Gray}\vrule}c!{\color{Gray}\vrule}c!{\color{Gray}\vrule}c!{\color{Gray}\vrule}Y!{\color{Gray}\vrule}c!{\color{Gray}\vrule}Y!{\color{Gray}\vrule}Y!{\color{Gray}\vrule}c!{\color{Gray}\vrule}c!{\color{Gray}\vrule}c!{\color{Gray}\vrule}c!{\color{Gray}\vrule}c}
  \toprule
    & 
    \rotatebox{270}{Access} &
    \rotatebox{270}{ArangoDB} &
    \rotatebox{270}{Azure DDB} &
    \rotatebox{270}{Azure SQL DB} &
    \rotatebox{270}{BlazingSQL} &
    \rotatebox{270}{Caché} &
    \rotatebox{270}{Interbase} &
    \rotatebox{270}{Cassandra} &
    \rotatebox{270}{CockroachDB} &
    \rotatebox{270}{CouchDB} &
    \rotatebox{270}{Couchbase} &
    \rotatebox{270}{DB/2} &
    \rotatebox{270}{Db4o} &
    \rotatebox{270}{Elasticsearch} &
    \rotatebox{270}{Firebase} &
    \rotatebox{270}{H2} \\ 
    \midrule
  \rowcolor{Gray}
  Access             &  &  & &  & & & & & & & & & & & & \\
  ArangoDB           &  &  & &  & & & & & & \citesupp{gunawan_2019} & & & & & & \\
  \rowcolor{Gray}
  Azure DDB          &  &  &  & \citesupp{baralis_2017} & & & & & & & & & & & & \\
  Azure SQL DB       &  &  & &  & & & & & & & & & &  & & \\
  \rowcolor{Gray}
  BlazingSQL         &  &  & &  & & & & & & & & & & & & \\
  Caché              &  &  & &  & & & & & & & & & & & & \\
  \rowcolor{Gray}
  Interbase          &  &  & &  & & & & & & & & & & & & \\
  Cassandra          &  &  & &  & & & & & & & & & & & & \citesupp{kabakus_2017}\\
  \rowcolor{Gray}
  CockroachDB        &  &  & &  & & & & & & & & & & & & \\
  CouchDB            &  & \citesupp{mavrogiorgos_2021} & & & & & & \citesupp{Li_2013} &  & & & & & & & \\
  \rowcolor{Gray}
  Couchbase          &  &  & & & & & & \citesupp{Li_2013,tang_2016} & & \citesupp{Li_2013} & & & & & & \\
  DB/2               & \citesupp{bassil_2012} &  & & & & & &  &  & & & & & & \\
  \rowcolor{Gray}
  Db4o               &  &  &  & & & & & & & & & & & & & \\
  Elasticsearch      &  &  &  & & & & & \citesupp{abramova_2014_exp,magdum_2018} & & & & & & & & \\
  \rowcolor{Gray}
  Firebase           &  &  &  & & & & & & & & &  & & & & \\
  H2                 &  &  &  & & & & & & & & &  & & & & \\
  \rowcolor{Gray}
  HBase              &  &  &  & & & & & \citesupp{Gandini_2014,swaminathan_2016,tang_2016} & & & & & & & & \\
  Hive               &  &  &  & & & & & & & & & & & & & \\
  \rowcolor{Gray}
  Hypertable         &  &  &  & & & & & \citesupp{Li_2013} & & \citesupp{Li_2013} & & & & & & \\
  MariaDB            &  &  &  & & & & & & & & & & & & & \\
  \rowcolor{Gray}
  memcached          &  &  &  & & & & & & & & & & & & & \\
  MemSQL             &  &  &  & & & & & & \citesupp{Kaur_2017,schreiner_2019}  & & & & & & & \\
  \rowcolor{Gray}
  MongoDB            & & \citesupp{mavrogiorgos_2021} & & & & & & \citesupp{araujo_2021,ceresnak_2019,Fraczek_2017,Gandini_2014,Li_2013,mahmood_2019,seghier_2021,tang_2016,van_der_Veen_2012,abramova_2013} &  & \citesupp{gunawan_2019,kumar_2017,Li_2013,mavrogiorgos_2021,truica_2015} & \citesupp{chopade_2017,pereira_2018,truica_2015} & & & \citesupp{abubakar_2014} & & \citesupp{kabakus_2017} \\
  MySQL Cluster      &  &  &  &  & & & & & & & & & & & & \\
  \rowcolor{Gray}
  MySQL              & \citesupp{bassil_2012,khanna_2018} &  & & & & & & \citesupp{rabl_2012} &  & \citesupp{gyorodi_2020} & & & \citesupp{Roopak_2013} & & & \\
  Neo4J              &  &  &  & & & & & \citesupp{marrero_2020} & & & & & & & & \\
  \rowcolor{Gray}
  NuoDB              &  &  &  & & & & & & \citesupp{Kaur_2017,schreiner_2019}  & & & & & & & \\
  OmniSciDB          &  &  &  & & & & & & & & & & & & & \\
  \rowcolor{Gray}
  Oracle DB          & \citesupp{bassil_2012,khanna_2018} & &   & & & \citesupp{Haiyan_2010} & & &  &  & & & & & & \\
  Oracle NoSQL       &  &  &  & & & & & & & & & & & & &  \\
  \rowcolor{Gray}
  OrientDB           &  &  &  & & & & & & & & & & & \citesupp{abubakar_2014} & & \\
  PG-Strom           &  &  &  & & & & & & & & & & & & & \\
  \rowcolor{Gray}
  PostgreSQL         &  &  &  & & & & & \citesupp{Fraczek_2017,van_der_Veen_2012} & & & & & & & & \\
  PostgresXL         &  &  &  & & & & & & & & & & & & & \\
  \rowcolor{Gray}
  RavenDB            &  &  &  & & & & & & & & & & & & & \\
  Redis              &  &  &  & & & & & \citesupp{abramova_2014_which,abramova_2014_exp,rabl_2012,seghier_2021,tang_2016} & & & \citesupp{pereira_2018} & & & \citesupp{abramova_2014_exp}  & & \citesupp{kabakus_2017} \\
  \rowcolor{Gray}
  RethinkDB          &  &  &  & & & & & & & & & & & & & \\
  Riak               &  &  &  & & & & & & & & & & & & & \\
  \rowcolor{Gray}
  Scalaris           &  &  &  & & & & & \citesupp{abramova_2014_exp} & & & &  & & \citesupp{abramova_2014_exp} & & \\
  SQL Server         & \citesupp{bassil_2012} & &  & & & & & \citesupp{Li_2013,Samanta_2018} &  & \citesupp{Li_2013} & & & & & & \\
  \rowcolor{Gray}
  SQLite             &  &  &  & & & & & & & & & & & & & \\
  Tarantool          &  &  &  & & & & & \citesupp{abramova_2014_exp} & & & & & & \citesupp{abramova_2014_exp}  & & \\
  \rowcolor{Gray}
  Voldemort          &  &  &  & & & & & \citesupp{rabl_2012} & & & & &  &  & & \\
  VoltDB             &  &  &  & & & & & & \citesupp{Kaur_2017,schreiner_2019}  & & & & & & & \\
  \bottomrule
  \end{tabularx}
\end{table}

\begin{table}
  \centering
  \scriptsize
  \caption*{Table~\ref{table-results-read-1}. (cont.)}
  \label{table-results-read-2}
  
  \setlength\tabcolsep{3pt} 
  \begin{tabularx}{\linewidth}{r|Y!{\color{Gray}\vrule}c!{\color{Gray}\vrule}c!{\color{Gray}\vrule}c!{\color{Gray}\vrule}c!{\color{Gray}\vrule}c!{\color{Gray}\vrule}Y!{\color{Gray}\vrule}c!{\color{Gray}\vrule}Y!{\color{Gray}\vrule}Y!{\color{Gray}\vrule}c!{\color{Gray}\vrule}c}
  \toprule
    & 
    \rotatebox{270}{HBase} &
    \rotatebox{270}{Hive} & 
    \rotatebox{270}{Hypertable} & 
    \rotatebox{270}{MariaDB} & 
    \rotatebox{270}{memcached} &
    \rotatebox{270}{MemSQL} &
    \rotatebox{270}{MongoDB} & 
    \rotatebox{270}{MySQL Cluster} &
    \rotatebox{270}{MySQL} &
    \rotatebox{270}{Neo4J} & 
    \rotatebox{270}{NuoDB} & 
    \rotatebox{270}{OmniSciDB} \\
    \midrule
  \rowcolor{Gray}
  Access             & & & & & & & & & & & & \\
  ArangoDB           & & & & & & & \citesupp{gunawan_2019} & & & & & \\
  \rowcolor{Gray}
  Azure DDB          & & & & & & & & & & & & \\
  Azure SQL DB       & & & & & & & & & & & & \\
  \rowcolor{Gray}
  BlazingSQL         & & & & & & & & & & & & \\
  Caché              & & & & & & & & & & & & \\
  \rowcolor{Gray}
  Interbase          & & & & & & & & & \citesupp{afolabi_2008} & & & \\
  Cassandra          & \citesupp{abramova_2014_which,abramova_2014_exp,hassan_2018,hendawi_2018,kashyap_2013,rabl_2012} & &  &  & \citesupp{kabakus_2017} & & \citesupp{abramova_2014_which,abramova_2013,araujo_2021,chakraborty_2021,hendawi_2018,kabakus_2017,marrero_2020,Samanta_2018,swaminathan_2016,baruffa_2020,klein_2015} & & \citesupp{baruffa_2020,ceresnak_2019,chakraborty_2021,marrero_2020,murazza_2016} & & & \\
  \rowcolor{Gray}
  CockroachDB        & & & & & & & & & & & & \\
  CouchDB            & & & & & & & & & \citesupp{truica_2015} & & & \\
  \rowcolor{Gray}
  Couchbase          & \citesupp{tang_2016} & & \citesupp{Li_2013} & & & & \citesupp{Li_2013,schmid_2015_perf} & & \citesupp{truica_2015} & & & \\
  DB/2               & & & & & & & & & \citesupp{bassil_2012} & & \\
  \rowcolor{Gray}
  Db4o               & & & & & & & & & \citesupp{kulshrestha_2014} & & & \\
  Elasticsearch      & \citesupp{abramova_2014_exp} & & & & & & \citesupp{abramova_2014_exp} & & \citesupp{Seda_2018} & & & \\
  \rowcolor{Gray}
  Firebase           & & & & & & & & & \citesupp{ohyver_2019} & & & \\
  H2                 & & & & & & & & & & & & \\
  \rowcolor{Gray}
  HBase              & & & & & & & \citesupp{swaminathan_2016,yassien_2016} & \citesupp{franke_2013} & \citesupp{Ding_2012,lee_2018,ribeiro_2017,yassien_2016} & & & \\
  Hive               & & & & & & & & & & & & \\
  \rowcolor{Gray}
  Hypertable         & & & & & & & & & & & & \\
  MariaDB            & & & & & & & & & & & & \\
  \rowcolor{Gray}
  memcached          & & & & & & & & & & & & \\
  MemSQL             & & & & & & & & & & & \citesupp{schreiner_2019} & \\
  \rowcolor{Gray}
  MongoDB            & \citesupp{abramova_2014_which,abramova_2014_exp,Gandini_2014,hendawi_2018,tang_2016} & \citesupp{chaudhary_2018} & \citesupp{Li_2013} &  & \citesupp{kabakus_2017} & & &  & \citesupp{aghi_2015,ameri_2014,ceresnak_2019,chakraborty_2021,chickerur_2015,damodaran_2016,deari_2018,Eyada_2020,Fatima_2016,fioravanti_2016,gomes_2021,gyorodi_2015,jandaeng_2015,jose_2020,kumar_2015,nyati_2013,Patil_2017,rautmare_2016,truica_2015,Wei_2011,yassien_2016} & \citesupp{fioravanti_2016,sharma_2018,Wiseso_2020} & & \\
  MySQL Cluster      & & & & & & & & & & & & \\
  \rowcolor{Gray}
  MySQL              & \citesupp{rabl_2012} & & & & & & \citesupp{aghi_2015,ameri_2014,deari_2018,marrero_2020} &  & & \citesupp{sholichah_2020} & & \\
  Neo4J              &  &  & & & & & \citesupp{marrero_2020} & & \citesupp{batra_2012,fioravanti_2016,Jaiswal_2013,marrero_2020,vicknair_2010,almabdy_2018} & & &  \\
  \rowcolor{Gray}
  NuoDB              & & & & & & \citesupp{Kaur_2017} & & & & & & \\
  OmniSciDB          & & & & & & & & & & & & \\
  \rowcolor{Gray}
  Oracle DB          & & & & & & & & & \citesupp{ceresnak_2019,khanna_2018} & & & \\
  Oracle NoSQL       & \citesupp{abramova_2014_exp} & & & & & & & & & & & \\
  \rowcolor{Gray}
  OrientDB           & \citesupp{abramova_2014_which,abramova_2014_exp} &  & & & & & & & & & & \\
  PG-Strom           & & & & & & & & & & & & \\
  \rowcolor{Gray}
  PostgreSQL         & & & & \citesupp{hairah_2021} & & & \citesupp{makris_2019,makris_2021,van_der_Veen_2012,schmid_2015_perf} & & \citesupp{murazza_2016,truica_2015,vershinin_2021} & \citesupp{sharma_2018,Wiseso_2020} & &  \\
  PostgresXL         &  & \citesupp{fotache_2016} & & & & & & & & & & \\
  \rowcolor{Gray}
  RavenDB            & & & & & & & & & & & & \\
  Redis              & \citesupp{abramova_2014_which,abramova_2014_exp,rabl_2012,tang_2016} &  & & & \citesupp{kabakus_2017} & & \citesupp{abramova_2014_which,abramova_2014_exp,kabakus_2017,mahmood_2019,seghier_2021,tang_2016} & & \citesupp{ceresnak_2019,marrero_2020,rabl_2012} & & & \citesupp{ceresnak_2019} \\
  \rowcolor{Gray}
  RethinkDB          & & & & & & & & & & & & \\
  Riak               & & & & & & & \citesupp{klein_2015} & & & & & \\
  \rowcolor{Gray}
  Scalaris           & \citesupp{abramova_2014_exp} & & & & & & \citesupp{abramova_2014_exp} & & & & & \\
  SQL Server         & & & & & & & \citesupp{Samanta_2018} & & \citesupp{Almeida_2015,bassil_2012,ceresnak_2019,saikia_2015,vershinin_2021} & & & \\
  \rowcolor{Gray}
  SQLite             & & & & & & & & & & & & \\
  Tarantool          & \citesupp{abramova_2014_exp} & & & & & & \citesupp{abramova_2014_exp} & & & & & \\
  \rowcolor{Gray}
  Voldemort          & \citesupp{abramova_2014_exp,rabl_2012} & & & & & & & & \citesupp{rabl_2012} & & & \\
  VoltDB             & \citesupp{rabl_2012} & & & & & \citesupp{Kaur_2017} & \citesupp{Fatima_2016} & & \citesupp{Fatima_2016} & & & \\
  \bottomrule
  \end{tabularx}
\end{table}

\begin{table}
  \centering
  \scriptsize
  \caption*{Table~\ref{table-results-read-1}. (cont.)}
  \label{table-results-read-3}
    \setlength\tabcolsep{3pt} 
  \begin{tabularx}{\linewidth}{r|Y!{\color{Gray}\vrule}c!{\color{Gray}\vrule}Y!{\color{Gray}\vrule}c!{\color{Gray}\vrule}Y!{\color{Gray}\vrule}c!{\color{Gray}\vrule}Y!{\color{Gray}\vrule}Y!{\color{Gray}\vrule}c!{\color{Gray}\vrule}c!{\color{Gray}\vrule}c!{\color{Gray}\vrule}Y!{\color{Gray}\vrule}c!{\color{Gray}\vrule}c!{\color{Gray}\vrule}c!{\color{Gray}\vrule}Y}
  \toprule
    &
    \rotatebox{270}{Oracle DB} & 
    \rotatebox{270}{Oracle NoSQL} & 
    \rotatebox{270}{OrientDB} &
    \rotatebox{270}{PG-Strom} &
    \rotatebox{270}{PostgreSQL} &
    \rotatebox{270}{PostgresXL} &
    \rotatebox{270}{RavenDB} &
    \rotatebox{270}{Redis} &
    \rotatebox{270}{RethinkDB} &
    \rotatebox{270}{Riak} &
    \rotatebox{270}{Scalaris} &
    \rotatebox{270}{SQL Server} &
    \rotatebox{270}{SQLite} &
    \rotatebox{270}{Tarantool} &
    \rotatebox{270}{Voldemort} &
    \rotatebox{270}{VoltDB} \\ 
    \midrule
  \rowcolor{Gray}
  Access             & &  &  & &  & & & & & & & & & & & \\
  ArangoDB           & \citesupp{lorincz_2020} &  &  & &  & & & & & & & & & & & \\
  \rowcolor{Gray}
  Azure DDB          & &  &  & &  & & & & & & & & & & & \\
  Azure SQL DB       & &  &  &  &  & & & & & & & & & & & \\
  \rowcolor{Gray}
  BlazingSQL         & & &  & \citesupp{suh_2022} &  & & & & & & & & & & & \\
  Caché              & \citesupp{Haiyan_2010} & &  & &  & & & & & & & & & & & \\
  \rowcolor{Gray}
  Interbase          & &  &  & &  & & & & & & & & & & & \\
  Cassandra          & \citesupp{baruffa_2020,ceresnak_2019} &  & \citesupp{abramova_2014_which,abramova_2014_exp} & & \citesupp{murazza_2016} & & \citesupp{Li_2013} & \citesupp{baruffa_2020,ceresnak_2019,kabakus_2017,mahmood_2019} & & \citesupp{klein_2015} & & \citesupp{baruffa_2020,ceresnak_2019} & & & & \citesupp{rabl_2012} \\
  \rowcolor{Gray}
  CockroachDB        & &  &  & &  & & & & & & & & & & & \\
  CouchDB            & &  &  & & \citesupp{truica_2015} & & \citesupp{Li_2013,nepaliya_2015} & & &  & & \citesupp{truica_2015} & & & & \\
  \rowcolor{Gray}
  Couchbase          & &  &  & & \citesupp{schmid_2015_perf,truica_2015} & & \citesupp{Li_2013} & & & & & \citesupp{Li_2013,truica_2015} & & & & \\
  DB/2               & \citesupp{bassil_2012} & &  & & & & & &  &  & \citesupp{bassil_2012} & & & & \\
  \rowcolor{Gray}
  Db4o               & &  &  &  & & & & & & & & & & & & \\
  Elasticsearch      & & \citesupp{abramova_2014_exp} & \citesupp{abramova_2014_exp} &  & & & & & & & & & & & \citesupp{abramova_2014_exp} & \\
  \rowcolor{Gray}
  Firebase           & &  &  &  & & & & & & & & &  & & & \\
  H2                 & &  &  &  & & & & & & & & &  & & & \\
  \rowcolor{Gray}
  HBase              & & &  &  & & & & & & & & & & & & \\
  Hive               & & &  &  & & & & & & & & & & & & \\
  \rowcolor{Gray}
  Hypertable         & &  &  &  & & & \citesupp{Li_2013} & & & & & \citesupp{Li_2013} & & & & \\
  MariaDB            & &  &  &  & & & & \citesupp{puangsaijai_2017} & & & & && & & \\
  \rowcolor{Gray}
  memcached          & &  &  &  & & & & & & & & && & & \\
  MemSQL             & &  &  &  & & & & & & & & & & & & \citesupp{schreiner_2019,oliveira_2017} \\
  \rowcolor{Gray}
  MongoDB            & \citesupp{ceresnak_2019,shetty_2019} &  & \citesupp{abramova_2014_which,abramova_2014_exp, abubakar_2014} & & \citesupp{agarwal_2016,Bartoszewski_2019,cheng_2019,Fraczek_2017,jung_2015,laksono_2018,schmid_2015_perf,sharma_2018,truica_2015,Wiseso_2020} & & \citesupp{Li_2013} & \citesupp{ceresnak_2019,mahmood_2019,abubakar_2014} & \citesupp{pereira_2018} &  & & \citesupp{aboutorabi_2015,ceresnak_2019,Li_2013,parker_2013,truica_2015,Xu_2014} & \citesupp{wang_2015} & & & \\
  MySQL Cluster      & &  &  &  &  & & & & & & & & & & & \\
  \rowcolor{Gray}
  MySQL              & \citesupp{poljak_2017} &  &  & & \citesupp{andjelic_2008,poljak_2017,rafamantanantsoa_2018,stancu-mara_2008} & & & & & & & \citesupp{truica_2015} & & & & \citesupp{rabl_2012} \\
  Neo4J              & \citesupp{khan_2017,khan_2019,lazarska_2019} &  &  & & & & & & & & & & \citesupp{wang_2015} & & & \\
  \rowcolor{Gray}
  NuoDB              & &  &  &  & & & & & & & & & & & & \citesupp{Kaur_2017} \\
  OmniSciDB          & &  &  & \citesupp{suh_2022} & & & & & & & & & & & & \\
  \rowcolor{Gray}
  Oracle DB          & & & & & \citesupp{poljak_2017} & & & & &  &  & & & & & \\
  Oracle NoSQL       & & & \citesupp{abramova_2014_exp} &  & & & & & & & & & & & & \\
  \rowcolor{Gray}
  OrientDB           & &  &  &  & & & & \citesupp{abubakar_2014} & & & & & & & & \\
  PG-Strom           & &  &  &  & & & & & & & & & & & & \\
  \rowcolor{Gray}
  PostgreSQL         & \citesupp{Zhou_2009} &  &  &  & & & & & & & & \citesupp{truica_2015,Zhou_2009} & & & & \\
  PostgresXL         & &  &  &  & & & & & & & & & & & & \\
  \rowcolor{Gray}
  RavenDB            & &  &  &  & & & & & & & & & & & & \\
  Redis              & & \citesupp{abramova_2014_exp} & \citesupp{abramova_2014_which,abramova_2014_exp} &  & & & & & & & & \citesupp{ceresnak_2019} & & & \citesupp{abramova_2014_exp} & \citesupp{rabl_2012} \\
  \rowcolor{Gray}
  RethinkDB          & &  &  &  & & & & & & & & & & & & \\
  Riak               & &  &  &  & & & & & & & & & & & & \\
  \rowcolor{Gray}
  Scalaris           & & \citesupp{abramova_2014_exp} & \citesupp{abramova_2014_exp} & & & & & & & & & &  & & \citesupp{abramova_2014_exp} & \\
  SQL Server         & \citesupp{bassil_2012,ceresnak_2019,ilic_202} &  & &  & \citesupp{vershinin_2021} & & \citesupp{Li_2013} & & & & & & & & & \\
  \rowcolor{Gray}
  SQLite             & &  &  &  & & & & & & & & \citesupp{ahmed_2017} & & & & \\
  Tarantool          & & \citesupp{abramova_2014_exp} & \citesupp{abramova_2014_exp} &  & & & & &  & & & & & & \citesupp{abramova_2014_exp} & \\
  \rowcolor{Gray}
  Voldemort          & &  & \citesupp{abramova_2014_exp} &  & & & & & & & & & &  &  & \citesupp{rabl_2012} \\
  VoltDB             & &  &  &  & & & & & & & & & \citesupp{wang_2015} & & & \\
  \bottomrule
  \end{tabularx}
\end{table}

\begin{table}
  \centering
  \footnotesize
  \caption{DBMS performance comparisons in \textit{write} operations; a table cell shows that according to the study or studies cited, the DBMS in the row outperformed the DBMS in the column; subscripts (d)elete, (i)nsert and (u)pdate refer to the operations tested; citations without subscripts refer either to all three write operations, or to undisclosed general write operations}
  \label{table-results-write-1}
  \setlength\tabcolsep{3pt} 
  \begin{tabularx}{\linewidth}{r|c!{\color{Gray}\vrule}c!{\color{Gray}\vrule}c!{\color{Gray}\vrule}c!{\color{Gray}\vrule}c!{\color{Gray}\vrule}c!{\color{Gray}\vrule}Y!{\color{Gray}\vrule}Y!{\color{Gray}\vrule}Y!{\color{Gray}\vrule}Y!{\color{Gray}\vrule}c!{\color{Gray}\vrule}c!{\color{Gray}\vrule}c!{\color{Gray}\vrule}c!{\color{Gray}\vrule}c}
  \toprule
    & 
    \rotatebox{270}{Access} &
    \rotatebox{270}{ArangoDB} &
    \rotatebox{270}{Azure DDB} &
    \rotatebox{270}{Azure SQL DB} &
    \rotatebox{270}{Caché} & 
    \rotatebox{270}{Interbase} & 
    \rotatebox{270}{Cassandra} &
    \rotatebox{270}{CockroachDB} &
    \rotatebox{270}{CouchDB} & 
    \rotatebox{270}{Couchbase} &
    \rotatebox{270}{DB/2} & 
    \rotatebox{270}{Db4o} &
    \rotatebox{270}{Elasticsearch} &
    \rotatebox{270}{Firebase} &
    \rotatebox{270}{H2} \\ 
    \midrule
  \rowcolor{Gray}
  Access             &  &  & & & & & & & & & & & & & \\
  ArangoDB           &  &  & & & & & & & \citesupp{gunawan_2019,mavrogiorgos_2021} & & & & & & \\
  \rowcolor{Gray}
  Azure DDB          &  &  &  & \citesupp{baralis_2017} & & & & & & & & & & & \\
  Azure SQL DB       &  &  & &  & & & & & & & & & & & \\
  \rowcolor{Gray}
  Caché              &  &  &  & & & & & & & & & & & & \\
  Interbase          &  &  & & & & & & & & & & & & & \\
  \rowcolor{Gray}
  Cassandra          &  &  & & & & & & & \citesupp{Li_2013}\textsubscript{di} & & & & \citesupp{abramova_2014_exp}\citesupp{magdum_2018}\textsubscript{i} & & \\ 
  CockroachDB        &  &  & & & & & & & & & & & & & \\
  \rowcolor{Gray}
  CouchDB            &  &  & & & & & & & & \citesupp{truica_2015} & & & & & \\
  Couchbase          &  &  & & & & & \citesupp{tang_2016}\citesupp{Li_2013}\textsubscript{di} & & \citesupp{Li_2013}\textsubscript{di} & & & & & & \\
  \rowcolor{Gray}
  DB/2               & \citesupp{bassil_2012}\textsubscript{du} & & & & & & & & & & & & & & \\
  Db4o               &  &  & & & & & & & & & & & & & \\
  \rowcolor{Gray}
  Elasticsearch      &  &  & & & & & \citesupp{magdum_2018}\textsubscript{du} & & & & & & & & \\
  Firebase           &  &  & & & & & & & & & & & & & \\
  \rowcolor{Gray}
  H2                 &  &  & & & & & \citesupp{kabakus_2017} & & & & & & & & \\
  HBase              &  &  & & & & & \citesupp{Gandini_2014,kashyap_2013,rabl_2012,tang_2016} & & & & & & \citesupp{abramova_2014_exp} & & \\
  \rowcolor{Gray}
  Hypertable         &  &  & & & & & & & & & & & & & \\
  MariaDB            &  &  & & & & & & & & & & & & & \\
  \rowcolor{Gray}
  memcached          &  &  & & & & & \citesupp{kabakus_2017} & & & & & & & & \citesupp{kabakus_2017} \\
  MemSQL             &  &  & & & & & & \citesupp{schreiner_2019}\citesupp{Kaur_2017}\textsubscript{iu} & & & & & & & \\
  \rowcolor{Gray}
  MongoDB            &  & \citesupp{gunawan_2019}\textsubscript{du} & & & & & \citesupp{Gandini_2014,seghier_2021,swaminathan_2016,tang_2016,van_der_Veen_2012} \citesupp{araujo_2021,ceresnak_2019,Li_2013}\textsubscript{di} \citesupp{ceresnak_2019}\textsubscript{u} & & \citesupp{gunawan_2019,kumar_2017}\citesupp{Li_2013,mavrogiorgos_2021}\textsubscript{di} & \citesupp{tang_2016,truica_2015}\citesupp{pereira_2018}\textsubscript{i} & & & & &  \\ 
  
  MySQL              & \citesupp{bassil_2012}\textsubscript{du} &  & & & & \citesupp{afolabi_2008}\textsubscript{i} & \citesupp{rabl_2012} & & \citesupp{gyorodi_2020}\textsubscript{u} & \citesupp{truica_2015}\textsubscript{u} & \citesupp{bassil_2012}\textsubscript{du} & \citesupp{Roopak_2013}\textsubscript{d} & & & \\
  \rowcolor{Gray}
  Neo4J              &  &  & & & & & & & & & & & & & \\
  NuoDB              &  &   & & & & & & \citesupp{schreiner_2019}\citesupp{Kaur_2017}\textsubscript{iu} & & & & & & & \\
  \rowcolor{Gray}
  Oracle DB          & \citesupp{bassil_2012}\textsubscript{du} &  & & & & & & & & & \citesupp{bassil_2012}\textsubscript{du} & & & & \\
  Oracle NoSQL       &  &   & & & & & & & & & & & & & \\
  \rowcolor{Gray}
  OrientDB           &  &   & & & & & & & & & & & \citesupp{abubakar_2014}\textsubscript{iu} & & \\
  PostgreSQL         &  &  &  & & & & \citesupp{Fraczek_2017} & & & \citesupp{truica_2015}\textsubscript{u} & & & & & \\
  \rowcolor{Gray}
  RavenDB            &  &  &  & & & & & & \citesupp{Li_2013}\textsubscript{i} & & & & & & \\
  Redis              &  &  &  & & & & \citesupp{abramova_2014_which,abramova_2014_exp,kabakus_2017,rabl_2012,seghier_2021,tang_2016} \citesupp{ceresnak_2019}\textsubscript{i} & & & \citesupp{tang_2016} & & & \citesupp{abramova_2014_exp} & & \citesupp{kabakus_2017} \\ 
  \rowcolor{Gray}
  RethinkDB          &  &  &  & & & & & & & \citesupp{pereira_2018}\textsubscript{i} & & & & & \\
  Riak               &  &  &  & & & & & & & & & & & & \\
  \rowcolor{Gray}
  Scalaris           &  &  &  & & & & & & & & & & \citesupp{abramova_2014_exp} & & \\
  SQL Server         & \citesupp{bassil_2012}\textsubscript{du} & &  & & & & \citesupp{Li_2013}\textsubscript{d} & & \citesupp{Li_2013}\textsubscript{di} & \citesupp{truica_2015}\textsubscript{du} & \citesupp{bassil_2012}\textsubscript{du} & & & & \\ 
  \rowcolor{Gray}
  SQLite             &  & &  & & & & & & & & & & & & \\
  Tarantool          &  & &  & & & & \citesupp{abramova_2014_exp} & & & & & & \citesupp{abramova_2014_exp} & & \\
  \rowcolor{Gray}
  Voldemort          &  &  &  & & & & \citesupp{rabl_2012} & & & & & & & & \\
  VoltDB             &  &  &  & & & & & \citesupp{schreiner_2019}\citesupp{Kaur_2017}\textsubscript{iu} & & & & & & & \\
  \bottomrule
  \end{tabularx}
\end{table}

\begin{table}
  \centering
  \footnotesize
  \caption*{Table~\ref{table-results-write-1}. (cont.)}
  \label{table-results-write-2}
  \setlength\tabcolsep{3pt} 
  \begin{tabularx}{\linewidth}{r|Y!{\color{Gray}\vrule}c!{\color{Gray}\vrule}c!{\color{Gray}\vrule}c!{\color{Gray}\vrule}c!{\color{Gray}\vrule}Y!{\color{Gray}\vrule}Y!{\color{Gray}\vrule}c!{\color{Gray}\vrule}Y!{\color{Gray}\vrule}c!{\color{Gray}\vrule}c!{\color{Gray}\vrule}}
  \toprule
    & 
    \rotatebox{270}{HBase} &
    \rotatebox{270}{Hypertable} &
    \rotatebox{270}{MariaDB} &
    \rotatebox{270}{memcached} &
    \rotatebox{270}{MemSQL} &
    \rotatebox{270}{MongoDB} &
    \rotatebox{270}{MySQL} &
    \rotatebox{270}{NuoDB} &
    \rotatebox{270}{Oracle DB} &
    \rotatebox{270}{Oracle NoSQL} &
    \rotatebox{270}{OrientDB} \\
    \midrule
  \rowcolor{Gray}
  Access             & & & & &   & & & & & & \\
  ArangoDB           & & & & &   & \citesupp{mavrogiorgos_2021}\textsubscript{du} & & & & & \\
  \rowcolor{Gray}
  Azure DDB          & & & & &   & & & & & & \\
  Azure SQL DB       & & & & &   & & & & & & \\
  \rowcolor{Gray}
  Caché              & & & & &   & & & & \citesupp{Haiyan_2010} & & \\
  Interbase          & & & & &   & & \citesupp{afolabi_2008}\textsubscript{du} & & & & \\
  \rowcolor{Gray}
  Cassandra          & \citesupp{Hajjaji_2018,hassan_2018,swaminathan_2016}\citesupp{hendawi_2018}\textsubscript{iu} & \citesupp{Li_2013}\textsubscript{di} & &  & & \citesupp{abramova_2014_which,abramova_2013,abramova_2014_exp,Fraczek_2017,Hajjaji_2018,kabakus_2017,klein_2015,Samanta_2018}\citesupp{hendawi_2018}\textsubscript{iu}\citesupp{araujo_2021}\textsubscript{u} & \citesupp{ceresnak_2019,murazza_2016} & & \citesupp{ceresnak_2019} & \citesupp{abramova_2014_exp} & \citesupp{abramova_2014_which,abramova_2014_exp} \\
  CockroachDB        & & & & &   & & & & & & \\
  \rowcolor{Gray}
  CouchDB            & & & & &   & \citesupp{truica_2015}\textsubscript{du}\citesupp{mavrogiorgos_2021}\textsubscript{u} & \citesupp{truica_2015} & & & &  \\
  Couchbase          & \citesupp{tang_2016} & \citesupp{Li_2013}\textsubscript{di} & & &   & \citesupp{Li_2013}\textsubscript{di} \citesupp{pereira_2018}\textsubscript{du} \citesupp{gyorodi_2020}\textsubscript{d} \citesupp{chopade_2017}\textsubscript{i} & \citesupp{truica_2015} \textsubscript{di} & & & & \\ 
  \rowcolor{Gray}
  DB/2               & & & & &   & & & & & & \\
  Db4o               & & & & &   & & \citesupp{kulshrestha_2014}\citesupp{Roopak_2013}\textsubscript{i} & & & & \\
  \rowcolor{Gray}
  Elasticsearch      & & & & &   & \citesupp{abramova_2014_exp} & & & & \citesupp{abramova_2014_exp} & \citesupp{abramova_2014_exp} \\
  Firebase           & & & & &   & & \citesupp{ohyver_2019} & & & & \\
  \rowcolor{Gray}
  H2                 & & & & &   & \citesupp{kabakus_2017} & & & & & \\
  HBase              & & & & &   & \citesupp{abramova_2014_which,abramova_2014_exp,Hajjaji_2018,yassien_2016}\citesupp{hendawi_2018}\textsubscript{iu} & \citesupp{Ding_2012,rabl_2012,yassien_2016} & & & \citesupp{abramova_2014_exp} & \citesupp{abramova_2014_which,abramova_2014_exp} \\
  \rowcolor{Gray}
  Hypertable         & & & & &   & & & & & & \\
  MariaDB            & & & & &   & & & & & & \\ 
  \rowcolor{Gray}
  memcached          & & & & &   & \citesupp{kabakus_2017} & & & & & \\
  MemSQL             & & & & &   & & & \citesupp{schreiner_2019}\citesupp{Kaur_2017}\textsubscript{u} & & & \\
  \rowcolor{Gray}
  MongoDB            & \citesupp{swaminathan_2016,tang_2016} & \citesupp{Li_2013}\textsubscript{di} & & &   & & \citesupp{Abdullah_2015,ceresnak_2019,chickerur_2015,damodaran_2016,deari_2018,Filip_2020,gomes_2021,gyorodi_2015,nyati_2013,rautmare_2016}\citesupp{aghi_2015,Eyada_2020,jose_2020,Patil_2017,truica_2015,Wei_2011,yassien_2016}\textsubscript{i} & & \citesupp{Boicea_2012,ceresnak_2019}\citesupp{shetty_2019}\textsubscript{di} & & \citesupp{abramova_2014_which,abramova_2014_exp}  \\ 
  MySQL              & \citesupp{Gandini_2014} & & \citesupp{Tongkaw_2016} & &  & \citesupp{Fatima_2016}\citesupp{andjelic_2008}\textsubscript{di}\citesupp{wang_2015}\textsubscript{i} & & & \citesupp{ceresnak_2019} & & \\
  \rowcolor{Gray}
  Neo4J              & & & & &   & & & & \citesupp{lazarska_2019} & & \\  
  NuoDB              & & & & & \citesupp{Kaur_2017}\textsubscript{i} & & & & & & \\
  \rowcolor{Gray}
  Oracle DB          & & & & &   & \citesupp{shetty_2019}\textsubscript{u} & \citesupp{bassil_2012}\textsubscript{du}\citesupp{poljak_2017}\textsubscript{iu} & & & & \\
  Oracle NoSQL       & & & & &   & \citesupp{abramova_2014_exp} & & & & & \citesupp{abramova_2014_exp} \\
  \rowcolor{Gray}
  OrientDB           & & & & &   & & & & & & \\
  PostgreSQL         & & & & &   & \citesupp{Fraczek_2017} & \citesupp{murazza_2016}\citesupp{truica_2015}\textsubscript{di} \citesupp{poljak_2017}\textsubscript{iu} & & & & \\
  \rowcolor{Gray}
  RavenDB            & & & & &   & & & & & & \\
  Redis              & \citesupp{tang_2016} & & \citesupp{puangsaijai_2017}\textsubscript{du} & &  & \citesupp{abramova_2014_which,abramova_2014_exp,kabakus_2017,seghier_2021,tang_2016}\citesupp{pereira_2018}\textsubscript{u} & \citesupp{ceresnak_2019,rabl_2012} & & \citesupp{ceresnak_2019} & \citesupp{abramova_2014_exp} & \citesupp{abramova_2014_which,abramova_2014_exp} \\
  \rowcolor{Gray}
  RethinkDB          & \citesupp{rabl_2012} & & & &   & & & & & & \\
  Riak               & & & & &   & \citesupp{klein_2015} & & & & & \\
  \rowcolor{Gray}
  Scalaris           & & & & &   & \citesupp{abramova_2014_exp} & & & & \citesupp{abramova_2014_exp} & \citesupp{abramova_2014_exp} \\
  SQL Server         & & \citesupp{Li_2013}\textsubscript{d} & & & & \citesupp{Samanta_2018} & \citesupp{truica_2015}\textsubscript{di}\citesupp{bassil_2012,saikia_2015}\textsubscript{du} & & \citesupp{ceresnak_2019}\citesupp{bassil_2012}\textsubscript{du} & &  \\ 
  \rowcolor{Gray}
  SQLite             & & & & &   & & & & & & \\
  Tarantool          & & & & &   & \citesupp{abramova_2014_exp} & & & & \citesupp{abramova_2014_exp} & \citesupp{abramova_2014_exp} \\
  \rowcolor{Gray}
  Voldemort          & \citesupp{rabl_2012} & & & &   & \citesupp{abramova_2014_exp} & \citesupp{rabl_2012} & & & & \citesupp{abramova_2014_exp} \\
  VoltDB             & & & & &   & \citesupp{Fatima_2016}\citesupp{wang_2015}\textsubscript{i} & \citesupp{Fatima_2016} & & & & \\
  \bottomrule
  \end{tabularx}
\end{table}

\begin{table}
  \centering
  \footnotesize
  \caption*{Table~\ref{table-results-write-1}. (cont.)}
  \label{table-results-write-3}
  \setlength\tabcolsep{3pt} 
  \begin{tabularx}{\linewidth}{r|Y!{\color{Gray}\vrule}c!{\color{Gray}\vrule}c!{\color{Gray}\vrule}c!{\color{Gray}\vrule}c!{\color{Gray}\vrule}c!{\color{Gray}\vrule}Y!{\color{Gray}\vrule}c!{\color{Gray}\vrule}c!{\color{Gray}\vrule}c!{\color{Gray}\vrule}c}
  \toprule
    &
    \rotatebox{270}{PostgreSQL} &
    \rotatebox{270}{RavenDB} & 
    \rotatebox{270}{Redis} &
    \rotatebox{270}{RethinkDB} &
    \rotatebox{270}{Riak} &
    \rotatebox{270}{Scalaris} &
    \rotatebox{270}{SQL Server} &
    \rotatebox{270}{SQLite} &
    \rotatebox{270}{Tarantool} &
    \rotatebox{270}{Voldemort} &
    \rotatebox{270}{VoltDB} \\
    \midrule
  \rowcolor{Gray}
  Access             & & &  & & & & & & & & \\
  ArangoDB           & & &  & & & & & & & & \\
  \rowcolor{Gray}
  Azure DDB          & & &  & & & & & & & & \\
  Azure SQL DB       & & &  & & & & & & & & \\
  \rowcolor{Gray}
  Caché              & & &  & & & & & & & & \\
  Interbase          & & &  & & & & & & & & \\
  \rowcolor{Gray}
  Cassandra          & \citesupp{murazza_2016,van_der_Veen_2012} & \citesupp{Li_2013}\textsubscript{di} & \citesupp{ceresnak_2019}\textsubscript{di} & & \citesupp{klein_2015} & \citesupp{abramova_2014_exp} & \citesupp{ceresnak_2019}\citesupp{Li_2013}\textsubscript{i} & & & \citesupp{abramova_2014_exp} & \citesupp{rabl_2012} \\
  CockroachDB        & & &  & & & & & & & & \\
  \rowcolor{Gray}
  CouchDB            & \citesupp{truica_2015} & &  & & & & \citesupp{truica_2015} & & & & \\
  Couchbase          & \citesupp{truica_2015}\textsubscript{d} & \citesupp{Li_2013}\textsubscript{di} &  & \citesupp{pereira_2018}\textsubscript{du} & & & \citesupp{Li_2013}\textsubscript{di}\citesupp{truica_2015}\textsubscript{i} & & & & \\
  \rowcolor{Gray}
  DB/2               & & &  & & & & & & & & \\
  Db4o               & & &  & & & & & & & & \\
  \rowcolor{Gray}
  Elasticsearch      & & &  & & & & & & & \citesupp{abramova_2014_exp} & \\
  Firebase           & & &  & & & & & & & & \\
  \rowcolor{Gray}
  H2                 & & &  & & & & & & & & \\
  HBase              & & &  & & & \citesupp{abramova_2014_exp} & & & & \citesupp{abramova_2014_exp} & \citesupp{rabl_2012} \\
  \rowcolor{Gray}
  Hypertable         & & \citesupp{Li_2013}\textsubscript{di} &  & & & & \citesupp{Li_2013}\textsubscript{di} & & & & \\
  MariaDB            & & &  & & & & & & & & \\
  \rowcolor{Gray}
  memcached          & & & \citesupp{kabakus_2017} & & & & & & & & \\
  MemSQL             & & &  & & & & & & & & \citesupp{schreiner_2019}\citesupp{Kaur_2017}\textsubscript{iu} \\
  \rowcolor{Gray}
  MongoDB            & \citesupp{jung_2015,truica_2015,van_der_Veen_2012} & \citesupp{Li_2013}\textsubscript{di} & \citesupp{ceresnak_2019} & \citesupp{pereira_2018}\textsubscript{di} & & & \citesupp{aboutorabi_2015,ceresnak_2019,truica_2015}\citesupp{Li_2013}\textsubscript{di}\citesupp{parker_2013}\textsubscript{u} & \citesupp{wang_2015} & & & \\
  
  MySQL              & \citesupp{truica_2015}\textsubscript{u}& &  & & & & \citesupp{ceresnak_2019}\citesupp{saikia_2015}\textsubscript{i}\citesupp{truica_2015}\textsubscript{u} & \citesupp{wang_2015} & & & \citesupp{rabl_2012} \\
  \rowcolor{Gray}
  Neo4J              & &  &  & & & & & & & & \\
  NuoDB              & & &  & & & & & & & & \citesupp{Kaur_2017}\textsubscript{iu} \\
  \rowcolor{Gray}
  Oracle DB          & \citesupp{poljak_2017}\textsubscript{iu} & &  & & & & & & & & \\
  Oracle NoSQL       & & &  & & & & & & & \citesupp{abramova_2014_exp} & \\
  \rowcolor{Gray}
  OrientDB           & & & \citesupp{abubakar_2014}\textsubscript{iu} & & & & & & & & \\
  PostgreSQL         & & &  & & & & \citesupp{truica_2015}\textsubscript{i} & & & & \\
  \rowcolor{Gray}
  RavenDB            & & &  & & & & & & & & \\
  Redis              & & &  & &  & \citesupp{abramova_2014_exp} & \citesupp{ceresnak_2019} & & & \citesupp{abramova_2014_exp,rabl_2012} & \citesupp{rabl_2012} \\
  \rowcolor{Gray}
  RethinkDB          & & &  & & & & & & & & \\
  Riak               & & &  & & & & & & & & \\
  \rowcolor{Gray}
  Scalaris           & & &  & & & & & & & \citesupp{abramova_2014_exp} & \\
  SQL Server         & \citesupp{truica_2015}\textsubscript{d} & \citesupp{Li_2013}\textsubscript{di} &  & & & & & & & & \\
  \rowcolor{Gray}
  SQLite             & & &  & & & & & & & & \\
  Tarantool          & & &  & & &  \citesupp{abramova_2014_exp} & & & & \citesupp{abramova_2014_exp} & \\
  \rowcolor{Gray}
  Voldemort          & & &  & & & & & & & & \citesupp{rabl_2012} \\
  VoltDB             & & &  & & & & & \citesupp{wang_2015} & & & \\
  \bottomrule
  \end{tabularx}
\end{table}

\end{document}